\documentclass[a4paper,12pt]{article}

\usepackage[a4paper,textwidth=420pt]{geometry}
\usepackage{jheppub}

\usepackage[T1]{fontenc}          % Modern font encoding
\usepackage[utf8]{inputenc}       % UTF-8 support

\usepackage{amsmath, amssymb, amsfonts, mathrsfs}
\usepackage{physics}
\usepackage{slashed}
\usepackage{cancel}
\usepackage{bbold}
\usepackage{bm}

\usepackage{graphicx, subcaption, float, wrapfig}
\usepackage{epsfig, epsf}
\usepackage{tikz}
\usepackage{tikz-feynman}
\usetikzlibrary{math, arrows, matrix, positioning, decorations.pathreplacing}
\usepackage{feynmp-auto}

\usepackage{multirow}
\usepackage{array}

\definecolor{MyOrange}{HTML}{d55e00}
\definecolor{MyGreen}{HTML}{009e73}
\definecolor{MyBlue}{HTML}{0072b2}
\definecolor{MyMagenta}{HTML}{d28cb3}
\usepackage[normalem]{ulem} % for underlining/emphasis with \uline

\usepackage{setspace}
\usepackage{indentfirst}
\edef\restoreparindent{\parindent=\the\parindent\relax}
\usepackage{parskip}
\restoreparindent

\tikzset{
    vector/.style={decorate, decoration={snake}, draw},
    fermion/.style={postaction={decorate},
        decoration={markings,mark=at position .55 with {\arrow{>}}}},
    fermionbar/.style={draw, postaction={decorate},
        decoration={markings,mark=at position .55 with {\arrow{<}}}},
    fermionnoarrow/.style={},
    gluon/.style={decorate,
        decoration={coil,amplitude=4pt, segment length=5pt}},
    scalar/.style={dashed, postaction={decorate},
        decoration={markings,mark=at position .55 with {\arrow{>}}}},
    scalarbar/.style={dashed, postaction={decorate},
        decoration={markings,mark=at position .55 with {\arrow{<}}}},
    scalarnoarrow/.style={dashed,draw},
    vectorscalar/.style={loosely dotted,draw=black, postaction={decorate}},
}

\newcommand{\HL}{H_L} 
\newcommand{\HR}{H_R} 
\newcommand{\X}{X} 
\newcommand{\Y}{\tilde{X}}
\newcommand{\x}{x}
\newcommand{\y}{\tilde{x}}

\newcommand{\cmat}{\Omega}

\newcommand{\xD}{x_{D}} 
\newcommand{\xDelta}{x_{\Delta}} 
\newcommand{\yU}{\tilde{x}_{U}} 
\newcommand{\yQ}{\tilde{x}_{\mathcal{Q}}} 
\newcommand{\yT}{\tilde{x}_{T}} 
\newcommand{\yS}{\tilde{x}_{S}} 

\newcommand{\MD}{M_{D}} 
 
\newcommand{\MU}{\tilde{M}_{U}} 
 
\newcommand{\MT}{\tilde{M}_{T}} 
\newcommand{\MS}{\tilde{M}_{S}} 

\newcommand{\EqRef}[1]{Eq.\,(\textcolor{orange}{\ref{#1}})}
\newcommand{\FigRef}[1]{Fig.\,(\textcolor{orange}{\ref{#1}})}
\newcommand{\AppRef}[1]{App.\,(\textcolor{red}{\ref{#1}})}
\newcommand{\SecRef}[1]{Sec.\,(\textcolor{red}{\ref{#1}})}

\newcommand{\GeV}{{\rm GeV}}
\newcommand{{\HPU}}{Higgs Parity unification}

\title{\Large \bf 
A Flavor of $SO(10)$ Unification with a Spinor Higgs
}

\author[1,2]{Juanca Carrasco-Martinez,}\emailAdd{jc.carrasco@berkeley.edu }

\author[1,2]{~Lawrence J. Hall,}\emailAdd{ljh@berkeley.edu}

\author[3,4,5]{~Keisuke
Harigaya,}\emailAdd{kharigaya@uchicago.edu }

\author[1,2]{Kevin Langhoff}\emailAdd{klanghoff@berkeley.edu}

\affiliation[1]{Leinweber Institute for Theoretical Physics, Department of Physics, University of California, Berkeley, CA 94720, USA}

\affiliation[2]{Theoretical Physics Group, Lawrence Berkeley National Laboratory, Berkeley, CA 94720, USA}

\affiliation[3]{Department of Physics, University of Chicago, Chicago, IL 60637, USA}

\affiliation[4]{Enrico Fermi Institute, Leinweber Institute for Theoretical Physics, and Kavli Institute for Cosmological Physics, University of Chicago,
Chicago, IL 60637, USA}

\affiliation[5]{Kavli Institute for the Physics and Mathematics of the Universe (WPI),
The University of Tokyo Institutes for Advanced Study,
The University of Tokyo, Kashiwa, Chiba 277-8583, Japan}

\abstract{We investigate \emph{{\HPU}}---a realization of $SO(10)$ grand unification based on the Higgs Parity mechanism in which the Standard Model (SM) Higgs resides in a spinor representation. The theory has an intermediate left-right symmetric stage where the $SU(2)_R$ symmetry breaking scale is fixed by the vanishing of the SM Higgs quartic coupling. The strong $CP$ problem is solved by parity. Gauge coupling unification successfully predicts $\alpha_s(M_Z)$ to within 1\%. The spinor Higgs naturally leads to a seesaw origin for SM flavor observables. We identify a novel mechanism where large mixing of third generation fermions with additional heavy vector-like fermions accounts for the anarchical nature of the PMNS matrix and the lack of hierarchy in the neutrino mass spectrum, relative to the up-quarks. A fit to quark and lepton masses and mixings, with a minimal parameter set, predicts
1) A testable relation between the top quark mass and $\alpha_s(M_Z)$ which is about $(1-2)\sigma$ from current best fit values, 
2) The order of magnitude of the baryon asymmetry of the universe, via leptogenesis from second-generation right-handed neutrino decays.
3) The proton decay and the neutron EDM are likely observable in next generation experiments, and
4) A normal ordered neutrino mass spectrum where $0\nu \beta \beta$ decay and the mass of the lightest neutrino are out of reach of next generation experiments. 
}

\begin{document}

\maketitle

\section{Introduction}

An $SO(10)$ theory at very high energy offers a remarkable prospect: the Standard Model (SM) gauge symmetries are unified into a simple Lie group and quarks and leptons of a generation are unified into a single spinor representation, $\psi(\bm{16})$. This idea, now 50 years old \cite{Georgi:1974my, Fritzsch:1974nn}, still provides the simplest understanding of the SM gauge group and fermion gauge quantum numbers. 

However, while unification of gauge couplings in the SM appears plausible \cite{Georgi:1974yf}, predicted correlations are not numerically convincing and require large threshold corrections. Alternatively, if $SO(10)$ first breaks to an intermediate gauge group and to the SM at some lower scale, gauge couplings can be unified, but the appearance of this unknown intermediate scale removes the possibility of a precise prediction among SM couplings. Finally, supersymmetric extensions of the SM yield precise gauge coupling unification if superpartners lie near the weak scale. Absence of experimental signs of weak-scale superpartners over the last few decades motivates considering alternatives to this remarkable solution to the hierarchy problem.

This paper studies an $SO(10)$ unified theory with a precise prediction of gauge coupling unification, \emph{{\HPU}} \cite{Hall:2019qwx}. The symmetry breaking is 
\begin{align}
    SO(10) \times CP \;\; 
    &
    \longrightarrow \;\; SU(3) \times SU(2)_L \times SU(2)_R \times U(1)_{B-L} \times \mathcal{P} \nonumber \\
    &\overset{v_R}\longrightarrow \;\; SU(3) \times SU(2)_L \times  U(1)_Y.
    \label{eq:SO10breaking}
\end{align}
The Higgs Parity framework \cite{Hall:2018let} fixes the intermediate scale, $v_R$, to be the where the Higgs quartic coupling vanishes and is computed using only measured SM parameters. Using PDG central values \cite{ParticleDataGroup:2024cfk} and ignoring threshold corrections, gauge coupling unification predicts the QCD coupling strength in terms of the top quark pole mass
\begin{align} \label{eq:alphaspred}
    \alpha_s(M_Z) \, = \, 0.1175 \; + 0.0008 \left( \frac{m_t - 172.6 \, \mbox{GeV}}{0.29 \, \mbox{GeV}} \right). 
\end{align}
This agrees with the measured value of $0.1180 \pm 0.0009$ within $1\sigma$\footnote{There are subtleties with interpreting hadron collider measurements of the top quark mass as they arise from fits to parameters in Monte Carlo generators and are not strictly related to any specific renormalization scheme. Additionally, inherent ambiguities from infrared QCD effects limit theoretical uncertainties on this measurement by $\sigma(m_t) \sim \Lambda_{\rm QCD}$. These factors alter only the interpretation, not the prediction. Future $e^+e^-$ colliders can directly measure the top mass in well-defined renormalization schemes (e.g.\ $\overline{\text{MS}}$) via a $t\bar{t}$ threshold scan, achieving $\sigma_{m_t} \approx 50~\text{MeV}$. 
}.

Over the last several decades, particle physics experiments discovered the SM quarks and leptons and measured their masses and mixings with increasing precision. However, we lack a Standard Model of Flavor, even a provisional one that could be discarded if a better theory appears. In unifying quarks and leptons of a generation into a single spinor field, $SO(10)$ offers the prospect of progress in flavor. Conventional $SO(10)$ scenarios for flavor embed the SM Higgs in a tensor representation, $\phi_T$, transforming as {\bf 10},  {\bf 120}, or {\bf 126} with Yukawa couplings $y_{ij}\psi_i \psi_j \phi_T$. However, a single Yukawa interaction cannot describe nature as it forces all SM Yukawa matrices to be proportional to a single flavor matrix $y_{ij}$. Therefore, additional interactions are required for a successful theory of flavor. For example, the SM Higgs may have components in several tensor multiplets or there may be higher-dimensional interactions involving $SO(10)$ breaking scalar fields \cite{Georgi:1978fu, Georgi:1979dq,Georgi:1979ga,Harvey:1980je, Harvey:1981hk, Dimopoulos:1991yz, Dimopoulos:1991za, Anderson:1993fe, Dutta:2009ij, BhupalDev:2011gi}. Furthermore, large Majorana masses for $\bar{\nu}$ are required, for example, from interactions with $\phi_T$ in a {\bf 126}. 

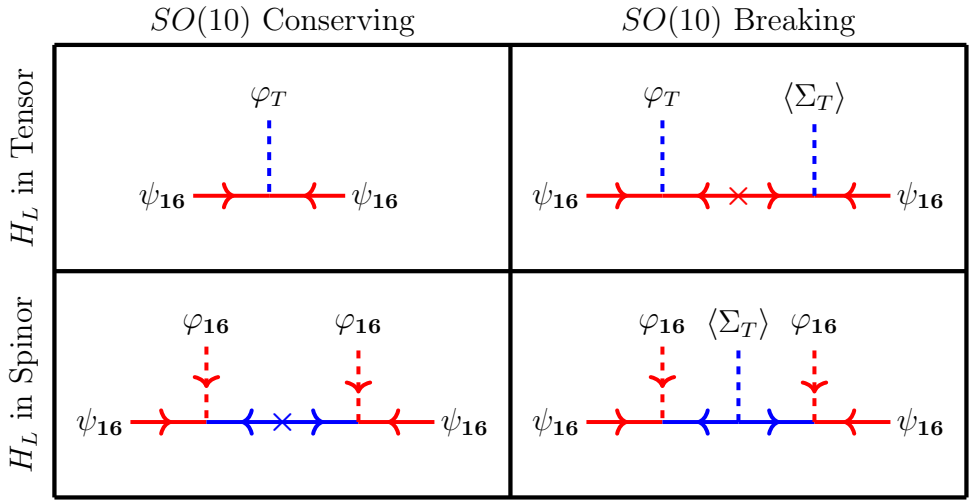
\begin{figure}[t]
    \centering
    \begin{tikzpicture}[line width=1.5 pt, scale=1.]
        %   
        % Draw the frame
        %
        \draw[ultra thick] (-2,1.5)-- (10,1.5);
        \draw[ultra thick] (-2,1.5)-- (-2,-4.5);
        \draw[ultra thick] (10,1.5)-- (10,-4.5);
        \draw[ultra thick] (-2,-1.5)-- (10,-1.5);
        \draw[ultra thick] (-2,-4.5)-- (10,-4.5);
        \draw[ultra thick] (4,-4.5)-- (4,1.5);
        %
        % Draw diagram (1,1)
        %
        \begin{scope}[xshift=2em, yshift = -.5cm]
            \draw[scalarnoarrow,blue] (90:1)--(0,0) ;
            \node at (90:1.3) {$\varphi_T$};
            \draw[fermion,red] (180:1)--(0,0);
            \draw[fermion,red] (0:1)--(0,0);
            \node at (180:1.4) {$\psi_{\bm{16}}$};
            \node at (0:1.4) {$\psi_{\bm{16}}$};
        \end{scope}
        %
        % Draw diagram (2,1)
        %
        \begin{scope}[xshift=6.0cm,yshift = -0.5cm]
            \draw[scalarnoarrow,blue] (90:1)--(0,0) ;
            \node at (90:1.3) {$\varphi_T$};
            \draw[fermion,red] (180:1)--(0,0);
            \draw[fermion,red] (0:1)--(0,0);
            \node at (180:1.4) {$\psi_{\bm{16}}$};
            \draw[fermion,red] (1,0)--(2,0);
            \draw[fermionbar,red] (2,0)--(3,0);
            \draw[scalarnoarrow,blue] (2,0)--(2,1) ;
            \node at (2,1.3) {$\expval{\Sigma_T}$};
            \node at (1,0) {\textcolor{red}{$\bm{\times}$}};
            \node at (3.4,0) {$\psi_{\bm{16}}$};
        \end{scope}
        %
        % Draw diagram (2,1)
        %
        \begin{scope}[xshift=0.0cm,yshift = -3.5cm]
            \draw[scalar,red] (90:1)--(0,0) ;
            \node at (90:1.3) {$\varphi_{\bm{16}}$};
            \draw[fermion,red] (180:1)--(0,0);
            \draw[fermion,blue] (0:1)--(0,0);
            \node at (180:1.4) {$\psi_{\bm{16}}$};
            \draw[fermion,blue] (1,0)--(2,0);
            \draw[fermionbar,red] (2,0)--(3,0);
            \draw[scalarbar,red] (2,0)--(2,1) ;
            \node at (2,1.3) {$\varphi_{\bm{16}}$};
            \node at (1,0) {\textcolor{blue}{$\bm{\times}$}};
            \node at (3.4,0) {$\psi_{\bm{16}}$};
        \end{scope}
        %
        % Draw diagram (2,2)
        %
        \begin{scope}[xshift=6.0cm,yshift = -3.5cm]
            \draw[scalar,red] (90:1)--(0,0) ;
            \node at (90:1.3) {$\varphi_{\bm{16}}$};
            \draw[fermion,red] (180:1)--(0,0);
            \draw[fermion,blue] (0:1)--(0,0);
            \node at (180:1.4) {$\psi_{\bm{16}}$};
            \draw[fermion,blue] (1,0)--(2,0);
            \draw[fermionbar,red] (2,0)--(3,0);
            \draw[scalarbar,red] (2,0)--(2,1) ;
            \draw[scalarnoarrow,blue] (1,0)--(1,1) ;
            \node at (1,1.3) {$\expval{\Sigma_{T}}$};
            \node at (2,1.3) {$\varphi_{\bm{16}}$};
            \node at (3.4,0) {$\psi_{\bm{16}}$};
        \end{scope}
        \node at (1,1.8) {$SO(10)$ Conserving};
        \node at (7,1.8) {$SO(10)$ Breaking};
        \node[rotate = 90] at (-2.4,0) {$\HL$ in Tensor};
        \node[rotate = 90] at (-2.4,-3) {$\HL$ in Spinor};
    \end{tikzpicture}
    \caption{
    Diagrams in the top (bottom) row generate SM flavor observables for a SM Higgs embedded in a tensor (spinor) representation of $SO(10)$. Diagrams in the right column show how $SO(10)$ violation enters flavor observables through an $SO(10)$ breaking VEV, $\expval{\Sigma_T}$. We draw particles in spinor representations in \textbf{\textcolor{red}{red}} and particles in tensor representations in \textbf{\textcolor{blue}{blue}}. Diagrams in the bottom row generate only flavor observables of the up and neutrino sector; down and charged-lepton flavor observables arise from diagrams where $\phi$ is replaced with $\phi^\dagger$.}
    \label{fig:flavdiags}
\end{figure}

Unlike conventional $SO(10)$ theories, {\HPU} {\it requires} the SM Higgs to lie in a spinor representation. The lowest-dimensional possibilities are {\bf 16}, {\bf 144} and {\bf 540}; in this paper we restrict ourselves to the simplest possibility, $\phi_{16}$. This dramatically changes the form of the flavor sector interactions, as illustrated diagrammatically in \FigRef{fig:flavdiags}. Diagrams leading to $SO(10)$ conserving and breaking contributions to SM flavor observables are shown for scenarios where the SM Higgs is embedded into a tensor and spinor representation. Much of this paper focuses on building a complete realistic $SO(10)$ theory of flavor with a Higgs in $\phi_{16}$. 

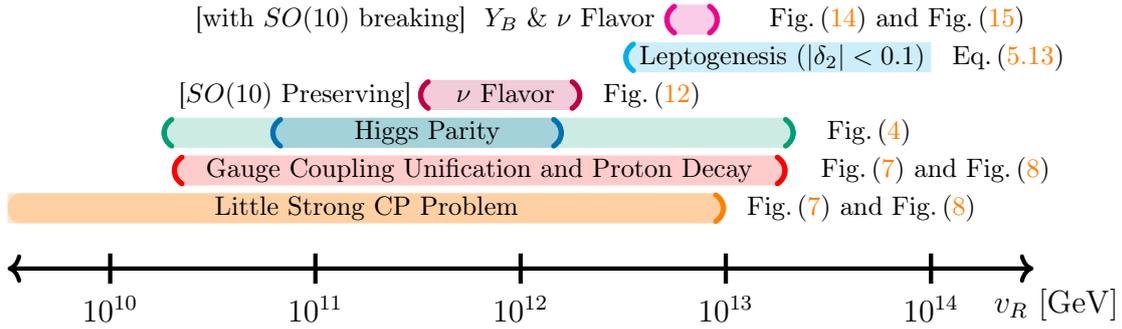
\begin{figure}[t] 
    \begin{tikzpicture}
        \def\s{2.7}
        \def\BarThickness{0.2}

        % Axis
        \draw[ultra thick,<->] (9.5*\s,-2.5) -- (14.5*\s,-2.5)
            node[below,xshift=0.3cm,yshift=-0.1cm] {$v_R~[\GeV]$};
        \foreach \i in {10,...,14} {
            \draw[ultra thick] (\i*\s,-2.3) -- (\i*\s,-2.7) node[below] {$10^{\i}$};
        }

        % -----------------------------
        % Gauge Coupling Unification and Proton Decay
        % -----------------------------
        \def\vRminUni{10.3}
        \def\vRmaxUni{13.3}
        \def\heightUni{-1.2}
        \def\colUni{red}

        \fill[opacity=0.2,\colUni,rounded corners=1ex]
            (\vRminUni*\s,\heightUni-\BarThickness) rectangle (\vRmaxUni*\s,\heightUni+\BarThickness);
        \draw[-), line width=0.7mm, \colUni] (\vRmaxUni*\s,\heightUni) -- ({(\vRmaxUni+0.001)*\s},\heightUni);
        \draw[(-, line width=0.7mm, \colUni] (\vRminUni*\s,\heightUni) -- ({(\vRminUni+0.001)*\s},\heightUni);
        \node at ({(\vRminUni+\vRmaxUni)/2*\s},\heightUni) {\footnotesize Gauge Coupling Unification and Proton Decay};
        \node[anchor=west, align=left] at ({\vRmaxUni*\s+0.3},\heightUni) {\footnotesize \FigRef{fig:Unification_Quality_Low_Mass} and \FigRef{fig:Unification_Quality_High_Mass}};

        % -----------------------------
        % Higgs Parity (2σ Band)
        % -----------------------------
        \def\vRminHP{10.25}
        \def\vRmaxHP{13.34}
        \def\heightHP{-0.7}
        \def\colHP{MyGreen}

        \fill[opacity=0.2, \colHP, rounded corners=1ex]
            (\vRminHP*\s,\heightHP-\BarThickness) rectangle (\vRmaxHP*\s,\heightHP+\BarThickness);
        \draw[-), line width=0.7mm, \colHP] (\vRmaxHP*\s,\heightHP) -- ({(\vRmaxHP+0.001)*\s},\heightHP);
        \draw[(-, line width=0.7mm, \colHP] (\vRminHP*\s,\heightHP) -- ({(\vRminHP+0.001)*\s},\heightHP);
        \node[anchor=west, align=left] at ({(0.1+\vRmaxHP)*\s},\heightHP) {\footnotesize \FigRef{fig:Higgs_Quartic_Running}};

        % -----------------------------
        % Higgs Parity (1σ Band)
        % -----------------------------
        \def\vRminHPP{10.78}
        \def\vRmaxHPP{12.21}
        \def\heightHPP{-0.7}
        \def\colHPP{MyBlue}

        \fill[opacity=0.2, \colHPP, rounded corners=1ex]
            (\vRminHPP*\s,\heightHPP-\BarThickness) rectangle (\vRmaxHPP*\s,\heightHPP+\BarThickness);
        \draw[-), line width=0.7mm, \colHPP] (\vRmaxHPP*\s,\heightHPP) -- ({(\vRmaxHPP+0.001)*\s},\heightHPP);
        \draw[(-, line width=0.7mm, \colHPP] (\vRminHPP*\s,\heightHPP) -- ({(\vRminHPP+0.001)*\s},\heightHPP);
        \node at ({(.1+\vRminHPP+\vRmaxHPP)/2*\s},\heightHPP) {\footnotesize Higgs Parity};

        % -----------------------------
        % Little Strong CP Problem
        % -----------------------------
        \def\vRminCP{9.5}
        \def\vRmaxCP{13}
        \def\heightCP{-1.7}
        \def\colCP{orange}

        \fill[opacity=0.2, \colCP, rounded corners=1ex]
            (\vRminCP*\s,\heightCP-\BarThickness) rectangle (\vRmaxCP*\s,\heightCP+\BarThickness);
        \fill[opacity=0.2, \colCP]
            (\vRminCP*\s,\heightCP-\BarThickness) --
            ({\vRmaxCP*\s-0.15},\heightCP-\BarThickness) to[out=20,in=-20]
            ({\vRmaxCP*\s-0.15},\heightCP+\BarThickness) --
            (\vRminCP*\s,\heightCP+\BarThickness) -- cycle;
        \draw[-), line width=0.7mm, \colCP] (\vRmaxCP*\s,\heightCP) -- ({(\vRmaxCP+0.001)*\s},\heightCP);
        \node at ({(\vRminCP+\vRmaxCP)/2*\s},\heightCP) {\footnotesize Little Strong CP Problem};

        \node[anchor=west, align=left] at ({(3.6+\vRminCP+\vRmaxCP)/2*\s},\heightCP) {\footnotesize \FigRef{fig:Unification_Quality_Low_Mass} and \FigRef{fig:Unification_Quality_High_Mass}};

        % -----------------------------
        % Neutrino Masses
        % -----------------------------
        \def\vRminNu{11.5}
        \def\vRmaxNu{12.3}
        \def\heightNu{-0.2}
        \def\colNu{purple}

        \fill[opacity=0.2, \colNu]
            ({\vRmaxNu*\s-0.15},\heightNu-\BarThickness) --
            ({\vRminNu*\s+0.15},\heightNu-\BarThickness) to[out=160,in=200]
            ({\vRminNu*\s+0.15},\heightNu+\BarThickness) --
            ({\vRmaxNu*\s-0.15},\heightNu+\BarThickness) to[out=20,in=-20] cycle;
        \draw[(-, line width=0.7mm, \colNu] (\vRminNu*\s,\heightNu) -- ({(\vRminNu+0.001)*\s},\heightNu);
        \draw[-), line width=0.7mm, \colNu] (\vRmaxNu*\s,\heightNu) -- ({(\vRmaxNu+0.001)*\s},\heightNu);
        \node at ({(-1.3+\vRminNu+\vRmaxNu )/2*\s},\heightNu) {\footnotesize [$SO(10)$ Preserving]~~~~ $\nu$ Flavor};
        \node[anchor=west, align=left] at ({(.9+\vRminNu+\vRmaxNu )/2*\s},\heightNu) {\footnotesize \FigRef{fig:Flavor_Fit_MUN}};

        % -----------------------------
        % Leptogenesis
        % -----------------------------
        \def\vRminLept{12.5}
        \def\vRmaxLept{14.0}
        \def\heightLept{0.3}
        \def\colLept{cyan}

        \fill[opacity=0.2, \colLept]
            (\vRmaxLept*\s,\heightLept-\BarThickness) --
            ({\vRminLept*\s+0.15},\heightLept-\BarThickness) to[out=160,in=200]
            ({\vRminLept*\s+0.15},\heightLept+\BarThickness) --
            (\vRmaxLept*\s,\heightLept+\BarThickness) -- cycle;
        \draw[(-, line width=0.7mm, \colLept] (\vRminLept*\s,\heightLept) -- ({(\vRminLept+0.001)*\s},\heightLept);
        \node at ({(-1.1 + \vRminLept+\vRmaxLept)/2*\s},\heightLept) {\footnotesize \hphantom{(w/o $\nu$ Flavor)}~~~~~~~Leptogenesis ($|\delta_2| < 0.1$)};
        \node[anchor=west, align=left] at ({(0.05+\vRmaxLept)*\s},\heightLept) {\footnotesize \EqRef{eq:Y_B_HPG}};

        % -----------------------------
        % Leptogenesis and nu Flavor
        % -----------------------------
        \def\vRminBoth{12.7}
        \def\vRmaxBoth{12.97}
        \def\heightBoth{0.8}
        \def\colBoth{magenta}

        \fill[opacity=0.2, \colBoth]
            ({\vRmaxBoth*\s-0.15},\heightBoth-\BarThickness) --
            ({\vRminBoth*\s+0.15},\heightBoth-\BarThickness) to[out=160,in=200]
            ({\vRminBoth*\s+0.15},\heightBoth+\BarThickness) --
            ({\vRmaxBoth*\s-0.15},\heightBoth+\BarThickness) to[out=20,in=-20] cycle;
        \draw[(-, line width=0.7mm, \colBoth] (\vRminBoth*\s,\heightBoth) -- ({(\vRminBoth+0.001)*\s},\heightBoth);
        \draw[-), line width=0.7mm, \colBoth] (\vRmaxBoth*\s,\heightBoth) -- ({(\vRmaxBoth+0.001)*\s},\heightBoth);
        \node at ({(-2.6+\vRminBoth+\vRmaxBoth )/2*\s},\heightBoth) {\footnotesize [with $SO(10)$ breaking] ~$Y_B$ \& $\nu$ Flavor};
        \node[anchor=west, align=left] at ({(.65+\vRminBoth+\vRmaxBoth )/2*\s},\heightBoth) {\footnotesize \FigRef{fig:Flavor_Fit_MUN_with_phase} and \FigRef{fig:AB_Changes}};

    \end{tikzpicture}
    \caption{Approximate ranges of parity breaking scale, $v_R$, from a variety of experimental and theoretical constraints. The prediction from Higgs Parity is shown for $1\sigma$ and $2\sigma$ uncertainties in $m_t$ and $\alpha_s$. Ranges corresponding to successfully accounting for neutrino flavor observables, leptogenesis, and both simultaneously are defined under different assumptions explained during their derivation. }
    \label{fig:scales}
\end{figure}

\begin{figure}[h]
    \centering
    \includegraphics[width=1\linewidth]{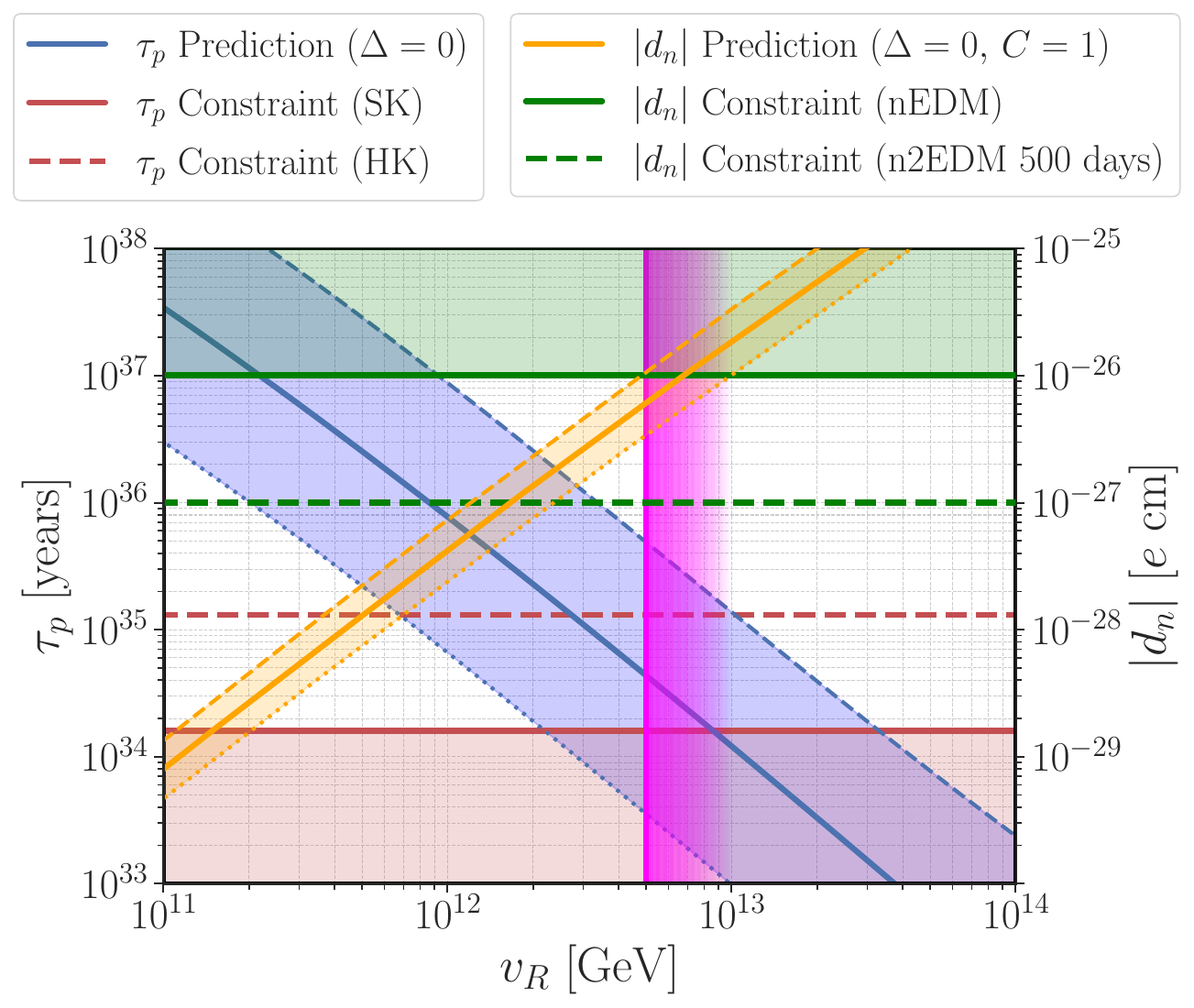}
    \caption{Predictions for the proton lifetime (\textbf{\textcolor{MyBlue}{blue}}, left axis) and the neutron EDM (\textbf{\textcolor{MyOrange}{orange}}, right axis) from Higgs Parity unification in the high mass scenario described in \SecRef{sec:The Higgs Parity Grand Unification Model}. In the low mass scenario, these predictions both decrease by about a factor of 2. The shaded band around each prediction corresponds to uncertainties from additional threshold corrections to gauge coupling unification of size $\Delta = 5/(8\pi^2)$ defined in \EqRef{eq:Delta}. Dashed (dotted) boundaries of these bands correspond to the largest (smallest) possible value of $M_{\rm GUT}$---defined as the mass of heavy gauge bosons that lead to proton decay. The \textbf{\textcolor{red}{red}} solid (dashed) lines show bounds (projections) on proton decay from Super-Kamiokande \cite{Super-Kamiokande:2020wjk} (Hyper-Kamiokande \cite{Hyper-Kamiokande:2018ofw}). The \textbf{\textcolor{MyGreen}{green}} solid (dashed) lines show bounds (projections) on the neutron EDM \cite{Abel:2020pzs,n2EDM:2021yah}. The \textbf{\textcolor{magenta}{magenta}} region shows values of $v_R$, obtained in \SecRef{sec:Leptogenesis}, that can simultaneously explain neutrino flavor observables and the baryon asymmetry, as extracted from the fit of \FigRef{fig:AB_Changes}. The shading across this band illustrates that greater tuning is needed at higher $v_R$ between $SO(10)$ breaking and preserving contributions to the second generation mass of $\Y$.  } 
    \label{fig:Union_Jack}
\end{figure}

Remarkably, as summarized in \FigRef{fig:scales},  {\HPU} simultaneously accounts for gauge coupling unification, proton decay constraints, the strong $CP$ problem, the baryon asymmetry of the universe, and SM flavor.  To simultaneously account for SM flavor and the baryon asymmetry, the parity breaking scale $v_R$ lies in the preferred range of $(6 \times 10^{12} - 10^{13})$ GeV. This requirement makes the exciting prediction that the proton lifetime and the neutron electric dipole moment are likely observable at next generation experiments shown in \FigRef{fig:Union_Jack}.

\subsubsection*{\emph{Outline}}

In \SecRef{sec:The Higgs Parity Grand Unification Model}
we define {\HPU}, review the Higgs Parity mechanism, and compute $v_R$ from measured SM parameters. We study gauge coupling unification, proton decay, and $CP$ violation in the strong sector.  

In \SecRef{sec:The Seesaw Origin of Flavor Observables} we study the seesaw origin of quark and lepton masses. This presents challenges for SM flavor observables, addressed in \SecRef{sec:directbtau}, where third generation mass eigenstates arise not from a seesaw, but directly from large mixing with vector-like fermions. Large mixing, or ``direct masses'', for the third generation provides an understanding of differences in hierarchies for neutrino and up-quark sectors.

In \SecRef{sec:Leptogenesis} we study leptogenesis from decays of second generation right-handed neutrinos. $SO(10)$ symmetry and having a single source of CP violation highly restricts parameter space for flavor and leptogenesis. Conclusions are drawn in \SecRef{sec:Conclusions}.  

\section{Higgs Parity Unification} \label{sec:The Higgs Parity Grand Unification Model}

In this section, we define {\HPU} and describe the process through which gauge symmetry is spontaneously broken. We illustrate the following:
\begin{enumerate}
    \item The Higgs parity mechanism identifies the scale where parity breaks spontaneously to be the scale at which the SM Higgs quartic vanishes.
    \item Gauge coupling unification is realized without tuning additional parameters.
    \item Predicted proton decay is consistent with current bounds but may be observed at future experiments. 
    \item Strong $CP$ violation is suppressed by parity to be below current bounds but may be observed at future experiments.
\end{enumerate}
We leave a detailed discussion of flavor to the following sections.

\subsection{Definition of Higgs Parity Unification} \label{subsec: Definition of the Higgs Parity Grand Unification Model}

We consider an $SO(10)$ gauge theory with an exact $CP$ symmetry. The fermion matter content consists of three flavor copies of
\begin{align}
    \text{Fermions = }\{\psi(\bm{16}),~ X(\bm{10}),~ \Y(\bm{45})\}.
\end{align}
The field $\psi$ contains all SM fermions and a SM singlet. $\X$ and $\Y$ are heavy vector-like fermions and the tilde is notation to identify the field in the $\bm{45}$; we later use a tilde to denote its masses and couplings. Integrating out these heavy vector-like fermions plays a crucial role in generating Standard Model flavor observables.

The scalar field content consists of
\begin{align}
    \text{Scalars = }\{ \varphi(\overline{\bm{16}}),~\Sigma(\bm{45}),~\mathcal{S}(\bm{54})\}.
\end{align}
The $CP$ odd scalar $\Sigma$ spontaneously breaks $SO(10)$ to $SU(3)_C \times SU(2)_L \times SU(2)_R\times U(1)_{\rm B-L}$ and $CP$ to $\mathcal{P}$, where $\mathcal{P}$ is both spacetime parity and the transformation $SU(2)_L \leftrightarrow SU(2)_R$. The $CP$ even scalar $\mathcal{S}$ gets a Pati-Salam singlet VEV, i.e. it would break $SO(10)$ to $SU(4) \times SU(2)_L \times SU(2)_R$ in the absence of $\Sigma$. Dynamics of these scalars is complicated and we do not describe the origin of these VEVs \cite{Bertolini:2009es}.

Color neutral components of $\varphi$ decompose into $SU(2)_R$ and $SU(2)_L$ doublets, $\HR$ and $\HL$, which eventually further break the gauge symmetry as
\begin{align}\label{eq:SSB}
    SO(10)\times CP&\overset{\expval{\Sigma}}{\longrightarrow} SU(3)_C \times SU(2)_L \times SU(2)_R\times U(1)_{\rm B-L}\times \mathcal{P}
    \\
    &\overset{\expval{ \HR}}{\longrightarrow} SU(3)_C \times SU(2)_L \times U(1)_{Y}
    \\ 
    &\overset{\expval{ \HL}}{\longrightarrow} SU(3)_C \times U(1)_{Q}.
\end{align}
At the first stage of symmetry breaking, fermions decompose into left-right symmetric representations (from now on called the ``3221'' theory) as shown in Table \eqref{tab:3221_representations}.
\begin{table}[h] % 'h' places the table approximately here
    \centering
    \small
    \renewcommand{\arraystretch}{1.4}
    \begin{tabular}{|c||c|c|c|c||c|c||c|c|c|c|c|c||c|c||} \hline 
                 & $q$           & $\ell$   & $\bar{q}$      & $\bar{\ell}$ & $\Delta$  & $D$ &$G$       & $T_L$       & $T_R$ & $S$ & $\mathcal{Q}$ & $U$            & $\HL$          & $\HR$ \\ \hline 
    $SU(3)_C$    & $\bm{3}$      & $\bm{1}$ & $\bar{\bm{3}}$ & $\bm{1}$     & $\bm{1}$  & $\bm{3}$       & $\bm{8}$  & $\bm{1}$  & $\bm{1}$  & $\bm{1}$       & $\bm{3}$     & $\bm{3}$     & $\bm{1}$ & $\bm{1}$     \\
    $SU(2)_L$    & $\bm{2}$      & $\bm{2}$ & $\bm{1}$       & $\bm{1}$     & $\bm{2}$  & $\bm{1}$       & $\bm{1}$  & $\bm{3}$  & $\bm{1}$  & $\bm{1}$       & $\bm{2}$     & $\bm{1}$     & $\bm{2}$ & $\bm{1}$     \\
    $SU(2)_R$    & $\bm{1}$      & $\bm{1}$ & $\bm{2}$       & $\bm{2}$     & $\bm{2}$  & $\bm{1}$       & $\bm{1}$  & $\bm{1}$  & $\bm{3}$  & $\bm{1}$       & $\bm{2}$     & $\bm{1}$     & $\bm{1}$ & $\bm{2}$       \\
    $U(1)_{B-L}$ & $\frac{1}{3}$ & $-1$     & $\frac{-1}{3}$ & $1$          & $0$       & $\frac{-2}{3}$ & $0$       & $0$       & $0$       & $0$            &$\frac{-2}{3}$& $\frac{4}{3}$& $1$      & $-1$         \\ \hline 
    $SO(10)$     & \multicolumn{4}{c||}{$\psi(\bm{16})$}& \multicolumn{2}{c||}{$\X (\bm{10})$} & \multicolumn{6}{c||}{$\Y(\bm{45})$}& \multicolumn{2}{c||}{$\varphi(\overline{\bm{16}})$} \\
    \hline
    \end{tabular}
    \normalsize
    \caption{Decomposition of $SO(10)$ gauge eigenstates to $3221$ gauge eigenstates.}
    \label{tab:3221_representations}
\end{table}

Fields $\mathcal{Q},\, U,\, D$ also have fields in their conjugate representations denoted $\bar{\mathcal{Q}},\, \bar{U},\, \bar{D}$. Hypercharge corresponds to the unbroken generator $Y = T^3_R +(B-L)/2$ after $SU(2)_R$ is broken by $\expval{\HR} = (v_R,\, 0)^\top$. Colored scalars in $\varphi$ are assumed to acquire a mass of order the unification scale, so the only light scalars are $H_{L,R}$. %
\footnote{The mass splitting between the colored scalars and $H_{L,R}$ can be achieved by tuning the coupling of $\Sigma$ and $\mathcal{S}$ to $\varphi$, as in the standard doublet-triplet splitting. Taking the colored-scalar masses to be much below the unification scale requires extra fine-tuning, since quantum corrections generate a mass squared of $\mathcal{O}(v_{\bf 45,54}^2/(16\pi^2))$ of the colored scalars. }

For $SU(2)_R$ singlets, the same symbol is used for both $U(1)_Y$ and $U(1)_{B-L}$ gauge eigenstates. Non-singlet $SU(2)_R$ states decompose into distinct SM gauge eigenstates as shown in Table \eqref{tab:321_representations}.

\begin{table}[h] % 'h' places the table approximately here
    \centering
    \small
    \renewcommand{\arraystretch}{1.4}
    \newcolumntype{P}[1]{>{\centering\arraybackslash}p{#1}}
    \begin{tabular}{|P{1.2cm}||P{.65cm}|P{.65cm}||P{.65cm}|P{.65cm}||P{.65cm}|P{.65cm}||P{.65cm}|P{.65cm}|P{.65cm}||P{.65cm}|P{.65cm}||}
\hline 
    &$\bar{d}$&$\bar{u}$&$\bar{e}$&$\bar{\nu}$&$\bar{L}$&$L$&$\bar{E}$&$N$&$E$&$Q$&$\mathscr{Q}$       \\ \hline 
    $SU(3)_C$ &$\bar{\bm{3}}$&$\bar{\bm{3}}$&$\bm{1}$&$\bm{1}$&$\bm{1}$&$\bm{1}$&$\bm{1}$&$\bm{1}$&$\bm{1}$&$\bm{3}$&$\bm{3}$\\
    $SU(2)_L$ &$\bm{1}$&$\bm{1}$&$\bm{1}$&$\bm{1}$&$\bm{2}$&$\bm{2}$&$\bm{1}$&$\bm{1}$&$\bm{1}$&$\bm{2}$&$\bm{2}$\\
    $U(1)_Y$ &$\frac{1}{3}$&$-\frac{2}{3}$&$1$&$0$&$\frac{1}{2}$&$-\frac{1}{2}$&$1$&$0$&$-1$&$\frac{1}{6}$&$-\frac{5}{6}$\\ \hline 
    $3221$     & \multicolumn{2}{c||}{$\bar{q}(\bm{3},\bm{1},\bm{2})_{-\frac{1}{3}}$}& \multicolumn{2}{c||}{$\bar{\ell}(\bm{1},\bm{1},\bm{2})_{1}$} & \multicolumn{2}{c||}{$\Delta(\bm{1},\bm{2},\bm{2})_{0} $}& \multicolumn{3}{c||}{$T_R(\bm{1},\bm{1},\bm{3})_{0}$} & \multicolumn{2}{c||}{$\mathcal{Q}(\bm{3},\bm{2},\bm{2})_{-\frac{2}{3}}$} \\
    \hline
    \end{tabular}
    \normalsize
    \caption{Decomposition of $3221$ gauge eigenstates into $321$ gauge eigenstates.}
    \label{tab:321_representations}
\end{table}

The SM Yukawa couplings arise from the following interactions:\footnote{Terms like $\psi X \varphi^\dagger$ have gauge indices contracted like $\varphi \mathcal{C} \Gamma^A \psi X_A$, but this clutters notation and distracts from our focus on flavor. We therefore drop such structure from our notation.} 
\begin{align}
    \mathscr{L}_0 = 
    \psi^{i} \x^{ij} \X^j \varphi^\dagger + \X^i M^{ij} \X^j + \psi^i \y^{ij} \Y^j \varphi + \Y^i \tilde{M}^{ij} \Y^j +
       {\rm h.c.}
    \label{eq:Lleading}
\end{align}
Additionally, $SO(10)$ breaking mass contributions for $X$ and $\Y$ are generated from
\begin{align} \label{eq:XMassesSigma}
    \mathscr{L}_{\cancel{SO(10)},\,M} = (i\, \kappa^{ij}\Sigma  + \beta^{ij} \mathcal{S}) \; \X^i \X^j \,   +(i\, \tilde{\kappa}^{ij} \Sigma + \tilde{\beta}^{ij}\mathcal{S} )\; \Y^i \Y^j  + {\rm h.c.}
\end{align}
$CP$ constrains all coefficients to be real as defined. $SO(10)$ invariance fixes $\kappa,\, \tilde{\kappa}$ to be anti-symmetric and  $\beta,\,\tilde{\beta}$ to be symmetric. When scalars get VEVs, normalized as $\Tr(\expval{\mathcal{S}}^2) = v_{\bm{54}}^2$ and $\Tr(\expval{\Sigma}^2) = v_{\bm{45}}^2$, masses are modified as shown in Table \eqref{tab:Xmass corrections}. Notably, $CP$ violating corrections from $\expval{\Sigma}$ do not affect the lepton sector obtained from integrating out $\Delta,\, S,\, T_{L,R}$ at this level. Therefore $CP$ is an accidental symmetry of the lepton sector.
\begin{table}[t]
%\centering
\renewcommand{\arraystretch}{1.4}
\begin{subtable}[t]{0.45\textwidth}
\centering
\begin{tabular}{|c|c|c|}
\hline
 & $\Delta M_D$ & $\Delta M_\Delta$ \\
\hline
$i\kappa \frac{v_{\bm{45}}}{\sqrt{6}}$ & 1 & 0 \\
$\beta \frac{v_{\bm{54}}}{\sqrt{15}}$  & 1 & $-\frac{3}{2}$ \\
\hline
\end{tabular}
\end{subtable}
\hspace{-.8cm}
\begin{subtable}[t]{0.45\textwidth}
\centering
\begin{tabular}{|c|c|c|c|c|c|}
\hline
 & $\Delta \tilde{M}_G$ & $\Delta \tilde{M}_U$ & $\Delta \tilde{M}_S$ & $\Delta\tilde{M}_{\mathcal{Q}}$ & $\Delta \tilde{M}_T$ \\
\hline
$i\tilde{\kappa} \frac{v_{\bm{45}}}{\sqrt{6}}$ & 0 & 1 & 0 & $\frac{1}{2}$ & 0 \\
$\tilde{\beta} \frac{v_{\bm{54}}}{\sqrt{15}}$  & 1 & 1 & 1 & $-\frac{1}{4}$ & $-\frac{3}{2}$ \\
\hline
\end{tabular}
\end{subtable}
\caption{Mass corrections to components of $\X$ and $\Y$ fields from $\expval{\bm{45}}$ and $\expval{\bm{54}}$.}
\label{tab:Xmass corrections}
\end{table}

$SO(10)$-invariant mass eigenvalues of $M$ and $\tilde{M}$ may lie well below the cutoff scale $\Lambda$ of the unified theory due to approximate flavor symmetries. These symmetries also suppress $SO(10)$-breaking mass terms, preventing large mass corrections. Therefore, $SO(10)$-breaking mass terms may introduce corrections of order $\mathcal{O}(v_{\bm{45},\bm{54}}/\Lambda) \lesssim 1$.

When $SO(10)$ breaks, gauge bosons in $(\bm{3},\bm{1},\bm{1})_{4/3}$ and $(\bm{3},\bm{2},\bm{2})_{-2/3}$ representations acquire masses through $\mathcal{L} \supset \frac{1}{2}g_{\rm GUT}^2\Tr\left([A_\mu,\expval{\Sigma}]^2\right) + \frac{1}{2}g_{\rm GUT}^2\Tr\left([A_\mu,\expval{\mathcal{S}}]^2\right)$. 
\begin{align} 
    g_{\rm GUT}^{-2} M^2_{(\bm{3},\bm{1},\bm{1})_{4/3}} &= \frac{2}{3} v_{\bm{45}}^2  \label{eq:311_Mass}\\
    g_{\rm GUT}^{-2}M^2_{(\bm{3},\bm{2},\bm{2})_{-2/3}} &= \frac{1}{6} v_{\bm{45}}^2  + \frac{5}{12} v_{\bm{54}}^2. \label{eq:322_Mass}
\end{align}
Gauge coupling unification depends on threshold corrections sensitive to the ratio
\begin{align}\label{eq:r}
    r \equiv \frac{M_{(\bm{3},\bm{1},\bm{1})_{4/3}}}{M_{(\bm{3},\bm{2},\bm{2})_{-2/3}}} = 2\left(1 + \frac{5}{2} \left( \frac{v_{\bm{54}}}{v_{\bm{45}}} \right)^2\right)^{-\frac{1}{2}}.
\end{align}
At dimension-5 we consider only operators which violate $CP$ at low energy\footnote{The notation of the previous footnote breaks down for  dimension-5 operators. Operators with $\X$ have two different gamma matrix structures; $\psi \mathcal{C} \Gamma^A  \Sigma^{AB} \X^B \varphi^\dagger$ and $\psi \mathcal{C} \Gamma^{ABC}  \Sigma^{AB} \X^C \varphi^\dagger$. Operators with $\Y$ have $\psi  \Sigma^{AB} \Y^{AB} \varphi$, $\psi \Gamma^{AB}  \Sigma^{BC} \Y^{CA} \varphi$ and $\psi \Gamma^{ABCD}  \Sigma^{AB} \Y^{CD} \varphi$. These all have unique Wilson coefficients. We don't require this level of detail. }:
\begin{align} \label{eq:Dim-5 Terms}
    \mathscr{L}_{\Sigma,5} = i \frac{c^{ij}}{\Lambda} \; \psi^i  \Sigma \X^j \, \varphi^\dagger   + i \frac{ \tilde{c}^{ij}}{\Lambda} \; \psi^i \Sigma \Y^j \, \varphi   + {\rm h.c.}
\end{align}
where $c^{ij}$ and $\tilde{c}^{ij}$ are real. These contribute $SO(10)$ and $CP$-violating to Yukawa couplings of \EqRef{eq:Lleading}, thus breaking the accidental $CP$ symmetry of the lepton sector.

A key feature is the theory lacks renormalizable terms bilinear in $\psi$ to generate SM Yukawa interactions. Dimension 5 operators $\psi^2 \phi^2/\Lambda$ induce Yukawa couplings no larger than $v_R/\Lambda$; requiring $\Lambda \gtrsim M_{\rm GUT}$ renders them too small to account for second and third generation masses. This motivates the introduction of $\X$ and $\Y$.

SM Yukawa couplings are then generated via a \textit{seesaw mechanism} by integrating out $\X$ and $\Y$, as discussed in \SecRef{sec:The Seesaw Origin of Flavor Observables}, similar to \textit{universal seesaw} models~\cite{Rajpoot:1987fca,Davidson:1987mh,Davidson:1987tr}. In particular, $y_t \sim \mathcal{O}(1)$ suggests that some of these fields have masses near $v_R$, necessitating care in integrating out heavy states; after $SU(2)_R$ breaking, mass eigenstates are linear combinations of gauge eigenstates of $\X$, $\Y$, and $\psi$.

Crucially, the light SM fermions may predominantly arise from $\X$ and $\Y$ components, allowing SM Yukawa couplings to be \textit{directly generated} through interactions in \EqRef{eq:Lleading}. We focus on this possibility for the third generation, referring to it as a \textit{direct third generation}. This structure plays a central role in explaining SM flavor observables and is detailed in \SecRef{sec:directbtau}.

\subsection{The Higgs Parity Mechanism}\label{subsec:Higgs Parity Mechanism}

The $SO(10)$ theory described above is a UV completion of the Higgs parity framework \cite{Hall:2018let,Hall:2019qwx}. With the Higgs fields that spontaneously break $SU(2)_L \times SU(2)_R\cross \mathcal{P}$ transforming as $\HL(\bm{2},\bm{1})$ and $\HR(\bm{1},\bm{2})$, the scale of parity breaking is determined to be where the SM Higgs quartic coupling vanishes. This prediction for $v_R$ has important consequences for gauge coupling unification and proton decay, which we discuss in the subsequent subsections. Details of the calculation of the running of the Higgs quartic described below are given in \AppRef{app:Gauge Coupling Unification in Higgs Parity GUTs}.

\begin{figure}
    \centering
    \includegraphics[width=\linewidth]{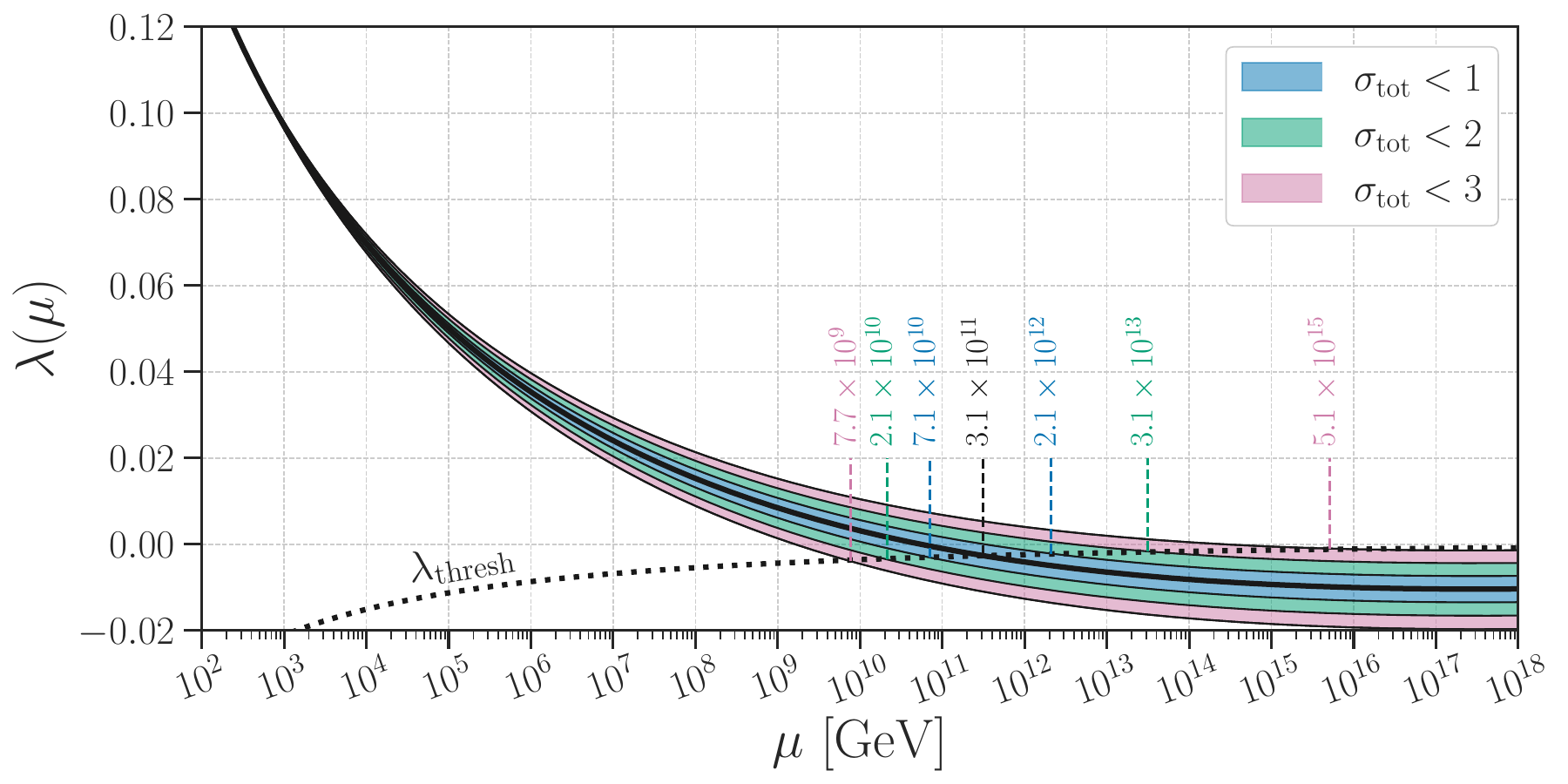}
    \caption{RG evolution of SM Higgs quartic coupling, $\lambda(\mu)$. The \textbf{\textcolor{black}{black}} central solid line gives running for $m_t=172.57$ GeV and $\alpha_s(M_Z)=0.1179$; the current best fit values from \cite{ParticleDataGroup:2024cfk}. The dotted black line gives $\lambda_{\rm thresh}(\mu)$, defined in \EqRef{eq:lambda_thresh}; $v_R$ is determined by the intersection of $\lambda(\mu)$ and $\lambda_{\rm thresh}(\mu)$. The \textbf{\textcolor{MyBlue}{blue}}, \textbf{\textcolor{MyGreen}{green}}, and \textbf{\textcolor{MyMagenta}{pink}} regions show the range of evolution when $m_t$ and $\alpha_s(M_Z)$ give $\sigma_{\rm tot}<1,2,3$, where $\sigma_{\rm tot}$ is defined in \EqRef{eq:sigma_tot}. Numerical values of $v_R$ are shown for $\sigma_{\rm tot} = 0,1,2,3$. }
    \label{fig:Higgs_Quartic_Running}
\end{figure}

First, we note $\mathcal{P}$ symmetry in \EqRef{eq:SSB} fixes the renormalizable potential of the color neutral Higgs fields in the 3221 theory to
\begin{align}
    V_{3221}\left(\HL, \HR\right)=-m^2\left(\left|\HL\right|^2+\left|\HR\right|^2\right)+\lambda\left(|\HL|^2+\left|\HR\right|^2\right)^2+\lambda^{\prime}|\HL|^2\left|\HR\right|^2 .
\end{align}
This symmetry is spontaneously broken when $\HR$ gets a VEV, $v_R^2 = \frac{m^2}{2\lambda}$. Expanding about this minimum  and integrating out the heavy mode gives an effective potential
\begin{align} \label{eq:V_LE}
    V_{321}(\HL)=\lambda^{\prime} v_R^2|\HL|^2-\lambda^{\prime}\left(1+\frac{\lambda^{\prime}}{4 \lambda}\right)|\HL|^4 .
\end{align}
For $H_L$ to have a VEV at the electroweak scale, the quadratic term must be small; $\lambda' \sim -v_L^2/v_R^2$. This implies the SM Higgs quartic coupling essentially vanishes at $v_R$, the scale where $\HR$ is integrated out to yield the above effective potential. More precisely, the 1-loop Coleman-Weinberg effective potential \cite{Coleman:1973jx} gives threshold corrections \cite{Hall:2019qwx,Dunsky:2019api} as described in \AppRef{app:Threshold Corrections}. These are
\begin{align} \label{eq:lambda_thresh}
    \lambda_{\rm thres} \left(v_R\right) & = \frac{3 y_t^4}{8 \pi^2} \ln \frac{y_t}{e}+\frac{3g^4}{64 \pi^2}  \ln \frac{e^{2/3} \sqrt{2}}{g} +\frac{3 \left(g^2+{g'}^2\right)^2 }{256 \pi^2} \log \frac{2 e^{4/3} \left(g^4-{g'}^4\right)}{g^4 \left(g^2+{g'}^2\right)}.
\end{align}
The scale $v_R$ of $SU(2)_R$ symmetry breaking is determined\footnote{There is additional model dependence from mass spectra of $\Y$ states discussed in \AppRef{app:Threshold Corrections}.} as shown in \FigRef{fig:Higgs_Quartic_Running}. 

\begin{figure}
    \centering
    \includegraphics[width=0.8\linewidth]{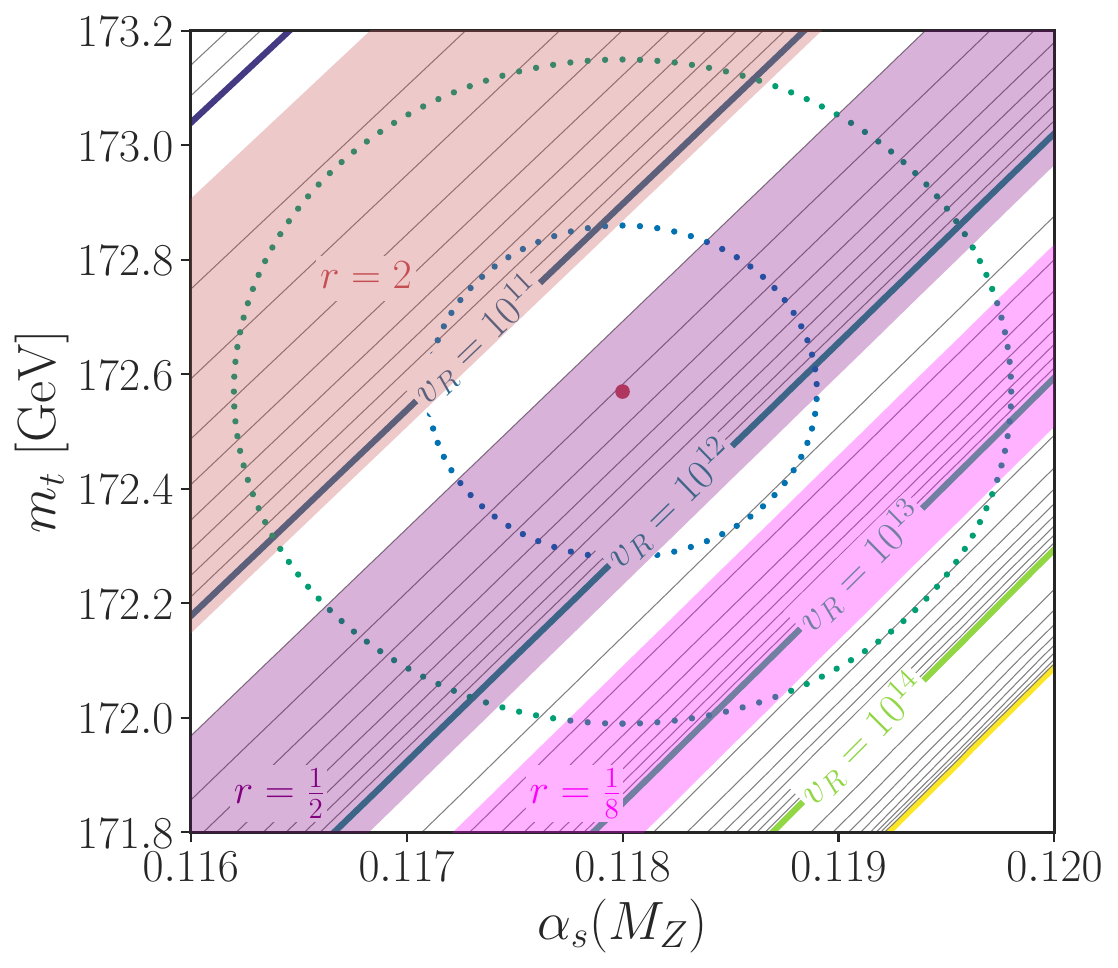}
    \caption{Solid lines show contours for the Higgs Parity prediction of $v_R$ as a function of the observed values of the top quark pole mass and the strong coupling evaluated at the $Z$ boson mass. Dotted circles are contours $\sigma_{\rm tot} = 1,2$. Shaded bands correspond to the values of $\alpha_s(M_Z)$ and $m_t$ that lead to gauge coupling unification for $r=2, 1/2, 1/8$, with additional threshold corrections limited to
    $\Delta < 5/(8\pi^2)$.}
    \label{fig:mt-alphascorrelation}
\end{figure}

Higgs quartic running, and therefore the prediction of $v_R$, is sensitive to measured values of $m_t$, $\alpha_s(M_Z)$ (and, to a lesser extent, $m_h$). Using the calculation of the dependence of the $\overline{MS}$ SM parameters at the scale $\mu = 200$ GeV to the $m_t$ pole mass and $\alpha_s(M_Z)$ given in \cite{Alam:2022cdv}, we calculate deviations in the prediction of $v_R$. We define 
\begin{align} \label{eq:sigma_tot}
    \sigma_{\rm tot} \equiv \sqrt{\left(\frac{m_t - m_{t,0}}{\sigma_{m_t}}\right)^2 + \left(\frac{\alpha_s(M_Z) - \alpha_{s,0}(M_Z)}{\sigma_{\alpha_s(M_Z)}}\right)^2}
\end{align}
where $m_{t,0} = 172.57$ GeV, $\alpha_{s,0}(M_Z) = 0.1179$, $\sigma_{m_t} = 0.29$ GeV, $\sigma_{\alpha_s(M_Z)} = 0.0009$ are best fit values and 1$\sigma$ uncertainties obtained from the PDG \cite{ParticleDataGroup:2024cfk}. Results are shown in \FigRef{fig:Higgs_Quartic_Running} and \FigRef{fig:mt-alphascorrelation}. Despite the high accuracy of experimental determinations of $\alpha_s(M_Z)$ and $m_t$, $v_R$ varies from $(2\times 10^{10} - 3\times 10^{13})$ GeV at $2\sigma$.

In the Higgs Parity mechanism, hierarchy problems beyond that of the electroweak scale are absent despite the existence of an intermediate energy scale $v_R$~\cite{Hall:2018let}. In fact, small $m^2 \sim v_R^2$ in comparison with the cutoff scale $\Lambda^2$ requires fine-tuning of $O(m^2/\Lambda^2)$, and $\lambda'$ requires fine-tuning of $O(v_L^2/v_R^2)$, so the total fine-tuning is $O(v_L^2/\Lambda^2)$, which is the same as the electroweak fine-tuning.

\subsection{Gauge Coupling Unification and Proton Decay} \label{subsec:Gauge Coupling Unification and Proton Decay}

\begin{figure}[t]
    \centering
    \includegraphics[width=1\linewidth]{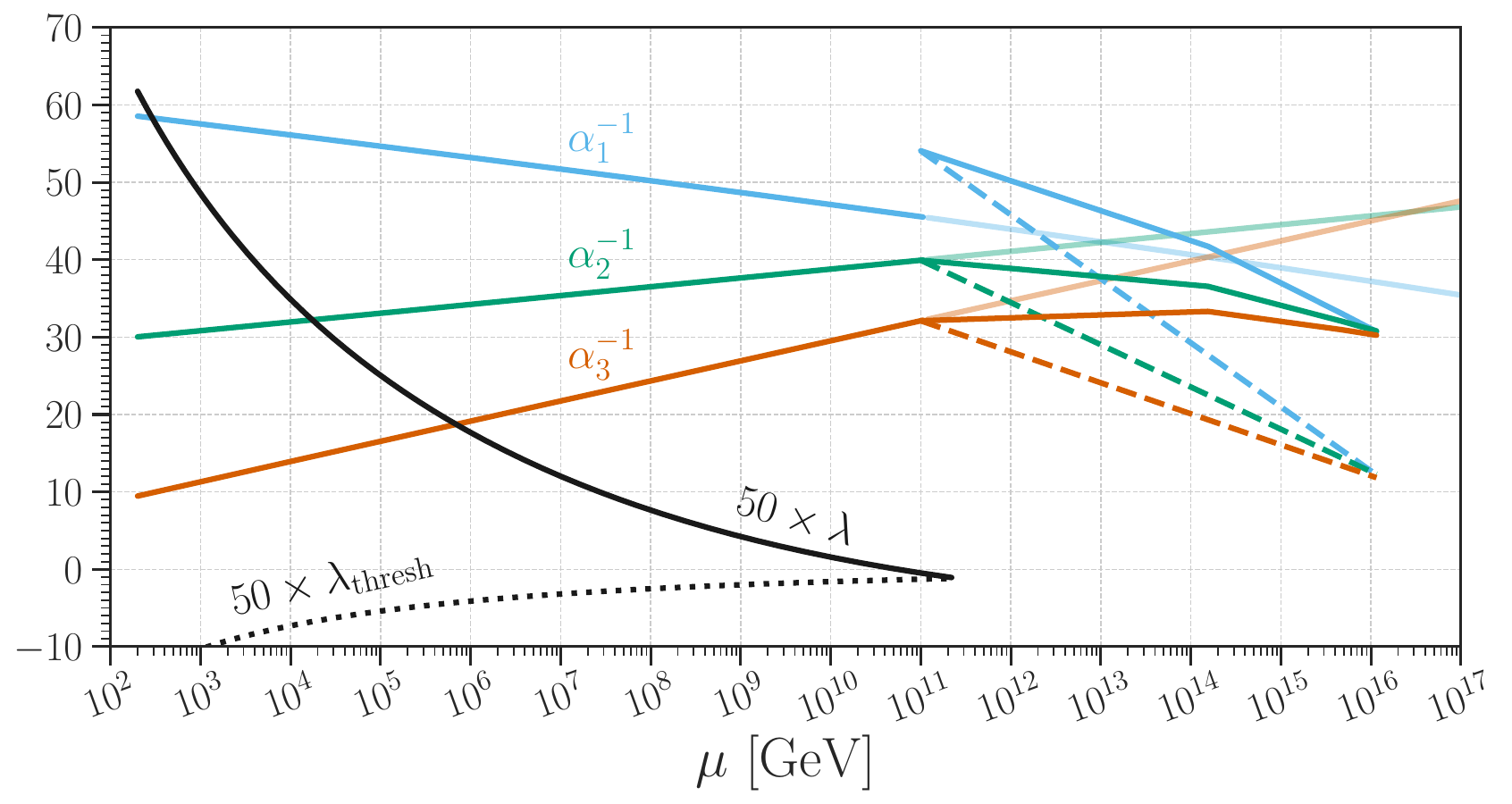}
    \caption{Running of gauge couplings and SM Higgs quartic in the Higgs Parity framework. Unification of gauge couplings occurs near $2 \times 10^{16}$ GeV. Colored lines show $\alpha_1^{-1}$ (\textbf{\textcolor{MyBlue}{blue}}), $\alpha_2^{-1}$ (\textbf{\textcolor{MyGreen}{green}}), and $\alpha_3^{-1}$ (\textbf{\textcolor{MyOrange}{orange}}). Dashed (solid) lines correspond to low (high) mass scenarios for heavy vector-like fermions. Faint lines show SM-only running. The Higgs quartic coupling is shown in \textbf{black}, with the dashed black line indicating the matching value from threshold corrections in \EqRef{eq:lambda_thresh}.}
    \label{fig:Gauge_Coupling_Unification}
\end{figure}

The running of gauge couplings in the Higgs parity framework using the best fit value of $v_R$ is illustrated in \FigRef{fig:Gauge_Coupling_Unification}. At a qualitative level, unification of gauge couplings near $2 \times  10^{16}$ GeV is nearly perfect. On the other hand, unification of SM couplings requires large threshold corrections, as shown by the faint lines. Details of RG running is given in \AppRef{app:Gauge Coupling Unification in Higgs Parity GUTs}, and we summarize the key points here.

To calculate the quality with which gauge couplings unify, we run SM gauge couplings to high energy and match to 3221 gauge couplings at the mass of heavy $SU(2)_R$ gauge bosons $M_{W_R}= g_2v_R/\sqrt{2}$. When matching the two different $U(1)$ gauge couplings we normalize $U(1)$ generators consistently with an $SO(10)$ embedding; we call normalized couplings $g_{\tilde{Y}}$ and $g_{\widetilde{B-L}}$. The matching condition is
\begin{align}
    \frac{1}{g^2_{\tilde{Y}}} = \frac{2}{5} \frac{1}{g^2_{\widetilde{B-L}}} + \frac{3}{5} \frac{1}{g_2^2} - \frac{1}{40 \pi^2}.
\end{align}

Gauge coupling running above $v_R$ depends on the heavy vector-like fermion mass spectrum. If $SO(10)$-breaking mass terms of \EqRef{eq:XMassesSigma} are ignored, these particles only affect the quality of unification at two loops; we account for this, but it has no major qualitative impact. However, vector-like fermions have a significant impact on gauge coupling strength at unification, and therefore on proton decay. For this reason, it is important to understand possible mass spectra of these fields. As alluded to above and discussed further below, at least one mass eigenvalue of $\X$ and $\Y$ field must be near $v_R$. Further, we expect the second lightest mass eigenvalues to satisfy 
\begin{align}
    M_2\lesssim v_R /y_\mu(v_R),\qquad \tilde{M}_2 \lesssim v_R/ y_c(v_R) 
\end{align}
in order that $\x,\, \y$ couplings remain perturbative. The heaviest mass eigenstate could be at or above the $SO(10)$ breaking scale. Therefore, to get an idea of the possible range for $\alpha_{\rm GUT}$ we consider two scenarios
\begin{enumerate}
    \item \textbf{Low Mass:} All heavy vector-like fermions are around $v_R$.
    \item \textbf{High Mass:} $M_1, \tilde{M}_1 > M_{\rm GUT}$; $M_2 \approx \frac{v_R}{y_\mu(v_R)} $, $\tilde{M}_2 \approx \frac{v_R}{y_c(v_R)}$; $M_3 \approx \tilde{M}_3 \approx v_R$.
\end{enumerate}
The first (second) scenario results in the largest (smallest) possible value of $\alpha_{\rm GUT}$ and requires hierarchy in SM fermion masses to originate from $\x,\, \y$ couplings ($M,\, \tilde{M}$ masses). Gauge coupling running for these two scenarios is shown in \FigRef{fig:Gauge_Coupling_Unification}.

\begin{figure}
    \centering
    \includegraphics[width=\linewidth]{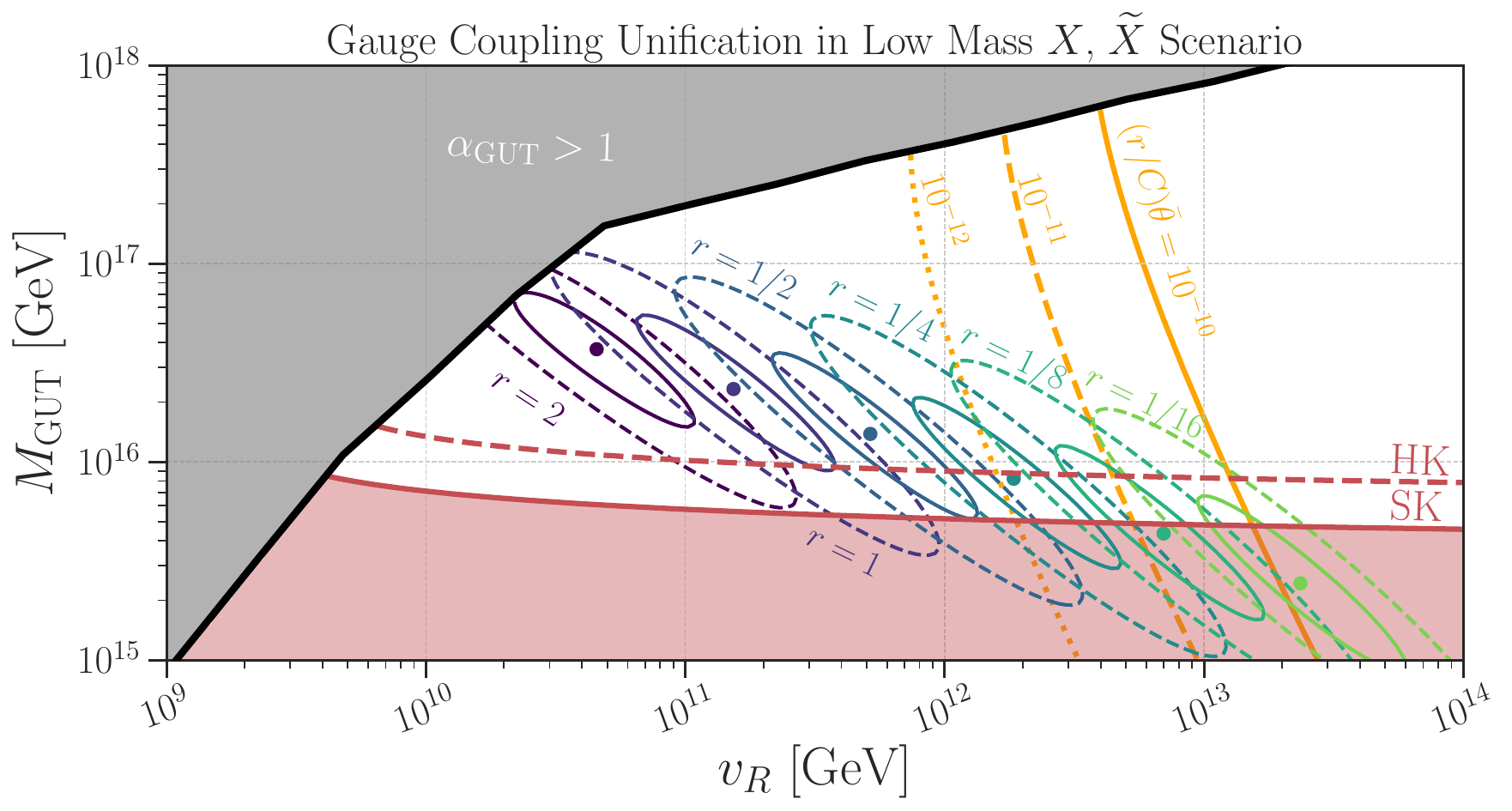}
    \caption{Determination of mass scales $M_{\rm GUT}$ and $v_R$ from precision gauge coupling unification in the scenario with low $X$ and $\tilde{X}$ masses. The solid dots show predictions for several values of $r$, defined in \EqRef{eq:r}. For each $r$, the prediction is enlarged from a dot to an ellipse by additional threshold corrections $\Delta$, defined in \EqRef{eq:Delta}. Solid  and dashed ellipses correspond to $\Delta = 5/(8\pi^2)$ $\Delta = 10/(8\pi^2)$ respectively. Solid and dashed {\textbf{\textcolor{red}{red}}} contours show constraints and projections from proton decay searches at Super-Kamiokande \cite{Super-Kamiokande:2020wjk} and Hyper-Kamiokande respectively \cite{Hyper-Kamiokande:2018ofw}. {\textbf{\textcolor{gray}{Gray}}} region corresponds to points where couplings become non-perturbative before unification which we quantify roughly as the region where $\alpha_{\rm GUT}>1$ using the approximation $\alpha_{\rm GUT}^{-1} \approx \frac{1}{3}\sum_{i=1}^3 \alpha_i^{-1}$. Contours of expected strong $CP$ angle are shown in {\textbf{\textcolor{orange}{orange}}}.  }
    \label{fig:Unification_Quality_Low_Mass}
\end{figure}

\begin{figure}
    \centering
    \includegraphics[width=\linewidth]{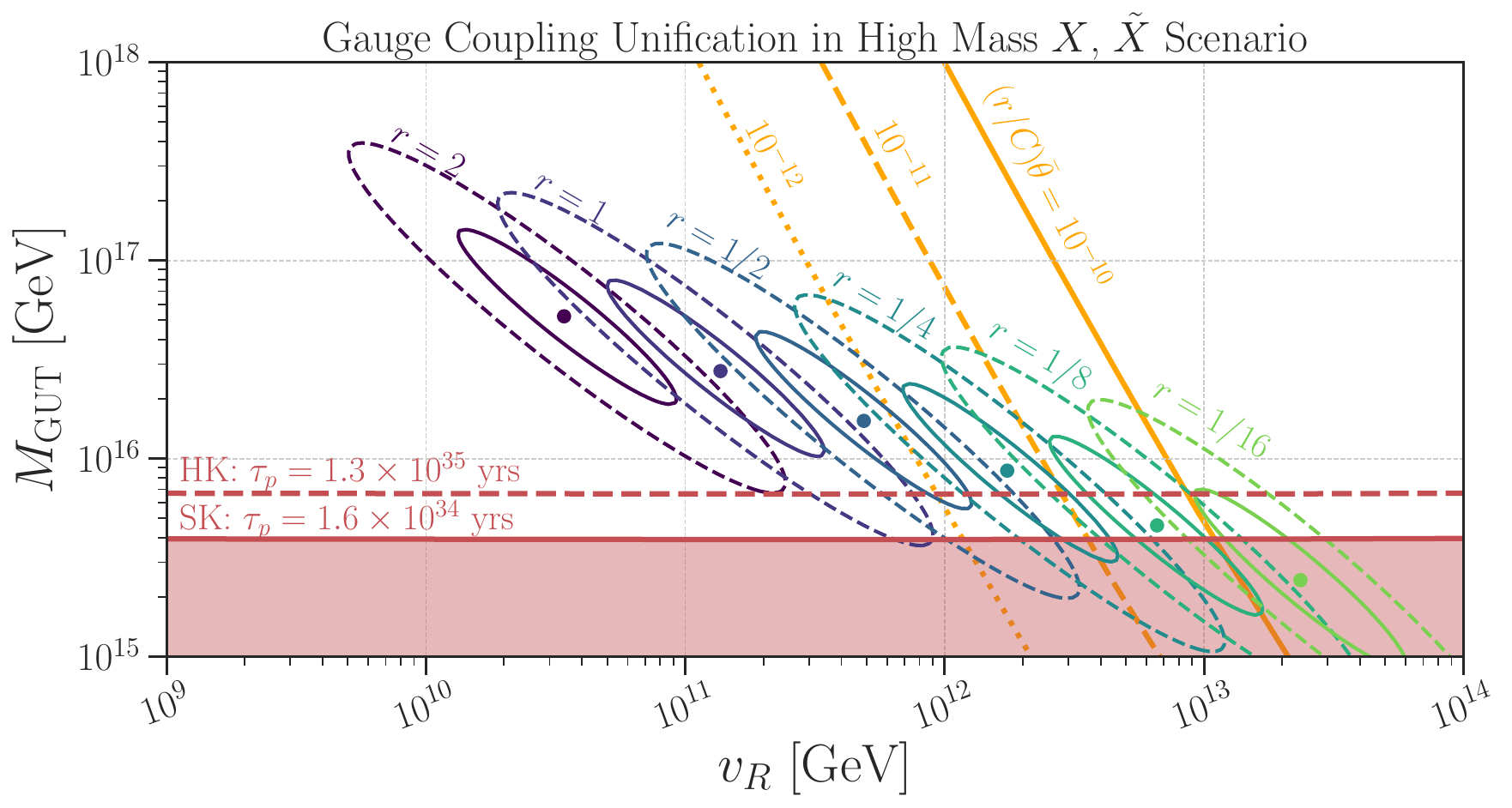}
    \caption{As in \FigRef{fig:Unification_Quality_Low_Mass}, except for the scenario with high $X$ and $\tilde{X}$ masses. }
    \label{fig:Unification_Quality_High_Mass}
\end{figure}

One-loop threshold corrections affect gauge coupling unification and depend on mass spectra of particles at the unified scale. Matching of 3221 gauge couplings to the unified coupling at scale $\mu$ is obtained using
\begin{align}
    g_{a}^{-2}(\mu) = g_{\rm GUT}^{-2}(\mu) - \Delta_{a,G} - \Delta_{a,S} - \Delta_{a,F},\qquad a = 1,2,3
\end{align}
where $\Delta_{a,G}$, $\Delta_{a,F}$ and $\Delta_{a,S}$ describe threshold correction from gauge bosons, fermions, and scalars \cite{Hall:1980kf,Weinberg:1980wa} and are described in \AppRef{app:Gauge Coupling Unification in Higgs Parity GUTs}. Scalar masses are challenging to calculate as they are only stabilized at 1-loop \cite{Bertolini:2009es}. We treat threshold corrections from these and any additional GUT scale fermions as unknowns. 

The gauge boson mass spectra is given in \EqRef{eq:311_Mass} and \EqRef{eq:322_Mass}. We match at scale $M_{\rm GUT}\equiv M_{(\bm{3},\bm{2},\bm{2})_{-2/3}}$, i.e., the mass of gauge bosons responsible for proton decay. Matching relations are
\begin{align}
    &\Delta_{3,G}(M_{\rm GUT})  =  \frac{1}{48 \pi^2} \left(5 -21 \log r \right), \\
    &\Delta_{2,G}(M_{\rm GUT}) = \frac{6}{48 \pi^2},\\
    &\Delta_{1,G}(M_{\rm GUT})=  \frac{1}{48 \pi^2}\left(8 - 84 \log r   \right),
\end{align}
where $r$ is defined in \EqRef{eq:r}. We quantify the quality of unification in terms of the amount of additional threshold corrections required using
\begin{align} \label{eq:Delta}
    \Delta &=  \left[\frac{1}{2}\sum_{ijk = 1}^3\varepsilon_{ijk}\left((g_j^{-2}(M_{\rm GUT}) + \Delta_{j,G}) - (g_k^{-2}(M_{\rm GUT}) + \Delta_{k,G})\right)^2 \right]^{1/2}.
\end{align}
This measure is independent of coupling strength $g_{\rm GUT}(M_{\rm GUT})$. Therefore, parameters that govern quality of unification are $v_R,\, r$, and $M_{\rm GUT}$. In \FigRef{fig:Unification_Quality_Low_Mass} and \FigRef{fig:Unification_Quality_High_Mass} we show contours of $\Delta$, representing the size of additional threshold corrections required for unification for both high and low mass scenarios. Additionally, we show regions constrained or to be constrained by proton decay, where the lifetime is calculated using the estimate $g_{\rm GUT}^{-2}(M_{\rm GUT}) = \frac{1}{3}\sum_{i=1}^3 g_i^{-2}(M_{\rm GUT})$ and results from \AppRef{app:Review of Proton Decay in Higgs Parity GUTs}.

Two important results emerge. First, precision gauge coupling unification can be achieved for $v_R = (10^{10} -10^{13})~{\rm GeV}$ with small additional threshold corrections. Remarkably, this range of $v_R$ agrees with the $2 \sigma$ range predicted by Higgs Parity shown in \FigRef{fig:Higgs_Quartic_Running}. Second, as $v_R$ is increased in the upper half of this range, it becomes progressively more likely for proton decay to be discovered at Hyper-Kamiokande.

\subsection{A Solution to The Strong $CP$ Problem} \label{subsec:strongCP}

{\HPU} naturally solves the strong $CP$ problem as we now describe. First, recall the SM Lagrangian contains a term 
\begin{align}
    \mathcal{L}_\theta = \frac{\alpha_s \theta }{4 \pi} \Tr (G^{\mu \nu} \tilde{G}_{ \mu \nu}).
\end{align}
The chiral anomaly relates $\theta$ and complex phases in quark masses leaving a single physical parameter invariant under field redefinitions
\begin{align} \label{eq:theta_bar}
    \bar{\theta} = \theta + {\rm arg\, det}(Y_u Y_d).
\end{align}
Using chiral perturbation theory, it can be shown that the neutron electric dipole moment is proportional to $\bar \theta$. The current upper limit on this dipole moment \cite{Abel:2020pzs} requires $\bar{\theta} \lesssim 4\times 10^{-11}$. This fine-tuning is called the \textit{strong-CP} problem.

In the $SU(3)\times SU(2)_L \times SU(2)_R\times U(1)_{B-L}\times \mathcal{P}$ theory the strong $CP$ problem is solved by setting $\bar \theta = 0$ above the $SU(2)_R$ symmetry breaking scale: ${\rm Tr} (G^{\mu \nu }\tilde{G}_{\mu \nu})$ term is parity odd giving $\theta = 0$ exactly and Yukawa couplings are hermitian giving ${\rm arg\, det}(Y_u Y_d) = 0$. Below the scale $v_R$, parity is spontaneously broken. (See \cite{Babu:1989rb,Babu:1988mw} for softly broken parity.) After integrating out heavy states, $\bar{\theta}$ is first generated at two loops and remains below current neutron EDM bounds~\cite{Hall:2018let, Hisano:2023izx}.

At tree-level,
the $3221$ symmetry allows the higher dimension operator 
\begin{align}\label{eq:Strong_CP_Quality}
    \mathcal{L}_6 \supset \frac{C}{ M_P^2}\left(|\HR|^2 - |\HL|^2\right)\Tr (G^{\mu \nu} \tilde{G}_{\mu \nu}),
\end{align}
where the scale of new physics is $\Lambda = M_P/\sqrt{C}$. This generates a contribution to $\bar \theta$ after parity is broken. The strong $CP$ problem is solved if $\sqrt{C}\,v_R \lesssim 10^{12}$ GeV. 

In the $SO(10)\times CP$ theory, \EqRef{eq:Strong_CP_Quality} is forbidden by $CP$. Contributions to $\bar{\theta}$ must involve $CP$ and $SO(10)$ breaking via $\expval{\Sigma}$ and parity breaking from $H_R \in \phi$. The leading operator which contributes to $\bar{\theta}$ is the dimension-7 operator\footnote{This is not altered if the theory contains a scalar multiplet in the 54 representation, even if it has a larger VEV than $\Sigma$, because it must be $CP$ even to preserve an unbroken parity symmetry.} 
\begin{align}\label{eq:Strong_CP_dim7}
    \mathcal{L}_7 \supset \frac{C}{ M_P^3} \;
    \varphi^\dagger \Sigma \varphi \; G \tilde{G}.
\end{align}
This generates \EqRef{eq:Strong_CP_Quality} 
with added suppression $v_{45}/M_P \sim 10^{-2} $. Thus, we expect 
\begin{align}
    \bar{\theta} &=\frac{4\pi C}{\alpha_s(v_R)}\frac{v_{45} v_R^2}{M_P^3}=10^{-10}\, C\, \left(\frac{v_{45}}{10^{16}~\rm GeV}\right)\left(\frac{v_R}{2\times 10^{13}~{\rm GeV}}\right)^2 .
\end{align}
We see {\HPU} resolves the strong $CP$ problem if $v_R\lesssim 10^{13}$ GeV. Corrections to $\bar{\theta}$ as a function of $v_R$ and $M_{\rm GUT}$ are shown in \FigRef{fig:Unification_Quality_Low_Mass} and \FigRef{fig:Unification_Quality_High_Mass}.

\section{The Seesaw Origin of Flavor Observables} \label{sec:The Seesaw Origin of Flavor Observables}

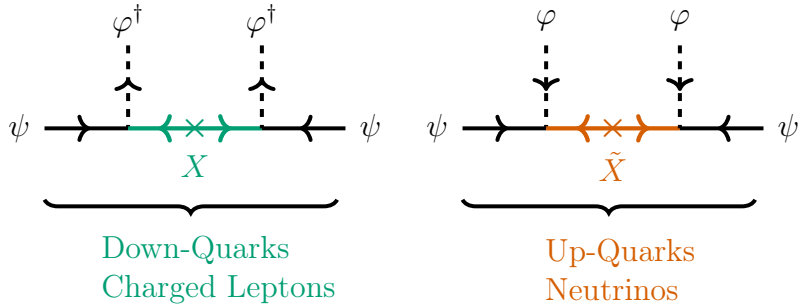
\begin{figure}[h!]
    \centering
    \begin{tikzpicture}[line width=1.5 pt, scale=1.1,baseline={(0,-2.2cm)}]
    \draw [decorate,decoration={brace,amplitude=5pt,mirror,raise=4ex}]
  (-1,-.2) -- (2.5,-.2) node[midway,yshift=-4em,xshift = 2em,text width=4cm]{\textcolor{MyGreen}{Down-Quarks Charged Leptons}};
    \draw[scalarbar] (90:1)--(0,0);
    \node at (90:1.3) {$\varphi^\dagger$};
    \draw[fermion] (180:1)--(0,0);
    \node at (180:1.3) {$\psi$};
    \draw[ultra thick,MyGreen,fermion] (0:.8)--(0,0);
    \node[yshift=-.5cm] at (0:.8) {\textcolor{MyGreen}{$\X$}};
    \begin{scope}[shift={(0:1.6)}]
        \draw[ultra thick,MyGreen,fermion] (0:-.8)--(0,0);
        \node at (0:-.8) {$\textcolor{MyGreen}{\bm{\times}}$};
        \draw[fermion] (0:1)--(0,0);
        \node at (0:1.3) {$\psi$};
        \draw[scalarbar] (90:1)--(0,0);
        \node at (90:1.3) {$\varphi^\dagger$};
    \end{scope}
    \begin{scope}[shift={(0:5.0)}]
    \draw [decorate,decoration={brace,amplitude=5pt,mirror,raise=4ex}]
  (-1,-.2) -- (2.5,-.2) node[midway,yshift=-4em,xshift=1em,text width=2.5cm]{\textcolor{MyOrange}{Up-Quarks  Neutrinos}};
    \draw[scalar] (90:1)--(0,0);
    \node at (90:1.3) {$\varphi$};
    \draw[fermion] (180:1)--(0,0);
    \node at (180:1.3) {$\psi$};
    \draw[ultra thick,MyOrange,fermion] (0:.8)--(0,0);
    \node[yshift=-.5cm] at (0:.8) {\textcolor{MyOrange}{$\Y$}};
    \begin{scope}[shift={(0:1.6)}]
        \draw[ultra thick,MyOrange,fermion] (0:-.8)--(0,0);
        \node at (0:-.8) {$\textcolor{MyOrange}{\bm{\times}}$};
        \draw[fermion] (0:1)--(0,0);
        \node at (0:1.3) {$\psi$};
        \draw[scalar] (90:1)--(0,0);
        \node at (90:1.3) {$\varphi$};
    \end{scope}
    \end{scope}
\end{tikzpicture}
    \caption{Feynman diagrams generating SM flavor observables at the $SO(10)$ level.}
    \label{fig:Feynman_Diagram_SO10}
\end{figure}

In this section we describe the seesaw origin of SM flavor observables assuming mass eigenstates of $\X$ and $\Y$ are heavier than $v_R$ and can safely be integrated out. Then all SM flavor observables arise from the diagrams of \FigRef{fig:Feynman_Diagram_SO10}.
Importantly, flavor observables of down-quarks and charged-leptons (up-quarks and neutrinos) have common origin from integrating out $\X$ ($\Y$). This is more constraining than $SU(5)$, where only $Y_d$ and $Y_e$ are related via $\bm{10}_F \bar{\bm{5}}_F \bm{5}_H^\dagger$, but less constraining than $SO(10)$ with a $\bm{10}_H$, where $Y_d$, $Y_u$, $Y_e$, and $Y_\nu$ arise from $\bm{16}_F \bm{16}_F \bm{10}_H$.

\subsection{Calculation of Standard Model Flavor Observables}

\subsubsection*{\emph{Down Quark Yukawa}}
The down-quark mass matrix is obtained by integrating out $D,\,\bar{D}\in \X$.
\begin{align}
    \mathscr{L}\supset q \xD \bar{D} \, \HL^\dagger + \bar{d} \xD^* D \, v_R  + \MD D \bar{D}.
\end{align}
Ignoring the $SO(10)$ origin of this term, $x_D$ is a complex matrix and $\MD$ is a Hermitian matrix.%
\footnote{  
Note the determinant of the mass matrix of $d\subset q$, $\bar{d}$, $D$, and $\bar{D}$ is real and contributions to $\bar{\theta}$ from quark masses is absent at leading order. We take VEVs of $H_L$, $H_R$ real by gauge transformation. If we instead take complex VEVs, the correction to $\bar{\theta}$ from the down sector is exactly canceled by that from the up sector. This is in contrast with $SU(2)_L$ breaking by $SU(2)_L\times SU(2)_R$ bi-fundamental Higgses~\cite{Beg:1978mt,Mohapatra:1978fy}, where phases in VEVs are physical and require suppression by extra symmetry to solve the strong CP problem~\cite{Kuchimanchi:1995rp,Mohapatra:1995xd,Kuchimanchi:2010xs,Mimura:2019yfi}.
}
We eventually match to the $SO(10)$ theory but keep them general momentarily to allow GUT breaking contributions. Assuming all eigenvalues of $m_D$ larger than $v_R$, we can integrate out $D,\, \bar{D}$ states using equations of motion to get
\begin{align}\label{eq:L_D}
    \mathscr{L}_D\supset -(q \HL)\left(\xD \MD^{-1} \xD^\dagger\right)(\bar{d}v_R).
\end{align}
Therefore, the down-quark yukawa is
\begin{align} \label{eq:Y_d_Approx}
    Y_d = \left(\xD \MD^{-1} \xD^\dagger\right) v_R.
\end{align}

\subsubsection*{\emph{Charged Lepton Yukawa}}
The charged-lepton yukawa is obtained by integrating out the $L,\,\bar{L} \in \Delta \in \X$.
\begin{align}\label{eq:L_L}
    \mathscr{L}_L \supset \ell \x_L \bar{L} \, v_R+ \bar{e} \x_L^* L \, \HL^\dagger + M_L L \bar{L}.
\end{align}
Integrating out $L$ and $\bar{L}$ results in the following charged lepton yukawa matrix:
\begin{align} \label{eq:Y_e_Approx}
    Y_e = \left(\x_L M_L^{-1} \x_L^\dagger\right) v_R.
\end{align}

\subsubsection*{\emph{Up Quark Yukawa}}
The up-quark yukawa is obtained by integrating out $U,\, \bar{U}\in \Y$ and $Q,\,\bar{Q} \in \Y$. 
\begin{align}\label{eq:L_UQ}
    \mathscr{L}_U \supset q \yU \bar{U} \, \HL + \bar{q} \yU^* U \, \HR + \MU U \bar{U},\\
    \mathscr{L}_{Q} \supset q \y_Q \bar{Q} \, \HR + \bar{q} \y_Q^* Q \, \HL + \tilde{M}_Q Q \bar{Q}.
\end{align}%
This results in the following up-quark yukawa matrix:
\begin{align} \label{eq:Y_u_Approx}
    Y_u = \left(\yU \MU^{-1} \yU^\dagger\right) v_R + \left(\y_Q \tilde{M}_Q^{-1} \y_Q^\dagger\right) v_R.
\end{align}
Note the large top Yukawa coupling requires one mass eigenvalue around $v_R$.

\subsubsection*{\emph{Neutrino Masses}}
Neutrino flavor observables arise by integrating out $S,\,T_L,$ and $T_R$ in $\Y$.
\begin{alignat}{3} \label{eq:L_ST}
    \mathscr{L}_{S,T_L,T_R} &=  \ell \yS S \HL&&+  \bar{\ell} \yS^*  S \HR&&+ \frac{\MS}{2} S^2\\
    &+   \ell \yT T_L \HL&&+ \bar{\ell}\yT^*  T_R \HR&&+ \frac{\MT}{2}  (T_L^2 + T_R^2).
\end{alignat}
Integrating out $S,\,T_L,$ and $T_R$ gives the dimension-5 operators
\begin{align} \label{eq:Weinberg_Ops}
    \mathscr{L}_{\nu,{\rm eff}} = 
    -\begin{pmatrix}
    \ell \HL & \bar{\ell} \HR    
    \end{pmatrix}
    \begin{pmatrix}
    \yS \MS^{-1} \yS^\top +   \yT \MT^{-1} \yT^\top & \yS \MS^{-1}\yS^\dagger \\
    \yS^* \MS^{-1}\yS^\top & \yS^* \MS^{-1}\yS^\dagger +   \yT^* \MT^{-1}\yT^\dagger
    \end{pmatrix}
    \begin{pmatrix}
    \ell \HL \\ \bar{\ell} \HR
    \end{pmatrix}.
\end{align}
Inserting VEVs for $\HR$, the mass matrix of the right-handed neutrinos is
\begin{align}
    \mathcal{M}_{\bar{\nu}} = \left(\yS^* \MS^{-1}\yS^\dagger +   \yT^* \MT^{-1}\yT^\dagger\right)v_R^2.
\end{align}
The off-diagonal $3 \times 3$ block gives the neutrino Yukawa coupling matrix 
\begin{align} \label{eq:Ynu}
    Y_\nu = \left(\yS \MS^{-1} \yS^\dagger \right)v_R.
\end{align}
The mass matrix for SM neutrinos after integrating out right handed neutrinos is 
\begin{align} \label{eq:Mnu}
    \mathcal{M}_{\nu} = \mathcal{M}_{\bar{\nu}}^*\left(\frac{v_L}{v_R}\right)^2 - Y_\nu \mathcal{M}_{\bar{\nu}}^{-1} Y_{\nu}^\top \; v_L^2.
\end{align}

\subsection{The $SO(10)$
and $CP$ invariant limit} \label{subsec:so10inv}

Consider the limit where $SO(10)$ and $CP$ breaking contributions and RG running can be ignored. Then every component of $\X$ ($\Y$) has mass matrix $M$ ($\tilde{M}$). Additionally $\xDelta = \xD = \x$ and $\yU = \yQ = \sqrt{2}\, \yT = -\frac{2}{\sqrt{3}}\yS = \y$.

It is worth briefly analyzing the Lagrangian at the $SO(10)$ level
\begin{align}
    \mathcal{L}_{SO(10)} \supset \psi \x \X \phi^\dagger + \psi \y \Y \phi + \X M \X + \Y \tilde{M} \Y .
\end{align}
We diagonalize $M$ ($\tilde{M}$) with a rotation on $\X$ ($\Y$). Then we rescale $\X$ and $\Y$ to set the mass matrix to $v_R \delta_{ij}$ (this assumes mass eigenvalues larger than $v_R$). This allows additional rotations on $\X$ and $\Y$ which along with a rotation on $\psi$ diagonalize $\y$ and sets $\x = R_L \widehat{\x}$ where $R_L$ is some rotation matrix. Hats denote diagonal matrices of singular values. This process converts the Lagrangian to 
\begin{align}\label{eq:SO10_Diagonal}
    \mathcal{L}'_{SO(10)} \supset \psi' R_L\widehat{\x} \X' \phi^\dagger + \psi' \,\widehat{\y} \Y' \phi + v_R \,\X'  \X' + v_R \,\Y' \Y' .
\end{align}
The primes denote fields after the transformations. We choose this basis in the remainder of this section, but remind the reader it relies on assuming $SO(10)$ and $CP$ breaking can be ignored and mass eigenvalues of $M$ and $\tilde{M}$ are well above $v_R$.

Down-quark and charged-lepton masses are obtained by integrating out $\X'$:
\begin{align} \label{eq:d/e_relation_Simplified}
    \mathcal{M}_e \approx \mathcal{M}_d \approx R_L \widehat{\x}^2 R_L^\top v_L.
    \end{align}
The relation of down-quark and charged-lepton masses is common to minimal SU(5) and SO(10) theories. The discrepancy with the SM must be accounted for.

Neutrinos and up-quarks get diagonal mass matrices from integrating out $\Y'$
\begin{align} \label{eq:Mu_Mnu_relation_Simplified}
    \frac{5}{2}\mathcal{M}_\nu \approx \mathcal{M}_u\left(\frac{v_L}{v_R}\right) \approx \frac{8}{5}\mathcal{M}_{\bar{\nu}}\left(\frac{v_L}{v_R}\right)^2 = 2\widehat{\y}^2 \frac{v_L^2}{v_R}.
\end{align}

We identify three problems with the predictions for quark and lepton masses and mixings in this $SO(10) \times CP$ invariant limit:
\begin{enumerate}
    \item Predicted down-quark to charged-lepton mass ratios from \EqRef{eq:d/e_relation_Simplified} are incorrect. The $b/\tau$, $s/\mu$, and $d/e$ mass ratios at $10^{12}$ GeV are 0.69, 0.24, and 2.56 respectively in the SM.
    \item \EqRef{eq:Mu_Mnu_relation_Simplified} relates up-quark masses to observed neutrino mass-squared differences. To remove uncertainty from $v_R$, we consider mass ratios
    \begin{align}
        \sqrt{\frac{\Delta m^2_{31}}{\Delta m^2_{21}}}
        \approx \frac{m_t}{ m_c}. 
    \end{align}
    At $10^{12}$ GeV, $m_t/m_c \simeq 300$ while $\Delta m^2_{31} / \Delta m^2_{21} \simeq 5.7$, so this prediction is off by more than a factor of 50.
    \item  \EqRef{eq:d/e_relation_Simplified} and \EqRef{eq:Mu_Mnu_relation_Simplified} predict identical $CP$ conserving PMNS and CKM matrices. Specifically $V_{\rm CKM} = V_{\rm PMNS} = R_L$.
\end{enumerate}

As expected, we must go beyond the $SO(10) \times CP$ invariant limit. We anticipate $SO(10) \times CP$ breaking can easily give $\mathcal{O}(1)$ corrections to the down-quark and charged-lepton mass ratios and lead to $\mathcal{O}(1)$ $CP$ violation in the CKM matrix. 

However, the second and third problems relating up-quark and neutrino sectors are much more puzzling. The heavy two generations of the up sector are much more hierarchical than the neutrino sector with $m_c/m_t \ll m_2/m_3$ and $V_{cb} \ll V_{23}$. Similarly, $V_{ub} \ll V_{13}$. 

In the next section, we present a simple, common origin for small values of $(m_c/m_t) / (m_2/m_3)$, $V_{cb}/V_{23}$, and $V_{ub}/V_{13}$. We assume $b$-quark and $\tau$-lepton are mixtures of $\psi_3$ and $X_3$, with dominant $X_3$ component. As a result, third-generation flavor observables scale linearly with $x$, in contrast to quadratic dependence in see-saw mechanism. We call this a \textit{direct origin} of flavor. Locating $\HL$ in a $\bm{16}$ of $SO(10)$ enables third-generation masses to arise from a qualitatively distinct mechanism compared to lighter generations.

\section{The Direct Third Generation} \label{sec:directbtau}

\begin{figure}[t]
  \centering
    \begin{tikzpicture}
    \draw[black, ultra thick, ->] (0,-1.) -- (0,4.5) ;
    \node[rotate = 90] at (-0.5,4) {Energy};
    \draw[blue, ultra thick] (.5,-1)--(2,-1)node[right]  {$M_3$};
    \draw[blue, ultra thick] (.5,2.2)--(2,2.2)node[right] {$M_2$};
    \draw[blue, ultra thick] (.5,3.5)--(2,3.5)node[right] {$M_1$};
    \draw[red, ultra thick] (.5,0)--(2,0)node[right] {$\x_{33}\, v_R$};
    \draw[blue, ultra thick] (4.5,1.)--(6,1.)node[right,yshift = 6pt]  {$\tilde{M}_3$};
    \draw[blue, ultra thick] (4.5,2.4)--(6,2.4)node[right] {$\tilde{M}_2$};
    \draw[blue, ultra thick] (4.5,4)--(6,4)node[right] {$\tilde{M}_1$};
    \draw[red, ultra thick] (4.5,0.8)--(6,0.8)node[right,yshift = -6pt] {$\y_{33}\, v_R$};
    \draw[black, ultra thick,dashed, MyGreen] (.5,.9)node[right,yshift = 6pt,xshift = 45pt]  {$v_R$}--(5.95,.9);
    \draw[black, ultra thick,dashed, MyGreen] (.5,4.3)node[right,yshift = 6.5pt,xshift = 45pt]  {$M_{\rm GUT}$}--(5.95,4.3);
    \end{tikzpicture}
  \caption{Schematic example of spectra for $\X_i$ (left) and $\Y_i$ (right) giving a ``direct'' origin for third generation SM Yukawa couplings.}
  \label{fig:M_X}
\end{figure}
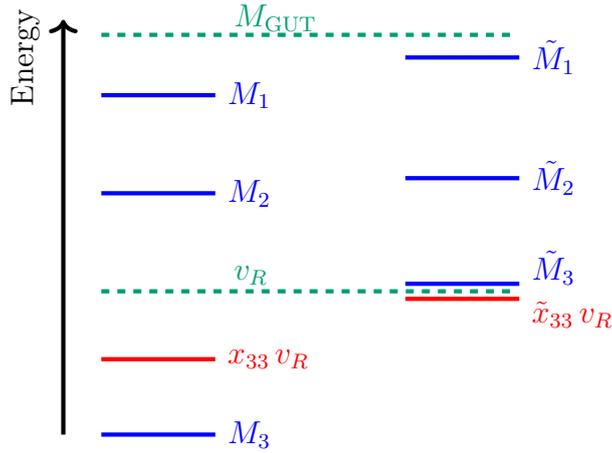

If SM Yukawa couplings arise from integrating out sufficiently heavy $X, \tilde{X}$ states, the previous section shows they are generically of the seesaw form
\begin{align}\label{eq:genericseesaw}
Y_{d,e} \propto   x \; M^{-1} \; x^\dagger \; v_R, \hspace{0.5in} Y_{u,\nu} \propto    \tilde{x} \; \tilde{M}^{-1} \; \tilde{x}^\dagger \; v_R.
\end{align}
A hierarchy in quark and lepton masses can originate from hierarchies in $x$, $\tilde{x}$ and/or eigenvalues of $M$, $\tilde{M}$. A particularly interesting case is a strong hierarchy in $M$ and $\tilde{M}$, as shown in \FigRef{fig:M_X}. In this scenario, the seesaw mechanism implies an inverted relationship between mass eigenvalues of $\X,\, \Y$ and SM fermions, with $M_1 \gg M_2 \gg M_3$ corresponding to generations (1,2,3).

This result holds only if eigenvalues of $M$ and $\tilde{M}$ are much larger than those of $x v_R$ and $\tilde{x} v_R$. Otherwise, SM fermions have sizable admixtures of $\X$ and $\Y$ components in addition to $\psi$. Upon decomposing $\psi$, $\X$, and $\Y$ into SM representations, each representation in $\psi$ appears once in $\X$ or $\Y$ (except SM singlets, as $\Y$ contains two). Consequently, mixing occurs within each representation. SM fermions--defined as massless states in absence of $SU(2)_L$ breaking--are linear combinations of gauge eigenstates in $\psi$ and $\X$ or $\Y$. We focus on mixing in down-quark and charged-lepton sectors and leave a detailed analysis of third generation mixing to \AppRef{app:General Discussion of Third Generation Mixing}. Surprisingly, mixing the third generation of $\Y$ has extremely weak effects on neutrino flavor observables and leptogenesis; therefore, little generality is lost in this assumption.

\subsection{A Direct Third Generation for Down-Quarks and Charge Leptons}

We consider the mass basis for components of $\X$ and assume only the third generation of $\X$ has mass eigenvalue near or below $v_R$. We rescale $\X_1$ and $\X_{2}$ to make the mass matrices equal to the third mass eigenvalue times the identity. This process modifies $\x$; we absorb this into the definition of $\x$. 

We integrate out the heaviest two mass eigenstates, as described previously, resulting in contributions to Yukawa couplings and neutrino masses of seesaw form. However, we treat the third generation differently. We consider the following terms in the Lagrangian after $SU(2)_R$ is broken:
\begin{align}
    \mathscr{L} \supset  
    \big[\bar{d}_i x^*_{Di3}   v_R  +  M_{D3}\bar{D}_3 \big] D_3
    +\big[\ell_i x_{Li3} v_R   + M_{L3} L_3 \big] \bar{L}_3
\end{align}
Heavy mass eigenstates are proportional to the terms in square brackets. Orthogonal directions in field space correspond to SM fermions. Heavy mass eigenstates consist mostly of states in $\X$ or $\psi$ depending on the relative size of $x_{i3}v_R$ and $M_3$.

\begin{figure}[t]
    \centering
    \begin{tikzpicture}
    \draw[ultra thick,scalarbar] (90:1)--(0,0);
    \node at (90:1.3) {$\HL^\dagger$};
    \draw[ultra thick,fermion] (180:1)--(0,0);
    \node at (180:1.3) {$q$};
    \draw[ultra thick,MyGreen,fermion] (0:1)--(0,0);
    \node[xshift=1.1cm] at (0:.8) {\textcolor{MyGreen}{$\bar{D}_3\approx \bar{d}_3'$}};
    \begin{scope}[shift={(0:6.0)}]
    \draw[ultra thick,scalarbar] (90:1)--(0,0);
    \node at (90:1.3) {$\HL^\dagger$};
    \draw[ultra thick, fermion, MyGreen] (180:1)--(0,0);
    \node at (180:1.8) {\textcolor{MyGreen}{$L_3\approx \ell_3'$}};
    \draw[ultra thick,fermion] (0:1)--(0,0);
    \node[xshift=0.6cm] at (0:.8) {$ \bar{e}_3$};
    \end{scope}
\end{tikzpicture}
    \caption{Diagrams generating $b$-quark mass (left) and $\tau$-lepton mass (right).}
    \label{fig:btaudirect}
\end{figure}
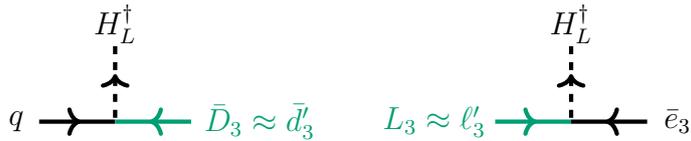

One possibility is $x_{33}v_R \gg M_3,\, x_{13}v_R,\, x_{23}v_R$, in which case heavy mass eigenstates are mostly $\psi_3$, and third-generation SM fermions are mostly $X_3$. We call this scenario a \textit{direct third generation}. Define
\begin{align} \label{eq:delta_epsilon}
    \delta_i = \frac{x_{i3} v_R}{\sqrt{|x_{i3}|^2 v_R^2 + M_3^2}}, \qquad
    \epsilon = \frac{M_3}{\sqrt{|x_{i3}|^2 v_R^2 + M_3^2}}, \qquad
    \sum_i |\delta_i|^2 + \epsilon^2 = 1.
\end{align}
The direct third generation limit corresponds to $\delta_3 \approx 1$ and $\delta_{1,2}, \epsilon \ll 1$. The transformation from gauge eigenstates to mass eigenstates for lepton doublets and down-quark singlets to leading order in $\delta_1$, $\delta_2$, and $\epsilon$ is
\begin{align}
&\begin{pmatrix}
        \ell_1'\\
        \ell'_2\\
        \ell'_3\\
        L'_3
    \end{pmatrix}
    \approx
    \begin{pmatrix}
        1&0 & -\delta_{L1}&0 \\
        0& 1&  -\delta_{L2} &0 \\
        0& 0& \epsilon_L& -1\\
        \delta_{L1}& \delta_{L2}& 1& \epsilon_L \\
    \end{pmatrix}
    \begin{pmatrix}
        \ell_1\\
        \ell_2\\
        \ell_3\\
        L_3
    \end{pmatrix},
    ~
    \begin{pmatrix}
        \bar{d}_1'\\
        \bar{d}'_2\\
        \bar{d}'_3\\
        \bar{D}'_3
    \end{pmatrix}
    \approx
    \begin{pmatrix}
        1&0 & -\delta_{D1}^*&0 \\
        0& 1&  -\delta_{D2}^* &0 \\
        0& 0& \epsilon_D^*& -1\\
        \delta_{D1}^*& \delta_{D2}^*& 1& \epsilon_D \\
    \end{pmatrix}
    \begin{pmatrix}
        \bar{d}_1\\
        \bar{d}_2\\
        \bar{d}_3\\
        \bar{D}_3
    \end{pmatrix}.
\label{eq:lLrotation}
\end{align}
The low energy Lagrangian is obtained by integrating out heavy mass eigenstates and writing $\ell_i$ and $\bar{d}_i$ in terms of the SM fermions $\ell'_i$ and $\bar{d}'_i$ through transformations
\begin{alignat}{3}
&\ell =  {\cmat}_\ell\,\ell' + {\rm heavy}, \hspace{0.5in} 
    &&\cmat_\ell \approx 
    \begin{pmatrix}
        1 & 0 & 0  \\
        0 & 1 & 0 \\
        \delta_{L1} & \delta_{L2} & \epsilon_L  
    \end{pmatrix} ,
    \label{eq:c_L}\\
&\bar{d} =  {\cmat}_{\bar{d}}\,\bar{d}'+ {\rm heavy} , \hspace{0.5in} 
    &&\cmat_{\bar{d}} \approx 
    \begin{pmatrix}
        1 & 0 & 0  \\
        0 & 1 & 0  \\
        \delta_{D1}^* & \delta_{D2}^* & \epsilon_D  
    \end{pmatrix},
    \label{eq:c_D}
\end{alignat}
where we perform a phase redefinition of $\ell_3$ ($\bar{d}_3$) to make $\epsilon_{L}$ ($\epsilon_{D}$) real and positive. The $\delta_{1}$ and $\delta_{2}$ 
terms may still have a phase generated from $\mathcal{L}_{\Sigma, 5}$ in \EqRef{eq:Dim-5 Terms}.

The above results modify down-quark and charged-lepton Yukawa couplings
\begin{align}
    Y_d &=   \x_D P_3 +   \frac{v_R}{M_{D3}} \x_D P_{12} x_D^\dagger  \Omega^{\bar{d}} \label{eq:Y_d_General},\\
    Y_e &=  P_3 x_L^\dagger  +  \frac{v_R}{M_{L3}}\Omega^{\ell \top } x_L P_{12} x_L^\dagger. \label{eq:Y_e_General}
\end{align}
The up-quark Yukawa and right handed neutrino mass are not modified. However, the neutrino Yukawa coupling and low energy neutrino mass become
\begin{align}
    Y_\nu &= \Omega_\ell^\top \left(\yS \MS^{-1} \yS^\dagger \right)v_R\\
    \mathcal{M}_{\nu} &= \Omega_\ell^\top \mathcal{M}_{\bar{\nu}}^* \Omega_\ell \left(\frac{v_L}{v_R}\right)^2 - Y_\nu \mathcal{M}_{\bar{\nu}}^{-1} Y_{\nu}^\top \; v_L^2.
\end{align}

\subsection{Flavor in the Minimal Down-Quark and Charged-Lepton Sectors} \label{sec:Flavor in the Minimal Down-Quark and Charged Lepton Sectors}

We first consider flavor in down-quark and charged-lepton sectors without $SO(10)$ breaking effects. From \EqRef{eq:Y_d_General} and \EqRef{eq:Y_e_General} we see charged-lepton and down-quark Yukawa couplings in the limit under consideration are
\begin{align}
    Y_d = Y_e^\top =   x P_{12}x^\top \cmat_\top +x\,P_3.    
\end{align}
Since we ignore $SO(10)$ breaking, $x$ and $\cmat$ are real and we drop the $D, L$ subscripts. The relation $Y_d = Y_e^\top$ raises two important points. First it falsely predicts identical masses for down-quarks and charged-leptons (up to running corrections). Second, unitary matrices acting on the left of $Y_d$—which contributes to the CKM matrix—can differ significantly from the one acting on the left of $Y_e$, which enters the PMNS matrix. We now illustrate this point concretely.

Coefficients of $x$ take the form
\begin{align}
    x = 
    \begin{pmatrix}
    x_1 & a x_2 & \delta_1 x_3\\
    b x_2 & x_2 & \delta_2 x_3\\
    c x_3 & d x_3 & x_3
    \end{pmatrix}
\end{align}
where we normalize certain entries proportional to $x_2$ and $x_3$. Notably, parameters $\delta_1$, $\delta_2$, and $\epsilon$ enter the neutrino observables via $\cmat_\ell$ and are constrained as we will see in \SecRef{sec:The Minimal Realistic Up Quark and Neutrino Sectors}. To reduce the number of independent parameters, we rotate the $(X_1, X_2)$ fields to eliminate one degree of freedom, $c$.

To perform a perturbative analysis we assume the following scales for parameters: 
\begin{enumerate}
    \item $\{a,\,b,\,d\} \sim \mathcal{O}(10^{0})$
    \item $\{ \delta_2,\,\epsilon\} \sim \mathcal{O}(10^{-1})$
    \item $\{\x_{2},\x_{3},\,\delta_1\} \sim \mathcal{O}(10^{-2})$
    \item $x_1 \sim \mathcal{O}(10^{-3})$
\end{enumerate}
This assumed structure of $\x$ need not hold in general\footnote{This guess can be motivated by approximate $U(1)$ flavor symmetries.}. However, it is largely consistent with observations; for example, assumed scales of $x_{1,2,3}$ are consistent with 
\begin{align} 
    y_{e,d} \approx x_1^2,~ y_{\mu,s} \approx x_2^2,~ {\rm and}~ y_{b,\tau} \approx x_3.
\end{align}
The CKM matrix, $V_{\rm CKM} = V_u^\dagger V_d$, is found using $V_u=I_{3\times 3}$ and $V_d$ from the diagonalization $V_d^\top Y_{d} Y_d^\top V_d =  {\rm diag}(y_d^2,\, y_s^2,\, y_b^2) $. To leading order we have
\begin{align} \label{eq:YdYdTdir}
    Y_d Y_d^\top =
x_3^2
\begin{pmatrix}
 \delta_1^2 & \delta_1 \delta_2 & \delta_1 \\
 \delta_1 \delta_2 & \delta_2^2 &
   \delta_2 \\
 \delta_1 & \delta_2 & 1 
\end{pmatrix}.
\end{align}
If sub-leading terms are included in $Y_d Y_d^\top$ this gives 
\begin{align}
&|V_{cb}| \approx \delta_2 + (\delta_2^3 - d x_2 \epsilon),\\
&|V_{ub}| \approx \delta_1  - ad x_2 \epsilon,\\
&|V_{us}| \approx \delta_1/\delta_2 +\frac{ \left(a^3 x_2^4+a x_2^3 \left(b^2 x_2+x_2\right)+a d \delta_2 x_2 x_3^2 \epsilon \right)}{\delta_2^2
   x_3^2} .    
\end{align}
The CKM matrix is relatively insensitive to $a,\,b,\, d$. At leading order this predicts
\begin{align} \label{eq:CKMpred}
|V_{us}| \approx |V_{ub}|/|V_{cb}|,   
\end{align}
which almost agrees with data: $|V_{us}| \simeq 2.4 \,|V_{ub}|/|V_{cb}|$.
This prediction is relatively robust because direct contributions to $Y_d$ are linear in $x_3$ while seesaw contributions are bilinear in $x_{i}$ so every entry in $Y_d Y_d^\top$ is dominated by direct contributions. 
%Discrepancy with data will be removed by adding SO(10) breaking as discussed in the next subsection.

The lepton rotation matrix $V_e$ in the PMNS matrix is determined by diagonalizing $V_{e}^\top  Y_{e} Y_e^\top V_{e} =  {\rm diag}(y_e^2,\, y_\mu^2,\, y_\tau^2) $. To leading order this matrix is
\begin{align}
   Y_e Y_e^\top =  \left(
\begin{array}{ccc}
 a^2 x_2^2 \left(\left(a^2+1\right) x_2^2+d^2
   x_3^2\right) & a x_2^2 \left(x_2^2
   \left(a^2+b^2+1\right)+d^2 x_3^2\right) & a d x_2
   x_3^2 \\
 a x_2^2 \left(x_2^2 \left(a^2+b^2+1\right)+d^2
   x_3^2\right)~& x_2^4
   \left(a^2+\left(b^2+1\right)^2\right)+d^2 x_2^2 x_3^2 ~& d
   x_2 x_3^2 \\
 a d x_2 x_3^2 & d x_2 x_3^2 & x_3^2 \\
\end{array}
\right). 
\end{align}
If sub-leading terms are included in $Y_e Y_e^\top$ this gives 
\begin{align} 
    &|V_{e23}| \approx d x_2 +  \frac{\delta_2}{x_3}\left(d^2 x_3^2 - (b^2+1) x_2^2\right) ,  \label{eq:V_e_23} \\ &|V_{e13}| \approx a d x_2+\frac{a \delta_2 x_2^2}{x_3},\label{eq:V_e_13}\\
    &|V_{e12}|\approx a\frac{  x_2^2
   \left(a^2+b^2+1\right)+d^2 x_3^2 }{x_2^2
   \left(a^2+b^4 + 2b^2 + 1\right)+d^2  x_3^2 }.\label{eq:V_e_12}
\end{align} 
We do not show sub-leading terms for $V_{e12}$ as the expression is lengthy. The main observation is these values have almost no relation to the CKM matrix elements calculated above. Direct contributions, so dominant in $Y_d Y_d^\top$, only enter the 33 entry of $Y_e Y_e^\top$, so $\delta_{1,2}$ contribute only at subleading order in the mixing angles of $V_e$. 

We outline a few limitations of the minimal $(d,e)$ sector, each of which can be addressed with modest extensions. First, without incorporating $SO(10)$-breaking effects, it predicts identical masses for down-type quarks and charged leptons, in tension with observation. Second, CKM mixing angles are only reproduced up to $\mathcal{O}(1)$ corrections.  Third, $CP$ violation is absent at this level, and thus the CKM phase is not predicted. Finally, as we will see in \SecRef{sec:The Minimal Realistic Up Quark and Neutrino Sectors}, some large mixing angles in $V_e$ are typically needed to account for flavor observables in the $(u,\nu)$ sectors. %While this naturally occurs for $|V_{e12}|$, $|V_{e13}|$ and $|V_{e23}|$ remain too small, typically $\lesssim \mathcal{O}(10^{-2})$.
%\KH{I think the last point is mainly addressed by $V_\nu$ larther than by $SO(10)$ breaking.} \KL{I think the point here is that additional large mixing angles are also required from the charged lepton rotation $V_e$ even after $V_\nu$ gets large angles.}

To address the first three points within a renormalizable $SO(10)$ framework, we can include $SO(10)$-breaking mass terms in \EqRef{eq:XMassesSigma}. Alternatively, it is natural to consider dimension-5 operators from \EqRef{eq:Dim-5 Terms}, given the proximity of GUT and Planck scales. These modify $\x_D$ and $\x_L$ and introduce imaginary parts. Crucially, they induce complex phases to $\delta_1$ and $\delta_2$ in $\cmat_\ell$ of \EqRef{eq:c_L}, transmitting $CP$ violation to the neutrino sector even if $CP$ violation involving $\Y$ are negligible. The last point may be addressed by considering less stringent assumptions on the scales of the parameters or through dimension-5 operators.

\subsection{Flavor in the Minimal Up-Quark and Neutrino Sectors} \label{sec:The Minimal Realistic Up Quark and Neutrino Sectors}

We now explore the minimal $(u,\nu)$ sector where $SO(10)$ and $CP$ violating terms are ignored, and we work in the approximation that the top quark Yukawa coupling arises from the seesaw mechanism. In this limit, the up quark Yukawa matrix is 
\begin{align} \label{eq:M_u_Approx}
    Y_u & =
    \begin{pmatrix}
    2\y_1^2& 0 & 0 \\
    0&2\y_2^2 & 0\\
    0 & 0 &2 \y_3^2   
    \end{pmatrix}.
\end{align}
However, generating $b$ and $\tau$ masses directly rather than from a seesaw, modifies the disastrous mass relations of \EqRef{eq:Mu_Mnu_relation_Simplified} to
\begin{align} 
\label{eq:Mu_Mnu_relation_Simplified_with_c} 
    \mathcal{M}_\nu & = \frac{2}{5} \, \cmat_\ell \, Y_u \, \cmat_\ell^\top \,  \frac{v_L^2}{v_R},
\end{align}
offering the prospect of a realistic and remarkably simple $(u,\nu)$ sector.

We work in a basis where $Y_u$ is real and diagonal so \EqRef{eq:Mu_Mnu_relation_Simplified_with_c} implies 
\begin{align} \label{eq:M_nu}
    \mathcal{M}_\nu = \frac{2}{5} \,y_t
    \begin{pmatrix}
        \frac{y_u}{y_t} +\delta_1^2  & \delta_1 \delta_2  & \delta_1 \epsilon \\
        \delta_1 \delta_2  & \frac{y_c}{y_t}+ \delta_2^2  & \delta_2 \epsilon \\
        \delta_1 \epsilon  & \delta_2 \epsilon  & \epsilon^2 
    \end{pmatrix} \frac{v_L^2}{v_R}  .
\end{align}
From \EqRef{eq:M_nu}, we see large mixing angles enter the PMNS matrix if $\delta_1,\, \delta_2$ and $\epsilon$ are of the same order of magnitude. We begin by ignoring complex phases in $\delta_1$ and $\delta_2$ arising from the $(d,e)$ sector, so $\mathcal{M}_\nu$ is diagonalized by a rotation $V _\nu$
\begin{align}
    V_\nu^\top \mathcal{M}_\nu V_\nu = \widehat{\mathcal{M}}_\nu.
\end{align}
This unitary matrix enters the PMNS matrix in the standard way
\begin{align}
U_{\rm PMNS} &= V_{e}^\dagger V_\nu .
\end{align}
We choose the following parametrization for the PMNS matrix:
\begin{align}
U_{\rm PMNS} &=
\left(\begin{array}{ccc}
1 & 0 & 0 \\
0 & c_{23} & s_{23} \\
0 & -s_{23} & c_{23}
\end{array}\right) \left(\begin{array}{ccc}
c_{13} & 0 & s_{13}  \\
0 & 1 & 0 \\
-s_{13}  & 0 & c_{13}
\end{array}\right) \left(\begin{array}{ccc}
c_{21} & s_{12} & 0 \\
-s_{12} & c_{12} & 0 \\
0 & 0 & 1
\end{array}\right),  
\end{align}
where $c_{ij} = \cos\theta^{(\rm PMNS)}_{ij}$ and $s_{ij} = \sin\theta^{(\rm PMNS)}_{ij}$.
\begin{figure}[t]
    \centering
    \includegraphics[width=1\linewidth]{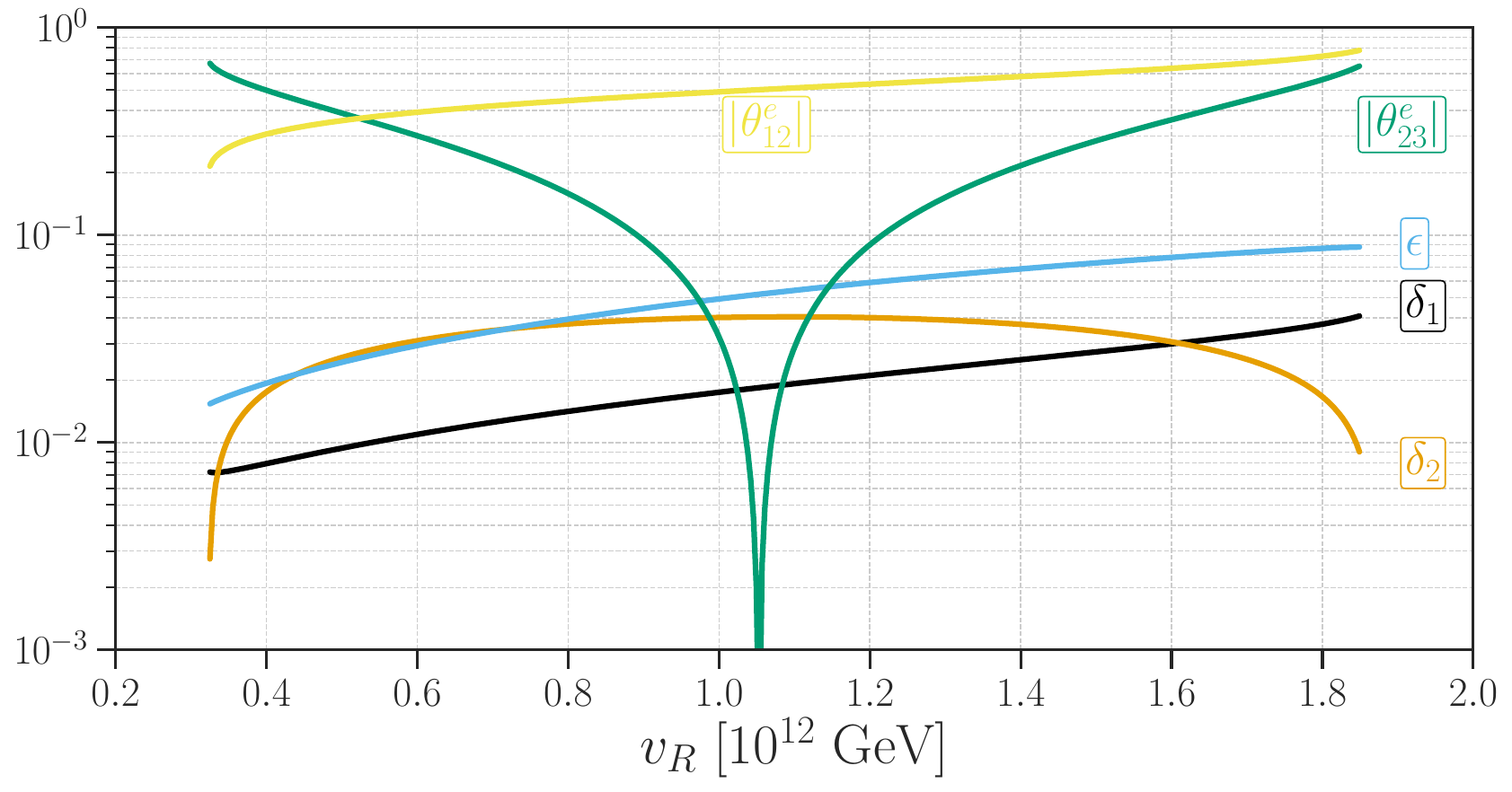}

    \caption{Parameters of the minimal realistic up-quark and neutrino sectors that give the observed PMNS angles and neutrino mass spectrum as a function of $v_R$. The plot cuts off for $v_R$ where no exact solutions exists. }
    \label{fig:Flavor_Fit_MUN}
\end{figure}

Neutrino angles and masses depend on 3 unfixed parameters $\{\delta_1,\, \delta_2,\, \epsilon\}$, 3 charged lepton angles, and $v_R$. We wish to fit to 5 observables $ \Delta m^2_{32},$ $ \Delta m^2_{21}, \theta_{23}^{({\rm PMNS})},$ $ \theta_{13}^{({\rm PMNS}},$ $ \theta_{12}^{({\rm PMNS})}$. We have 2 more parameters than observables. However, charged lepton angles are not completely free parameters as described in \SecRef{sec:Flavor in the Minimal Down-Quark and Charged Lepton Sectors}. Specifically, we expect $|V_{12}|\gg |V_{23}|\gtrsim |V_{13}|\approx  |V_{23}| |V_{12}|$. Therefore, parametrizing the charged lepton rotation matrix as $V_{e} = R^{(e)}_{23} R^{(e)}_{13}R^{(e)}_{12}$ we expect a small $\theta^e_{13}$. We use this as motivation to approximate $\theta^{e}_{13} = 0$ and perform the following fit for each value of $v_R$:

\begin{tikzpicture}
    \node at (0,1) {\textsc{Model Parameters (5)}};
    \draw[ultra thick] (-2.5,.7) -- (2.5,0.7);
    \node at (0,0.2) {$\{\delta_1, \delta_2, \epsilon,  \theta_{23}^{(e)},\theta_{12}^{(e)}\}$};

    \draw[ultra thick, <->] (3,.7) -- (4,0.7);
    
    \node at (8,1) {\textsc{Flavor Observables (5)}};
    \draw[ultra thick] (4.3,.7) -- (11.7,0.7);
    \node at (8,0.2) {$\{  \Delta m^2_{32}, \Delta m^2_{21}, \theta_{23}^{({\rm PMNS})}, \theta_{13}^{({\rm PMNS})}, \theta_{12}^{({\rm PMNS})}\}$};
\end{tikzpicture}

When fitting neutrino masses we take into account effects of running. Treating $m_3^{(\nu)}$ as an effective parameter, the SM running is \cite{Antusch:2005gp}
\begin{align}
    \frac{d\log m_3^{(\nu)}}{d\log \mu} 
    \approx 
    \frac{1}{16\pi^2}\times
    \begin{cases}
          6 y_t^2 - 3g_2^2,& \mu < m_{\bar{\nu}_3}\\
          6 y_t^2 - \frac{9}{2}g_2^2 - \frac{9}{10}g_1^2,& \mu > m_{\bar{\nu}_3}
    \end{cases}
\end{align}
For example, setting $m_{\bar{\nu}_3} \approx  v_R \approx 10^{12}~{\rm GeV}$ gives $m_3^{(\nu)}(v_R) \approx 1.25 \times m_3^{(\nu)}(M_W)$.

In \FigRef{fig:Flavor_Fit_MUN}, we show predictions for $\delta_1,\, \delta_2,\, \epsilon ,\, \theta^e_{12},\,\theta^e_{23}$ as a function of $v_R$. In this minimal $(u,\nu)$ sector, $\cmat_\ell$ of \EqRef{eq:c_L}—arising from direct contributions to $\tau$—provides a simple explanation for hierarchies $m_2/m_3 \gg m_c/m_t$, $V_{23} \gg V_{cb}$, and $V_{13} \gg V_{ub}$. 

Does this successful understanding of neutrino masses and mixings survive deviations away from the minimal $(u, \nu)$ sector? If so, how robust is the prediction for the range of $v_R$? For example, in the analysis above we assumed the top quark mass generation mechanism is purely seesaw; in reality, it is intermediate between seesaw and direct. Also there may be sizable effects from dimension-5 $SO(10)$ breaking operators \EqRef{eq:Dim-5 Terms} or from renormalizable $SO(10)$ breaking mass terms \EqRef{eq:XMassesSigma}; such effects are necessarily large in the $(d,e)$ sector. To study these effects we find it convenient to consider a scaling transformation.

\subsubsection*{\emph{A Scaling Transformation}}

Suppose \EqRef{eq:Mu_Mnu_relation_Simplified_with_c} relating neutrino and up quark masses is modified to
\begin{align} 
\label{eq:Mu_Mnu_relation_scaled} 
    \mathcal{M}_\nu & = \frac{2}{5} \, \cmat_\ell \, \begin{pmatrix}
    AB \; y_u & 0 & 0 \\
    0& AB \; y_c & 0\\
    0 & 0 & B \; y_t   
    \end{pmatrix} \, 
    \cmat_\ell^\top \,  \frac{v_L^2}{v_R}.
\end{align}
The minimal model being $A=B=1$. This results from the minimal model by scaling the $\tilde{x}_{1,2,3}^2$ parameters that reside in \EqRef{eq:M_nu} by
\begin{align} \label{eq:scaling}
    \y_{1,2}^2\rightarrow AB \, \y_{1,2}^2, \quad 
     \y_3^2 \rightarrow B\,  \y_3^2.
\end{align}
This modifies \EqRef{eq:M_nu} to
\begin{align} \label{eq:M_nuscaled}
    \mathcal{M}_\nu = \frac{2}{5} \,By_t
    \begin{pmatrix}
        A\frac{y_u}{y_t} +\delta_1^2  & \delta_1 \delta_2  & \delta_1 \epsilon \\
        \delta_1 \delta_2  & A\frac{y_c}{y_t}+ \delta_2^2  & \delta_2 \epsilon \\
        \delta_1 \epsilon  & \delta_2 \epsilon  & \epsilon^2 
    \end{pmatrix} \frac{v_L^2}{v_R}  .
\end{align}
Rescaling the values of $\delta_{1,2}$ and $\epsilon$ that gave the successful fit in the minimal model
\begin{align} \label{eq:paramscaling}
    \delta_{1,2} \rightarrow  A^{1/2}\,  \delta_{1,2},\quad 
     \epsilon \rightarrow A^{1/2} \, \epsilon, 
\end{align}
\EqRef{eq:M_nuscaled} becomes
\begin{align} \label{eq:M_nuscaledtwice}
    \mathcal{M}_\nu = \frac{2}{5} \,AB\,y_t
    \begin{pmatrix}
        \frac{y_u}{y_t} +\delta_1^2  & \delta_1 \delta_2  & \delta_1 \epsilon \\
        \delta_1 \delta_2  & \frac{y_c}{y_t}+ \delta_2^2  & \delta_2 \epsilon \\
        \delta_1 \epsilon  & \delta_2 \epsilon  & \epsilon^2 
    \end{pmatrix} \frac{v_L^2}{v_R}.
\end{align}
This yields the observed neutrino mass ratios and mixing angles; furthermore, the observed  normalization of the neutrino mass matrix results from the scaling
\begin{align} \label{eq:scalingvR}
    v_R \rightarrow  AB \, v_R. 
\end{align}
In summary, solutions for the minimal $(u,\nu)$ sector become solutions for the rescaled theory described by \EqRef{eq:Mu_Mnu_relation_scaled} provided model parameters are also rescaled according to \EqRef{eq:paramscaling} and \EqRef{eq:scalingvR}. Hence, our understanding that order unity neutrino mixing angles and $m_{\nu_2}/m_{\nu_3}$ can result from a direct $\tau$ lepton mass via the parameters $\delta_{1,2}$ and $\epsilon$ persists beyond the minimal $(u,\nu)$ sector. In fact, it applies to theories of the form \EqRef{eq:Mu_Mnu_relation_scaled} even if the $y_u$ and $y_c$ coefficients are not equal, because the $y_u$ contribution to the neutrino mass matrix is so small.

\subsubsection*{\emph{A Mixed Seesaw/Direct Top Yukawa Coupling}}

The top quark neither arises through pure seesaw nor purely directly. Rather, it involves a mixing parameter $t_{\Y} =s_{\Y}/c_{\Y} = \y_3 v_R / \tilde{M}_3$ within a factor of 2.5 of unity.  In \AppRef{app:General Discussion of Third Generation Mixing} we show that the neutrino mass matrix then takes the form of \EqRef{eq:Mu_Mnu_relation_scaled} with $AB=1$ and $B=1 + t^2_{\Y}$. Hence, the fit of \FigRef{fig:Flavor_Fit_MUN} applies with the range of $v_R$ unchanged but values of $\delta_{1,2}$ and $\epsilon$ decreased by a factor $\sqrt{1+ t^2_{\Y}}$.

\subsubsection*{\emph{SO(10) Breaking from $\Y \mathcal{S} \Y$}}

Dimension-5 operators may be negligible due to additional scale suppression; however, renormalizable mass corrections from $\tilde{X} (\Sigma, S) \tilde{X}$ may be quite large. The field $\Y$ has several 3221 components, $\Y_a$, each picking up mass corrections proportional to group theory factors $c_a$ shown in Table \eqref{tab:Xmass corrections}. The seesaw diagrams from $\Y_a$ exchange for up quark and neutrino masses receive corrections. Hence, relations between up and neutrino mass matrices takes the form \EqRef{eq:Mu_Mnu_relation_scaled}, except that coefficients of $y_u$ and $y_c$ need not be the same.  However, the $y_u$ contribution to the neutrino mass is very small, of order $10^{-6}$ eV, and can be ignored when computing $m_{\nu_{2,3}}$ and neutrino mixing angles, allowing us to use \EqRef{eq:Mu_Mnu_relation_scaled}.

We first describe how the scaling transformations \EqRef{eq:scaling} arise from the introduction of $SO(10)$ breaking masses induced by $\expval{\mathcal{S}}$ in \EqRef{eq:XMassesSigma}. Motivated by approximate flavor symmetries, we take these corrections to be diagonal in the same basis which diagonalizes $\tilde{M}$, so that $\tilde{M}_{ia} = \tilde{M}_i(1 + b_i c_a)$, where $b_i = \frac{v_{\bm{54}}}{\sqrt{15}} \,\tilde{\beta}_i/\tilde{M}_i$. Rescaling $\Y_i$ fields to set $\tilde{M}_i = v_R$, the seesaw diagrams give 
\begin{align}
    &Y_u^{ii} = \y_i^2 \left(\frac{1}{1+b_i} +\frac{1}{1-b_i/4} \right)\\
    &(\Omega_\ell^{-1}\mathcal{M}_\nu\Omega_\ell^{\top-1})^{ii} = \y_i^2 \Bigg[\left(\frac{3/4}{1+b_i} +\frac{1/2}{1-3b_i/2} \right) \notag \\&~~~~~~~~~~~~~~~~~~~~~~~~~~~~~~~~~ - \left(\frac{3/4}{1+b_i}\right)^2\left(\frac{3/4}{1+b_i} +\frac{1/2}{1-3b_i/2} \right)^{-1}\Bigg]\frac{v_L^2}{v_R}.
\end{align}
We define the function
\begin{align} \label{eq:Gamma_Function}
    \Gamma(b_i) = \frac{(\Omega_\ell^{-1}\mathcal{M}_\nu\Omega_\ell^{\top-1})^{ii}(b_i)}{ (\Omega_\ell^{-1}\mathcal{M}_\nu\Omega_\ell^{\top-1})^{ii}(0)}\, \, \frac{Y_u^{ii}(0)}{Y_u^{ii}(b_i)}
\end{align}
and plot $\Gamma(b)$ in \FigRef{fig:AB_Changes}. Ignoring contributions to neutrino masses from $y_u$, and therefore corrections from $b_1$, the scaling parameters are 
\begin{align} \label{eq:b_scaling}
     AB =\Gamma(b_2),~ B = \Gamma(b_3). 
\end{align}

\subsubsection*{\emph{SO(10) Breaking from $\Y \Sigma \Y$}}

We now consider how the above scaling transformations can arise from introduction of $SO(10)$ breaking masses induced by $\expval{\Sigma}$ in \EqRef{eq:XMassesSigma}. Importantly, $SO(10)$ breaking mass terms involving $\expval{\Sigma}$ do not affect neutrinos or charged leptons as $S,\, T_{L,R},$ and $ \Delta$ are not charged under $B-L$. However, $U,\,\bar{U}$ and $Q,\, \bar{Q}$ receive imaginary mass terms. We focus only on $U$ and $\bar{U}$ as an example. A series of rotations and rescalings as described previously leads to a Lagrangian
\begin{align}
    \mathscr{L}_U \; \supset \; q \,\y \,\bar{U} \, \HL + \bar{q} \, \y \, U \, \HR - v_R \, U_i\left(\delta_{ij} + i a_k \varepsilon_{ijk}\right) \bar{U}_j,
\end{align}
where $\y$ is real, diagonal, and includes no $SO(10)$ breaking effects. Coefficients $a_k$ contain effects of $SO(10)$-breaking mass terms. Corrections for $Q$ are a factor 2 smaller; we aim for a proof of principle and ignore this difference and write 
\begin{align}
    Y_u^{ij} = 2\y_i\left(\delta_{ij} + i a_k \varepsilon_{ijk}\right)^{-1} \y_j = \frac{2 \y_i\y_j}{1- |\vec{a}|^2}\left(\delta_{ij} - i \varepsilon_{ijk}a_k 
 - a_i a_j\right)
\end{align}
where no sum is made over the $i,j$ indices.
Ignoring corrections from $\y_1$
\begin{align}
    &y_c \approx 2\y_2^2 \left(\frac{1}{1 - a_3^2}\right) \label{eq:y_c_From_SO10_Breaking_Masses}\\
    &y_t \approx 2\y_3^2  \left(\frac{1-a_3^2}{1 - |\vec{a}|^2}\right).
\end{align}
The parameters of the scaling \EqRef{eq:scaling} that leads to \EqRef{eq:Mu_Mnu_relation_scaled} are
\begin{align}
    AB = 1-a_3^2,~ B = \frac{1-|\vec{a}|^2}{1-a_3^2}.
\end{align}

For both above scenarios, a successful fit to neutrino data is maintained and, from \EqRef{eq:scalingvR}, $v_R$ is scaled by a factor $AB$. If mass corrections push any mass below $v_R$, integrating out heavy states becomes invalid. For corrections from $\expval{\mathcal{S}}$, this prevents very large corrections from $b_3$ but allows large corrections from $b_2$. For corrections from $\expval{\Sigma}$ the largest correction to $AB$ is $1-a_3^2 = y_c/y_t$. Theoretically, $v_R$ may be lowered by around 2 orders of magnitude, illustrating the point that $SO(10)$ breaking may lead to significant corrections. Of course, extreme corrections are very fine-tuned, but the point remains that $\mathcal{O}(1)$ corrections appear quite plausible.

\section{Leptogenesis in Higgs Parity Unification} \label{sec:Leptogenesis}

In {\HPU}, all $CP$ violation originates from $\expval{\Sigma}$. Therefore, it is natural to ask if the baryon asymmetry of the universe can be explained in this model. While there are several potential sources of $CP$-violating decays—such as heavy states in $\X$ and $\Y$—any asymmetry they generate would be washed out by interactions with lighter right-handed neutrinos in thermal equilibrium. These have masses $\mathcal{M}_{\bar{\nu}} \sim \mathcal{M}_u (v_R/v_L)$, and may generate a baryon asymmetry through their out-of-equilibrium decays~\cite{Fukugita:1986hr}. In this section, we compute the resulting asymmetry.

A challenge arises from the Davidson-Ibarra bound \cite{Davidson:2002qv} which requires leptogenesis via lightest right-handed neutrino to have $M_1 \gtrsim 10^9$ GeV. This implies $v_R \gtrsim 10^{14}$ GeV, in tension with Higgs parity, gauge coupling unification, and proton decay.

We therefore consider leptogenesis via the second-lightest right-handed neutrino with mass $\mathcal{M}_{\bar{\nu}2} \sim 10^{10}~\GeV \times (v_R/10^{12}~\GeV)$. To ensure sufficient abundance of $\bar{\nu}_2$, reheating must occur with $T_{\rm RH}\gtrsim \mathcal{M}_{\bar{\nu}2}$. At such temperatures, $SU(2)_R$ gauge interactions thermalize $\bar{\nu}_{1,2}$. An estimate suffices to demonstrate this: for $T < v_R$, the interaction rate is approximated by $\Gamma \sim G_R^2 T^5$, analogous to standard weak interactions, with $G_R = 1/(\sqrt{2} v_R^2)$. Assuming radiation domination $H \sim T^2 / M_P$ and decoupling occurs when $\Gamma \sim H$. This corresponds to the temperature
\begin{align}
    T_{\rm dec} \sim (G_R^2 M_P)^{-1/3} \sim 10^{10}~{\rm GeV} \left(\frac{v_R}{10^{12}~{\rm GeV}}\right)^{4/3} \sim \mathcal{M}_{\bar{\nu}2} \left(\frac{v_R}{10^{12}~{\rm GeV}}\right)^{1/3}.
\end{align}
We assume the reheating temperature exceeds both $T_{\rm dec}$ and $\mathcal{M}_{\bar{\nu}2}$.

It is important to check any asymmetry generated by $\bar{\nu}_2$ is not washed out by $\bar{\nu}_1$. We investigate this requirement and then discuss the asymmetry generated by $\bar{\nu}_2$. 

\subsection{The Effects of $\mathbf{\bar{\nu}_1}$ on the Lepton Asymmetry} 

We now consider whether lepton-violating processes involving $\bar{\nu}_1$ wash out preexisting $B-L$ asymmetry. The analytical solution for $B-L$ asymmetry is \cite{Buchmuller:2004nz}
 \begin{align}
     Y_{B-L}(z) = Y_{B-L}(z_0)e^{-\int_{z_0}^z W(z')dz'} + Y^{(\bar{\nu}_1)}_{B-L}(z)
 \end{align}
$z = M_1/T$, $z_0 \gg 1$ sets the initial condition, $W(z)$ is a function describing washout, and $Y^{(\bar{\nu}_1)}_{B-L}$ is the $B-L$ asymmetry from decay of $\bar{\nu}_1$ in the situation where $Y_{B-L}(z_0) = 0$; wthichs is negligible by the Davidson-Ibarra bound and $Y_{B-L}(z_0)$ can be taken as the asymmetry generated by the decay of $\bar{\nu}_2$. Therefore, we must determine the fraction of the asymmetry generated from decays of $\bar{\nu}_2$ which is washed out, i.e. $\exp\left(-\int_{z_0}^z W(z')dz'\right)$. The dominant effect from inverse decays gives 
 \begin{align}
     &W(z) = \frac{1}{4}z^3 K_1(z) \, \frac{\tilde{m}_1}{m_*}
     \implies \int_{0}^\infty W(z')dz' \approx 1.18 \, \frac{\tilde{m}_1}{m_*}
 \end{align}
where, following notation of \cite{Giudice:2003jh}, we define
\begin{align}
    &m_*\equiv \frac{16 \pi^{5 / 2}}{3 \sqrt{5}} g_*^{1/2} \frac{v_L^2}{M_{\rm{Pl}}} \approx 10^{-3}~ \mathrm{eV},\quad{\rm and }\quad \tilde{m}_1 \equiv \frac{(Y_\nu^\dagger Y_\nu)_{11} v_L^2}{M_1}.
\end{align}
For example, if $\tilde{m}_1/m_*\lesssim 0.1$, $\bar{\nu}_1$ erases less then 10\% of the pre-existing asymmetry. We verify this condition when specializing to {\HPU}.

\subsection{The Lepton Asymmetry Produced by $\mathbf{\bar{\nu}_2}$ Decay}

\begin{figure}[h!]
    \centering
    \begin{tikzpicture}[line width=1.5 pt, scale=1.1,baseline={(0,-2.2cm)}]
    \draw[scalar] (0.333,0)--(1.666,1) node[midway,left, yshift = 0.3 cm] {$H$};
    \draw[fermion] (-1,0)--(0.333,0);
    \node at (180:1.3) {$\bar{\nu}'_2$};
    \draw[ultra thick,fermionbar] (0.333,0)--(1.666,-1)node[midway,left, yshift = -0.3 cm] {$\ell'_{i}$};
    \draw[ultra thick,fermion,red] (1.666,-1)-- (1.666,1) node[midway,right, xshift = 0.2 cm] {};
    \draw[scalarbar] (1.666,-1)--(3,-1);
    \draw[fermionbar] (1.666,1)--(3,1);
    \node at (3.3,-1) {$H$};
    \node at (3.3,1) {$\ell'_j$};
    \begin{scope}[xshift = 7.5cm, yshift = -1.5cm]
        \draw[ultra thick, -> ] (-3,0)--(-2,0);
        \node at (180:1.3) {$\bar{\nu}'_2$};
        \draw[fermion] (-1,0)--(0,0);
        \draw[fermionbar] (0,0) to[out=-90, in=-90,looseness=2] (1,0);
        \draw[scalar] (0,0) to[out=90, in=90,looseness=2] (1,0);
        \draw[fermionbar] (1,0)--(2,1);
        \draw[scalarbar] (1,0)--(2,-1);
        \filldraw[red] (1,0) circle (2.5pt);
        \node at (2.3,-1) {$H$};
        \node at (2.3,1) {$\ell'_j$};
        \node at (.5,-1) {$\ell'_i$};
        \node at (.5,1) {$H$};
    \end{scope}
     \begin{scope}[yshift = -3cm]
        \node at (180:1.3) {$\bar{\nu}'_2$};
        \draw[fermion] (-1,0)--(-0.2,0);
        \draw[fermionbar] (-0.2,0) to[out=-90, in=-90,looseness=2] (0.8,0);
        \draw[scalar] (-0.2,0) to[out=90, in=90,looseness=2] (0.8,0);
        \draw[fermionbar] (2.2,0)--(3,1);
        \draw[scalarbar] (2.2,0)--(3,-1);
        \draw[fermion,red] (.8,0)--(2.2,0) node[midway,above, yshift = 0.3 cm] {};
        \node at (3.3,-1) {$H$};
        \node at (3.3,1) {$\ell'_j$};
        \node at (.2,-1) {$\ell'_i$};
        \node at (.2,1) {$H$};
    \end{scope}
\end{tikzpicture}
    \caption{Diagrams contributing to leptogenesis. \textbf{\textcolor{red}{Red}} fermion propagators on the left-hand side correspond to propagators of neutral fermions heavier than $\bar{\nu}_2$. These are $S'_3$, $N'_3$, $\bar{\nu}'_3$, and $T_{Li}$ and can be integrated out to yield an effective dimension-5 operator, $(\mathcal{M}_{\nu,{\rm Lepto}}^{ij}/v_L^2)(\ell'_i \HL)(\ell'_j \HL)$, depicted by a \textbf{\textcolor{red}{red}} dot on the right. }
    \label{fig:Leptogenesis_Feynman_Diagram}
\end{figure}
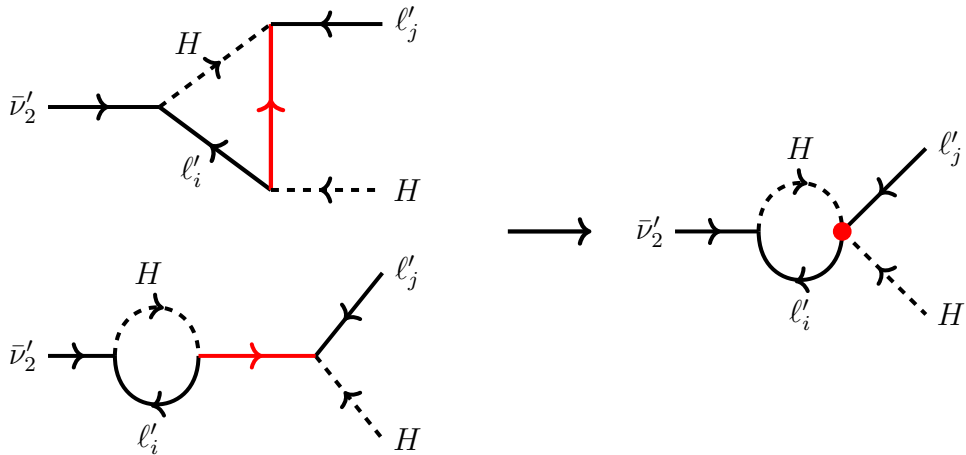

We first consider the lepton asymmetry produced by the decay of $\bar{\nu}_2$ in a general theory. The lepton asymmetry decay factor is \cite{Davidson:2008bu}
\begin{align} \label{eq:epsilon_2}
    \varepsilon_2 \equiv \frac{\Gamma\left(\bar{\nu}_2 \rightarrow \phi \ell \right)-\Gamma\left(\bar{\nu}_2 \rightarrow \bar{\phi} \bar{\ell}\right)}{\Gamma\left(\bar{\nu}_2 \rightarrow \phi \ell\right)+\Gamma\left(\bar{\nu}_2 \rightarrow \bar{\phi} \bar{\ell}\right)}  = \frac{3\mathcal{M}_{\bar{\nu}2}}{16 \pi v_L^2} \frac{\operatorname{Im}  \left(Y_{\nu,{\rm Lepto}}^{\dagger} \mathcal{M}_{\nu,{\rm Lepto}} Y_{\nu,{\rm Lepto}}^* \right)_{2 2} }{ (Y_{\nu,{\rm Lepto}}^{\dagger} Y_{\nu,{\rm Lepto}} )_{22}}.
\end{align}
This lepton asymmetry decay factor results from interference of tree-level amplitudes with 1-loop amplitudes illustrated in \FigRef{fig:Leptogenesis_Feynman_Diagram}, and $\mathcal{M}_{\nu,{\rm Lepto}}$ ($Y_{\nu,{\rm Lepto}}$) are effective left-handed neutrino mass (neutrino Yukawa coupling) after integrating out SM singlets above the scale $\mathcal{M}_{\bar{\nu}2}$, including the heaviest right handed neutrino.

At the time of $\bar{\nu}_2$ decay, lepton-number–violating interactions involving $\bar{\nu}_2$ remain active, placing the decay in the strong washout regime. The corresponding efficiency, which quantifies the extent of washout in this regime, is approximately
\cite{Giudice:2003jh}
\begin{align} \label{eq:eta2}
    \eta_2 \approx \left(\frac{0.55 \;  \rm meV}{\tilde{m}_2}\right)^{1.16}\quad {\rm for }~ \tilde{m}_2\gtrsim 1~\rm{meV},
\end{align}
where $\tilde{m}_2$ is the contribution to the light seesaw mass from $\bar{\nu}_2$ exchange
\begin{align} \label{eq:m2tilde}
    \tilde{m}_2 \equiv \frac{(Y_\nu^\dagger Y_\nu)_{22} \,v_L^2}{\mathcal{M}_{\bar{\nu}2}}.
\end{align}
The baryon asymmetry generated from this decay is 
\begin{align} \label{eq:Y_B}
    Y_{B} &= \frac{28}{79} \; \frac{135\, \zeta(3)}{4 \pi^4 g_*} \; \eta_2 \, \varepsilon_2 ,
\end{align}
where we use $g_* = 108.5$ accounting for SM degrees of freedom and $\bar{\nu}_1$.

\subsection{Leptogenesis from the Minimal $(u,\nu)$ Sector}

We now compute leptogenesis in the minimal $(u, \nu)$ sector of {\HPU}, as defined in \SecRef{sec:The Minimal Realistic Up Quark and Neutrino Sectors}, assuming that $CP$ violation arises solely from a phase in $\delta_{L2}$. (We take $\delta_{L1}$ to be real, as it does not contribute to leptogenesis.) Since $\delta_{L2} = x_{L23}/x_{L33}$, a sizable phase requires the imaginary part of $x_{L23}$—generated by dimension-5 operators—to be comparable in magnitude to its real part. Given that $x_{L33} \sim 10^{-2}$ and $|\delta_{L2}| \lesssim 10^{-1}$ (as we will see), this implies $\Im(x_{L23}) \sim 10^{-3}$, a value that is plausibly generated even by Planck-suppressed dimension-5 operators.

Using the results of {\HPU} we have
\begin{align}
    \tilde{m}_1 \approx \frac{9}{40}\frac{y_u v_L^2}{v_R} \approx 10^{-5}~{\rm eV}\; \left(\frac{v_R}{10^{11}~{\rm GeV}}\right)^{-1}.
\end{align}
For $v_R \gtrsim 10^{11}~\GeV$ we have $\tilde{m}_1 / m_* \lesssim 0.1$, implying that $\bar{\nu}_1$ erases less than 10\% of preexisting lepton asymmetry. For $v_R \gtrsim 10^{12}~\GeV$, which is our main focus, this washout is exponentially suppressed and can be neglected.

The lepton asymmetry decay factor for $\bar{\nu}_2$ from \EqRef{eq:epsilon_2} is\footnote{Since $\mathcal{M}_{\bar{\nu}2}$ is not too far from $v_R$, we use couplings obtained after matching at $v_R$. Hence, $y_{u,c,t}$ are values at $10^{12}~{\rm GeV}$ which is used a proxy for $v_R$, e.g. $y_{u,c,t}(10^{12}~{\rm GeV})$.} 
\begin{align} \label{eq:eps_2_HPG}
    \varepsilon_{2} & \approx  \frac{3\, y_c\, y_t\,|\delta_2|^2\, \sin (2 \phi_2 )}{64 \pi } \approx 1.2\times 10^{-7}\left(\frac{|\delta_2|^2 \sin(2\phi_2)}{10^{-2}}\right).
\end{align}
Furthermore, \EqRef{eq:eta2} and \EqRef{eq:m2tilde} give efficiency
\begin{align} \label{eq:eta_HPG}
    &\tilde{m}_2 \approx \frac{9}{40}\frac{y_c v_L^2 }{v_R} \approx 8~{\rm meV}\; \left(\frac{10^{12}~{\rm GeV}}{v_R}\right),\\
    \implies &\eta_2 \approx  6.2\times 10^{-2}\; \left(\frac{v_R}{10^{12}~{\rm GeV}}\right)^{1.16},\qquad 
\end{align}
in the strong washout regime, valid for $v_R \lesssim  10^{13}~{\rm GeV}$. Inserting these results into \EqRef{eq:Y_B}, the baryon asymmetry of the universe produced from decay of $\bar{\nu}_2$ in {\HPU} for $v_R \lesssim 10^{13}~{\rm GeV}$ is
\begin{align} \label{eq:Y_B_HPG}
    %Y_{B} &= \frac{28}{79}\times \frac{135 \zeta(3)}{4 \pi^4 g_*} \times \eta \times \varepsilon_2  \notag\\
    Y_B \; &\approx \;8.75\times 10^{-11} \; \left(\frac{v_R}{3.5\times 10^{12}~{\rm GeV}}\right)^{1.16} \; \left(\frac{|\delta_2|^2 \sin(2\phi_2)}{10^{-2}}\right).
\end{align}
For $v_R \gtrsim  10^{13}$ GeV, we are in the weak washout regime. In numerical fits below we use full results of \cite{Giudice:2003jh} for thermal leptogenesis valid beyond this point.

\subsection{Combined fit to Neutrino Masses and Leptogenesis} \label{subsec: Combined fit to Neutrino Masses and Leptogenesis}

In this subsection, we perform a combined fit to neutrino masses, mixing angles, and leptogenesis in the minimal $(u, \nu)$ sector, with a phase for $\delta_{2}$. The fit agrees with data to within an order of magnitude. Hence, SO(10) breaking corrections are required and discussed in the next subsection.
\begin{figure}[t]
    \centering
    \includegraphics[width=.95\linewidth]{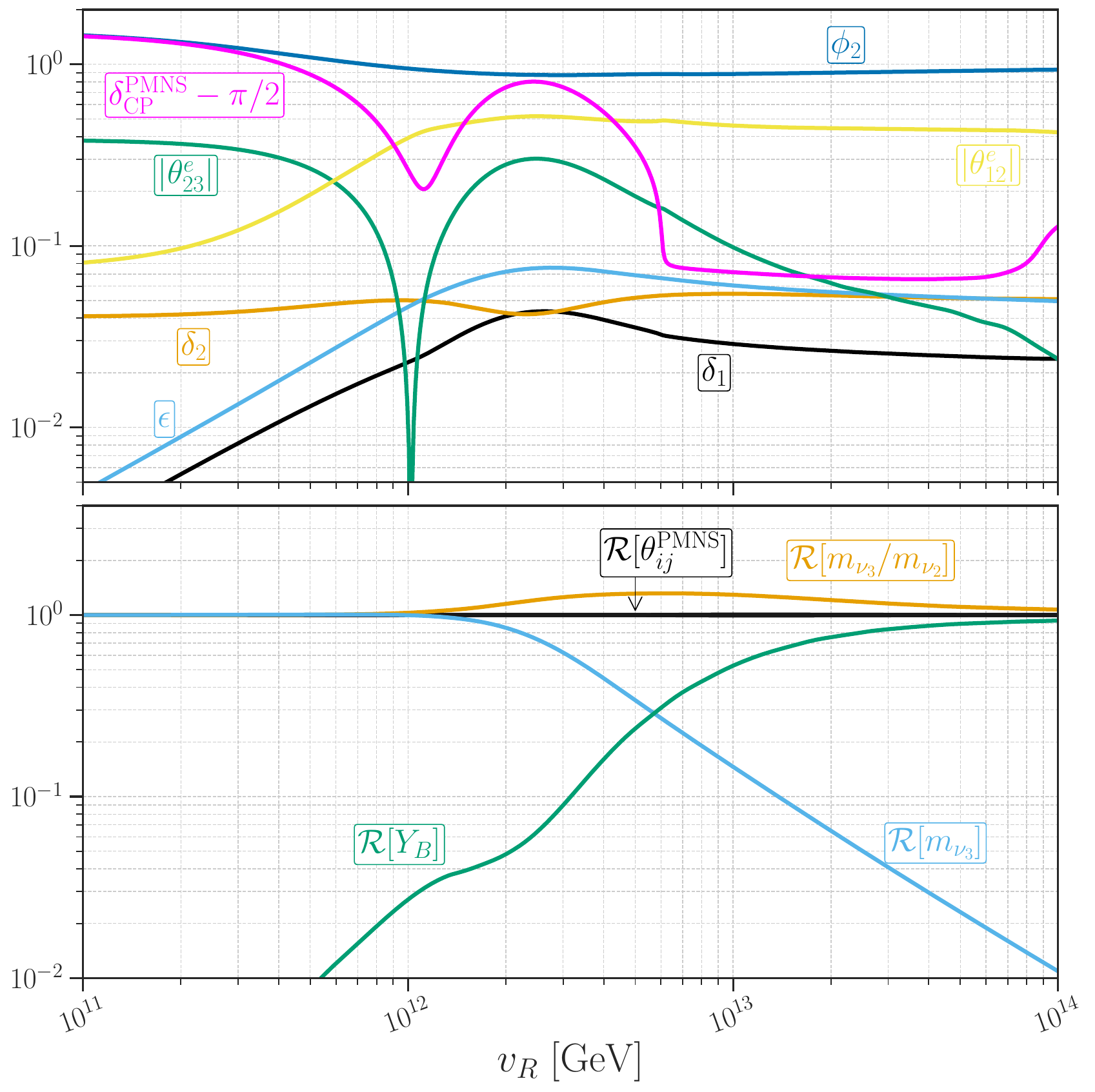}
    \caption{Upper plot: parameters minimizing $\mathcal{F} = \sum_{x}(\mathcal{R}[x] - 1)^2$; also, the $\cancel{CP}$ phase of the PMNS matrix. 
    Lower plot: ratios of fitted observables to measured values from \EqRef{eq:Fit_Errors} for fit of $\theta_{ij}^{\rm PMNS}$, $m_{\nu_3}/m_{\nu_2}$, $m_{\nu_3}$ and $Y_B$. }
    \label{fig:Flavor_Fit_MUN_with_phase}
\end{figure}
We define ratios of fitted to observed parameters for ($\theta^{\rm PMNS}_{ij},\,m_{\nu_3}/m_{\nu_2},\, m_{\nu_3}, Y_B$),
\begin{align} \label{eq:Fit_Errors}
    \mathcal{R}[x] = \frac{x_{\rm calc} }{x_{\rm obs}},
\end{align}
and minimize $\mathcal{F} = \sum_{x}(\mathcal{R}[x] - 1)^2$ over  $\delta_1,$ $\delta_2,$ $ \epsilon,$ $\phi_1$, $ \phi_2,$ $ \theta^{e}_{12},$ $ \theta^{e}_{23} $. Errors in the fit to observables are shown in \FigRef{fig:Flavor_Fit_MUN_with_phase}.

Although this fit has tension between $Y_B$ and neutrino masses, it has two very exciting features. First, the range $4 \times 10^{12}~{\rm GeV}\lesssim v_R \lesssim 10^{13}~{\rm GeV}$ where the fit works to within roughly an order of magnitude matches our expectations from Higgs Parity, gauge coupling unification, and proton decay. Second, the $\cancel{CP}$ phase of the PMNS matrix can theoretically be calculated in this fit and bridges the gap between the baryon asymmetry of the universe and $\cancel{CP}$ violation in the PMNS matrix. Additionally, getting within an order of magnitude for a 6 parameter fit to 6 observables with complicated dependence on the 6 parameters is actually rather remarkable. This warrants consideration of how this tension can be eased.

Before doing so, since the numerical fit gives little insight, we briefly consider a simplification involving only the two heavy generations. The neutrino mass matrix
\begin{align}
    \mathcal{M}_\nu = \frac{2}{5}\frac{v_L^2 y_t}{v_R}
    \begin{pmatrix}
    \tilde{y}_c e^{-2i\phi_2} +\delta_2^2~~  & \delta_2 \epsilon\\
    \delta_2 \epsilon & \epsilon^2
    \end{pmatrix},\qquad \tilde{y}_c \equiv y_c/y_t,
\end{align}
is diagonalized by a unitary transformation $V_\nu^\top \mathcal{M}_\nu V_\nu$ with 
\begin{align}
    V_\nu = 
    \begin{pmatrix}
     c_\nu & s_\nu e^{i\chi}\\
     -s_\nu e^{-i\chi} & c_\nu
    \end{pmatrix}
\end{align}
where $c_\nu,\, s_\nu\equiv \cos\theta_\nu,\, \sin \theta_\nu$. In this case, it is useful to consider the hermitian matrix
\begin{align}
    &\mathcal{M}_\nu^\dagger \mathcal{M}_\nu = \left(\frac{2}{5}\frac{v_L^2 y_t}{v_R}\right)^2\begin{pmatrix}
        A & C\\
        C^* & B
    \end{pmatrix},\\
    &A = \tilde{y}_c^2+\delta _2^4 + 2 \delta _2^2 \tilde{y}_c \cos (2 \phi_2 )+ \delta_2^2 \epsilon^2\\
    &B = \epsilon^2(\delta_2^2 +\epsilon^2)\\
    &C = \delta_2 \epsilon \left(\epsilon^2 + \delta_2^2 + \tilde{y}_c e^{2i\phi_2}\right).
\end{align}
The rotation angle, $\theta_\nu$, of the unitary $V_\nu$ is 
\begin{align}
    \tan(2\theta_\nu) = \frac{2|C|}{B-A}.
\end{align}
Given that $\theta_{23}^{\rm PMNS}\sim 45^\circ$, we expect $\tan2\theta_\nu\gg1$, unless charged-lepton rotation angles are large, hence we make approximate $A = B$. To maximize the baryon asymmetry we take $2\phi_2 \approx \pi/2$, so $A=B$ can be written as 
\begin{align}
    \epsilon^4 \approx \delta_2^4 + \tilde{y}_c^2.
\end{align}
In this case, the neutrino mass eigenvalues are 
\begin{align}
    m^{2}_{\nu2,3} &=\left(\frac{2}{5}\frac{v_L^2 y_t}{v_R}\right)^2\left(B \mp |C|\right).
\end{align}
The numerical solutions for $\delta_2$ and $\epsilon$ giving the correct mass ratio are
\begin{align}
    \delta_2 = 1.145\,\tilde{y}_c^{1/2} ~{\rm and}~\epsilon = 1.284\,\tilde{y}_c^{1/2}.
\end{align}
This gives the following third generation neutrino mass:
\begin{align} \label{eq:mnu3approx}
    {m_{\nu_3}} & \approx 3.1\times \frac{2}{5}\, \frac{v_L^2\, y_c}{v_R}   \approx {{m_{\nu_3,\rm obs}}} \left(\frac{1.3\times 10^{12}~{\rm GeV}}{v_R}\right).
\end{align}
Using the the above value of $\delta_2$ yields a lepton asymmetry decay factor 
\begin{align} \label{eq:eps_2approx}
    \varepsilon_{2} & \approx 1.3\times \frac{3}{64 \pi } \, y_c^2 \,.
\end{align}
Using this and the efficiency from 
\EqRef{eq:eta_HPG} the baryon asymmetry is
\begin{align} \label{eq:YBapprox}
    Y_{B} \; &\approx \; 0.04 \, {Y_{B,\rm obs}} \;\left(\frac{v_R}{1.3\times 10^{12}~{\rm GeV}}\right)^{1.16}\;,
\end{align}
where we assumed strong washout valid for $v_R \lesssim 10^{13}~{\rm GeV}$. These results align with the full numerical fit shown in \FigRef{fig:Flavor_Fit_MUN_with_phase}, where blue and green curves exhibit tension in simultaneously matching observed values of $m_{\nu_3}$ and $Y_B$ for a fixed $v_R$. The approximate result reveals the following parametric link between $m_{\nu_3}$ and $Y_B$:
\begin{align} \label{eq:YBapprox2}
m_{\nu_3} & \propto \frac{y_c}{v_R}, \qquad Y_{B}  \propto   y_c^{0.84}  v_R^{1.16},
\end{align}
The product is nearly independent of $v_R$,
\begin{align}
m_{\nu_3} Y_B \propto  y_c^{1.84} v_R^{0.16}.
\end{align}

Tension between $m_{\nu_3}$ and $Y_B$ can be eased or eliminated by adding $SO(10)$-breaking corrections. Since the tension is primarily between $m_{\nu_3}$ and $Y_B$, we return to scaling arguments of \EqRef{eq:scaling}, which leave PMNS angles and mass ratios undisturbed. For simplicity, we consider diagonal corrections from $\mathcal{S}(\bm{54})$ in \EqRef{eq:XMassesSigma} motivated by approximate flavor symmetries. 

In the strong washout regime where $v_R \lesssim 10^{13}$ GeV, rescaling gives
\begin{align}
    &m_{\nu_3}\rightarrow |AB|\; m_3^{(\nu)}\\
    &Y_B \rightarrow |AB|^{0.84}\; Y_B
\end{align}
These scaling parameters are determined from $b_i$ in \EqRef{eq:b_scaling}. However, we see that both observables depend only on the product $AB$ which depends only on $b_2$. 

\begin{figure}[t]
    \centering
    \includegraphics[width=1\linewidth]{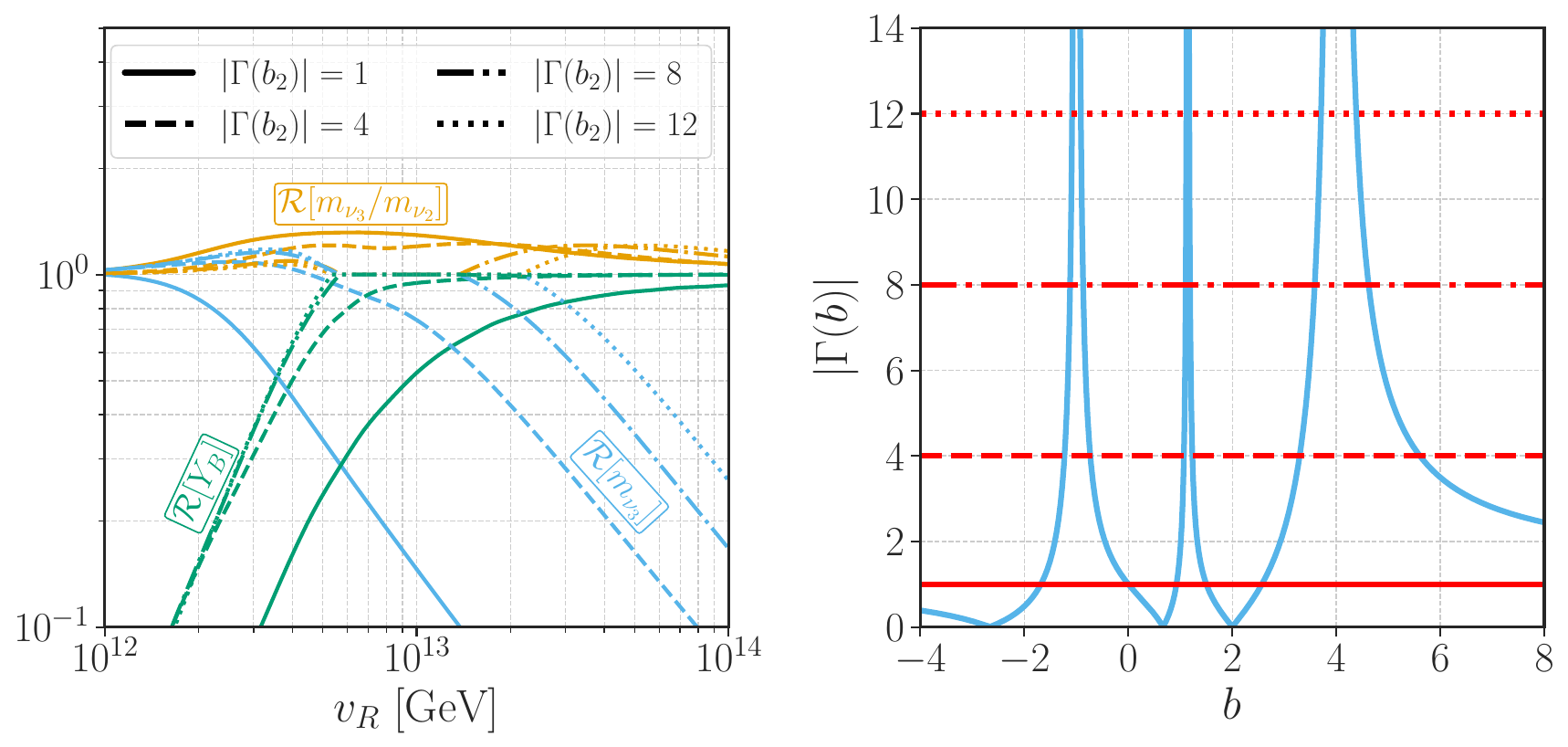}
    \caption{On the left we show the ratios of best fit values to measured values for the heavy neutrino masses and the baryon asymmetry for several values of $\Gamma(b_2)$. As $\Gamma(b_2)$ increases, the range of $v_R$ that allows a complete fit also increases; however, a very important feature is that the lower bound of this region does not decrease below $5\times 10^{12}$ GeV. On the right we plot the function $|\Gamma(b)|$ in \textcolor{MyBlue}{\textbf{blue}}. The values of $|\Gamma(b_2)|$ used on the left are shown as \textcolor{red}{\textbf{red}} horizontal lines. From this we can see that large values of $|\Gamma(b)|$ require tuning $b$ to either $-1$, $8/7$ or $4$. }
    \label{fig:AB_Changes}
\end{figure}

Therefore one approximate solution consistent with the above scaling is at $v_R = 6\times 10^{12}$ GeV--essentially where $\mathcal{R}[Y_B]$ and $\mathcal{R}[m_3^{(\nu)}]$ cross--if one has $|AB| \approx 4$ which occurs for several values of $b_2$, e.g. see \FigRef{fig:AB_Changes}. Except for the last solution, these are all relatively small corrections to the mass spectra. Interestingly, $\lim_{b_2\rightarrow \pm \infty}\Gamma(b_2) = 39/11\approx 3.5$ which suggests a possible solution is to assume the second mass eigenvalue of $\Y$ arises entirely from interactions with $\expval{S}$. In fact, since the fit depends extremely weakly on $b_3$, it is possible all mass eigenvalues of $\Y$ arise from $\expval{S}$.

The full fit is more complicated: once $|AB|\gtrsim 4$ the product $m_{\nu_3}\,Y_B$ becomes too large and additional parameters adjust to lower it. $Y_B$ is easily lowered by adjusting $\phi_2$ allowing solutions for $v_R\gtrsim 6\times 10^{12}$ GeV. However, little additional freedom exists to lower $m_{\nu_3}$. Therefore, solutions for $v_R$ much smaller than $5\times 10^{12}$ GeV are not accessible as can be seen in \FigRef{fig:AB_Changes} where the full numerical fit has been performed for different values of $\Gamma(b_2)$. Implications of this fit for the discovery prospects of proton decay and the neutron EDM are indicated by the magenta band in \FigRef{fig:Union_Jack}.

\section{Conclusions} \label{sec:Conclusions}

In this paper, we constructed a predictive unified $SO(10) \times CP$ theory and studied its remarkable consequences for gauge coupling unification, proton decay, quark and lepton masses and mixings, and leptogenesis. From a top-down perspective, key aspects of the theory are chosen to elegantly solve the strong CP problem. Simultaneous breaking of $CP$ and $SO(10)$ leaves an unbroken space-time parity. This forces $\theta = {\rm arg} \,{\rm det}(Y_u Y_d) = 0$ at tree-level without introducing extra symmetry, but {\it only} if Higgs doublets breaking $SU(2)_R$ and $SU(2)_L$ originate from a spinor of $SO(10)$. Such Higgs fields {\it necessarily} lead to Higgs Parity; the SM Higgs quartic coupling must vanish at the parity breaking scale $v_R$. \FigRef{fig:Higgs_Quartic_Running} and \FigRef{fig:mt-alphascorrelation} show $v_R$ to be predicted in the range $ (2\times 10^{10} - 3\times 10^{13})$ GeV at $2 \sigma$. 
Thus, solving the strong CP problem and Higgs Parity both require abandoning conventional embeddings of the SM Higgs into tensor representations of SO(10), and imply a qualitatively distinct mechanism for generating flavor observables.

Gauge coupling unification is inextricably linked to the Higgs Parity mechanism because the intermediate symmetry breaking scale, $v_R$, is determined by running of the SM quartic coupling, as shown in \FigRef{fig:Gauge_Coupling_Unification}. 
Furthermore, five SM observables are given in terms of four UV parameters, yielding a prediction for the QCD coupling in terms of the top quark mass, \EqRef{eq:alphaspred}, with a precision well within $1\sigma$ experimental uncertainties.
In the absence of additional threshold corrections, the mass of the gauge boson relevant for proton decay is predicted to be less than five times the limit set by the Super-Kamiokande experiment, as shown in \FigRef{fig:Unification_Quality_Low_Mass} and \FigRef{fig:Unification_Quality_High_Mass}. A signal at the Hyper-Kamiokande experiment is likely for $v_R \sim (10^{12} - 10^{13})$ GeV, the upper end allowed by Higgs Parity. 

Radiative corrections to $\bar{\theta}$ arise from integrating out states at scale $v_R$. These vanish at 1-loop, but are non-zero at 2-loops and are estimated to be within 1-2 orders of magnitude of the current experimental bound  \cite{Hall:2018let, Hisano:2023izx} and may be observable by future neutron EDM experiments. Contributions to $\bar{\theta}$ may also arise from higher-dimensional operators, offering the prospect of neutron EDM detection if $v_R \sim (10^{12} - 10^{13})$ GeV, as shown in \FigRef{fig:Unification_Quality_Low_Mass} and \FigRef{fig:Unification_Quality_High_Mass}. 

The flavor sector is unlike anything in conventional $SO(10)$ theories. Yukawa interactions $\psi \psi \varphi$ are forbidden when $\psi$ and $\varphi$ are both in spinor representations. In \SecRef{sec:The Seesaw Origin of Flavor Observables} we studied the simplest flavor sector
\begin{align}
    \mathscr{L}_0 = 
    \psi \,\x \,\X \;\varphi^\dagger + \X \,M \, \X + \psi \, \y \, \Y \; \varphi + \Y \, \tilde{M} \,\Y \, + \,
       {\rm h.c.}
    \label{eq:Lleading2}
\end{align}
Integrating out heavy $X(10)$ and $\tilde{X}(45)$ multiplets, the seesaw mechanism gives all Yukawa couplings of the SM   
\begin{align}
  Y_{d,e} &=  x \frac{1}{M} x^\dagger \; v_R, \qquad Y_{u,\nu} =  \tilde{x} \frac{1}{\tilde{M}} \tilde{x}^\dagger \; \left(2, \; \frac{3}{4} \right) \; v_R, 
    \label{eq:SMYuks}
\end{align}
and the right- and left-handed neutrino mass matrices 
\begin{align}
  (\mathcal{M}_{\bar{\nu}}, \mathcal{M}_{\nu}^*) &= \; \tilde{x} \frac{1}{\tilde{M}} \tilde{x}^\dagger  \left(\frac{5v_R^2}{4}, \frac{4 \,v_L^2}{5} \right). 
    \label{eq:nunubarmasses}
\end{align}
The equality of down-quark and charged-lepton Yukawa matrices is familiar from minimal $SU(5)$ schemes, and can be made realistic by adding $SO(10)$-breaking corrections. However, close relations between up-quark and neutrino sectors is a defining feature of Higgs Parity unification; deviations from pure seesaw are required to explain the lack of hierarchy in masses and mixings of neutrinos compared to up-quarks.  

In \SecRef{sec:directbtau} we studied consequences of the seesaw mechanism breaking down for the third generation. In particular, if $x_{33} v_R \geq M_3$ the $b$-quark and $\tau$-lepton arise from large mixing between $(\bar{d}_i, \bar{D}_3)$ and $(\ell_i, L_3)$ states contained in $(\psi_i, X_3)$. Although this mixing of quark singlets has no effect on the CKM matrix, the mixing of lepton doublets via $\cmat_\ell$ of \EqRef{eq:c_L} crucially affects both the PMNS matrix and neutrino masses, and changes the relation between neutrino and up-quark matrices
\begin{align}
 \mathcal{M}_\nu & = \frac{2}{5} \, \cmat_\ell \, Y_u \, \cmat_{\ell }^\top \,  \frac{v_L^2}{v_R},
 \qquad \cmat_\ell = \begin{pmatrix}
        1 & 0 & 0  \\
        0 & 1 & 0 \\
        \delta_1 & \delta_2 & \epsilon  
    \end{pmatrix}.
    \label{eq:numassesfromu}
\end{align}
Small parameters $(\delta_1, \delta_2, \epsilon)$ from the charged lepton sector are key to removing mass and mixing hierarchies from the neutrino sector. CP violation is not important here, so to demonstrate viability of this understanding of neutrino masses and mixings we perform a numerical fit in the minimal $(u, \nu)$ sector, given by the last two operators of \EqRef{eq:Lleading2} with mixings on lepton doublets described by $\cmat_\ell$. Eight observables $(m_{u,c,t}, \Delta m^2_{32}, \Delta m^2_{21}, \theta_{ij}^{({\rm PMNS})})$ were fit to 8 parameters $(\tilde{x}, \cmat_\ell, \theta^{e}_{23},\, \theta^{e}_{12})$ and results are shown in \FigRef{fig:Flavor_Fit_MUN}. A successful fit occurs only for a highly restricted region of $v_R \sim (0.3 - 2) \times  10^{12}$ GeV. CP violating phases and $SO(10)$ breaking corrections in the $(u,\nu)$ sector are expected to modify this range by order unity. For these values of $v_R$, a normal neutrino mass hierarchy is predicted with the mass of $\nu_1$ of order $10^{-6}$ eV, making the observation of $0\nu \beta \beta$ challenging. 

If the reheat temperature of the universe after inflation is above $m_{\bar{\nu}_2} \sim 10^{10}$ GeV, the cosmological baryon asymmetry of the universe may be created via thermal leptogenesis from $\bar{\nu}_2$ decays. Even with the minimal $(u,\nu)$ sector, CP violation results from phases $\phi_{1,2}$ of the parameters $\delta_{1,2}$ that describe mixing of $\ell_{1,2}$ with $L_3$. These CP violating phases originate from higher-dimensional operators in the $(d/e)$ sector through the VEV of $\Sigma$. In \SecRef{sec:Leptogenesis} we redid the fit to the minimal $(u,\nu)$ sector for up-quark masses, neutrino masses and mixings {\it and} the cosmological baryon asymmetry, including phases $\phi_{1,2}$ as free parameters. Remarkably, \FigRef{fig:Flavor_Fit_MUN_with_phase} shows that the order of magnitude of the observed baryon asymmetry is correctly predicted for values of $\phi_{1,2}$ that are order unity. Compared to the CP-conserving fit of \SecRef{sec:directbtau}, the preferred value of $v_R$ is increased to $(4-10) \times 10^{12}$ GeV. 
The predictions for $v_R$ and $Y_B$ can be understood analytically, as shown by \EqRef{eq:mnu3approx} and \EqRef{eq:YBapprox}.
In fact, the fit is not perfect; in this region, there is a tension between the precise values for $m_{\nu_3}$ and $Y_B$ that need correcting by a factor of 3-6.  Such corrections could arise, for example, from higher-order mass terms $\tilde{X} \Sigma \tilde{X}$, as shown by the successful fits of \FigRef{fig:AB_Changes}, where the range of $v_R$ is further increased to $(6-10) \times 10^{12}$ GeV.

The structure of the theory is so rigid that the $v_R > 4 \times 10^{12}$ GeV preferred by fits to neutrino masses and the baryon asymmetry has further consequences.  Gauge coupling unification implies that the proton lifetime is constrained to be close to the current bound, as shown in \FigRef{fig:Unification_Quality_Low_Mass} and \FigRef{fig:Unification_Quality_High_Mass}. This then leads to a preferred region in the $\alpha_s/m_t$ correlation of \FigRef{fig:mt-alphascorrelation}, shown by the band labeled $r=1/8$. As the experimental uncertainties are reduced, we expect $\alpha_s$ to increase and $m_t$ to decrease relative to the current central values.

In summary: Higgs Parity unification achieves precision gauge coupling unification, is consistent with current bounds on proton decay, and provides a solution to the strong $CP$ problem. With a minimal parameter set, it accounts for SM flavor observables, including neutrino masses and mixings,  and predicts the baryon asymmetry of the universe to within an order of magnitude.  The consistency between these various aspects of the theory is illustrated in \FigRef{fig:scales}, where the allowed ranges of $v_R$ for each aspect are compared. Central to this framework is a novel mechanism in which the third-generation fermions have a large mixing with heavy vector-like fermions. This naturally accounts for the very different hierarchies observed in the up quark and neutrino sectors. In \FigRef{fig:Union_Jack} we show predictions for the proton lifetime and for the neutron EDM from higher-dimensional operators. While uncertainties from threshold corrections to gauge coupling unification are larger for the former, the latter is subject to the uncertain value of an unknown order unity coefficient, $C$. The vertical magenta band highlights the region of the successful fit to neutrino flavor and the baryon asymmetry, and motivates continued experimental searches for both proton decay and the neutron EDM. 

While our focus has been to establish the viability of this framework and demonstrate its key features, several directions merit further investigation. In particular, a detailed numerical study that scans over parameters of $SO(10)$ breaking operators could sharpen predictions for the PMNS $CP$ violating phase, a loop-induced strong CP phase, and the rates for proton decay and neutrinoless double beta decay. Extensions incorporating dark matter also remain an open and promising area for exploration. Additionally, cosmological signatures may result from the spontaneous breaking of the left-right symmetry.

Given its minimal assumptions and its ability to explain such a large variety of physical phenomena, {\HPU} stands as a compelling alternative to conventional supersymmetric GUTs, and a rich target for both theoretical development and experimental tests.

\section*{Acknowledgment}
We thank Simon Knapen, Zoltan Ligeti, and Yasunori Nomura for useful discussions. This work was supported by the Office of High Energy Physics of the U.S. Department of Energy under contract DE-AC02-05CH11231 (L.H) and DE-SC0025242 (K.H), by the NSF grant PHY-2210390, and by World Premier International Research Center Initiative (WPI), MEXT, Japan (Kavli IPMU).

\newpage

\appendix

\section{Further Details on Gauge Coupling Unification} \label{app:Gauge Coupling Unification in Higgs Parity GUTs}

In this appendix, we give details of gauge coupling running and unification. 

\subsubsection*{\emph{Standard Model Running}}

\begin{figure}[t]
    \centering
    \includegraphics[width= \linewidth]{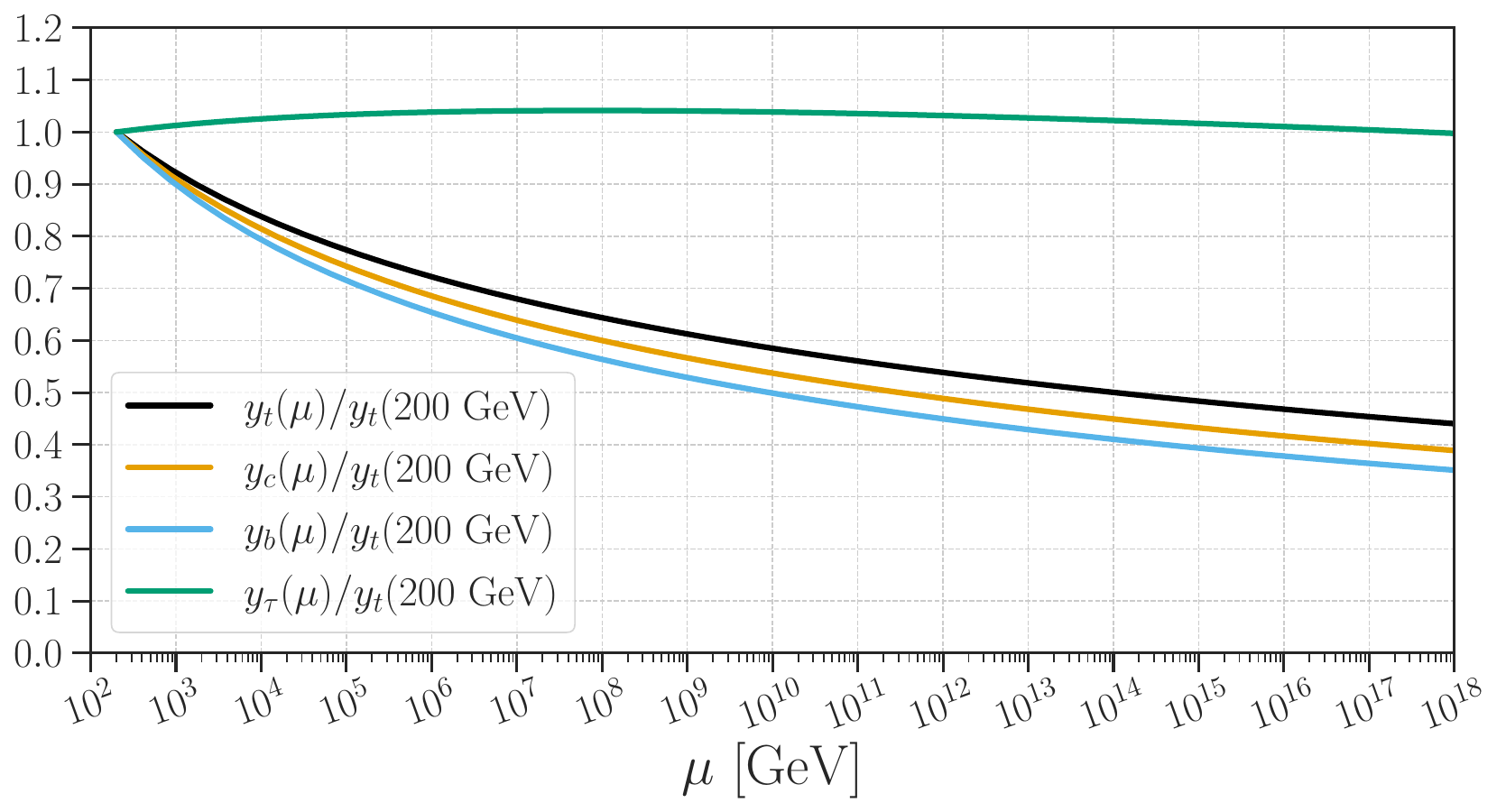}
    \caption{Two-loop Yukawa coupling running in $\overline{MS}$ relative to values at $200~\rm GeV$. }
    \label{fig:Yukawa_Running}
\end{figure}

Initial conditions for the SM couplings are calculated in $\overline{MS}$ scheme at $\mu = 200~{\rm GeV}$ \cite{Alam:2022cdv}. We normalize $U(1)$ charges consistent with an $SO(10)$ embedding, i.e. $\tilde{Y} = \sqrt{\frac{3}{5}}\, Y$ and $g_1 = \sqrt{\frac{5}{3}}\,g'$. We denote $SU(2)$ and $SU(3)$ gauge couplings $g_2$ and $g_3$. The gauge coupling RGEs are
\begin{align} \label{eq:beta_SM}
    &\frac{d g_i^{-2}}{d\log \mu} = \frac{1}{16\pi^2}a_i  + \frac{1}{(16\pi^2)^2} \left(b_{ij}  g_j^2 + c_i y_t^2\right)\\
    &a_i = \left(
\begin{array}{c}
 -\frac{41}{5} \\
 \frac{19}{3} \\
 14 \\
\end{array}
\right)
,
\qquad 
b_{ij} = \left(
\begin{array}{ccc}
 -\frac{199}{25} & -\frac{27}{5} & -\frac{88}{5} \\
 -\frac{9}{5} & -\frac{35}{3} & -24 \\
 -\frac{11}{5} & -9 & 52 \\
\end{array}
\right),\qquad 
c_i = \left(
\begin{array}{c}
 \frac{17}{5} \\
 3 \\
 4 \\
\end{array}
\right).
\end{align}
Defining $z_i = (g_1^2,\,g_2^2,\,g_3^2,\,y_t^2,\, \lambda)$, the two-loop RGE for the top-yukawa is
\begin{align}
    &\frac{d \log y_t}{d\log \mu} = \frac{1}{16\pi^2}a_i^{(y_t)}z_i + \frac{1}{(16\pi^2)}b_{ij}^{(y_t)}z_i z_j\\
&a_i^{(y_t)} =   
\begin{pmatrix}
    -\frac{17}{20}\\
    -\frac{9}{4}\\
    -8\\
    \frac{9}{2}\\
    0
\end{pmatrix}
,\qquad 
b_{ij}^{(y_t)} =   
\left(
\begin{array}{ccccc}
 \frac{1187}{600} & -\frac{9}{40} & \frac{19}{30} & \frac{393}{160} & 0 \\
 -\frac{9}{40} & -\frac{23}{4} & \frac{9}{2} & \frac{225}{32} & 0 \\
 \frac{19}{30} & \frac{9}{2} & -108 & 18 & 0\\
 \frac{393}{160} & \frac{225}{32} & 18 & -12 & -6 \\
  0 & 0 & 0 & -6 & 6
\end{array}
\right).
\end{align}
Running of Yukawa couplings are shown in \FigRef{fig:Yukawa_Running}. The Higgs quartic RGE is
\begin{align}
    \frac{d\lambda}{d\log \mu} &= \frac{1}{16\pi^2}\left[-\frac{9 g_1^2 \lambda }{5}-9 g_2^2 \lambda +\frac{27 g_1^4}{200}+\frac{9g_1^2 g_2^2}{20} +\frac{9 g_2^4}{8}+24 \lambda ^2+12 \lambda  y_t^2-6 y_t^4\right]\notag\\
    &+ \frac{1}{(16\pi^2)^2}\Big[-\frac{171}{100} g_1^4 y_t^2-\frac{8}{5} g_1^2 y_t^4+\frac{63}{10} g_2^2 g_1^2 y_t^2-32 g_3^2 y_t^4-\frac{9}{4} g_2^4 y_t^2-\frac{3411 g_1^6}{2000}\notag \\ &\qquad \qquad \quad \, -\frac{1677}{400}
   g_2^2 g_1^4-\frac{289}{80} g_2^4 g_1^2+\frac{305 g_2^6}{16}+30 y_t^6\Big]\notag\\
   &+ \frac{\lambda}{(16\pi^2)^2}\Big[ \frac{17}{2} g_1^2 y_t^2+\frac{45}{2} g_2^2 y_t^2+80 g_3^2 y_t^2+\frac{1887 g_1^4}{200}+\frac{117}{20} g_2^2 g_1^2-\frac{73 g_2^4}{8}-3 y_t^4\Big]\notag \\
   &+ \frac{\lambda^2}{(16\pi^2)^2}\Big[ \frac{108 g_1^2}{5}+108 g_2^2-144 y_t^2  - 312 \lambda \Big].
\end{align}

\subsubsection*{\emph{Left-Right Symmetric Model Matching and Running}}

Consider $SU(2)_R \times U(1)_{\tilde{X}}\rightarrow U(1)_{\tilde{Y}}$ where $\tilde{X}$, $\tilde{Y}$ are generators of the $U(1)$ gauge groups normalized consistently with the GUT embedding and related as
\begin{align}
    \tilde{Y} = c_X \tilde{X} + c_R T_R^3.
\end{align}
We match 321 and 3221 gauge couplings at scale $\mu = M_{W_R}= g_2v_R/\sqrt{2}$ using
\begin{align}
    \frac{1}{g^2_{\tilde{Y}}} = c_X^2 \frac{1}{g^2_{\tilde{X}}} + c_R^2\left( \frac{1}{g_2^2} - \frac{C_2(\bf{3})}{48 \pi^2}\right) \longrightarrow  \frac{1}{g^2_{\tilde{Y}}} = \frac{2}{5} \frac{1}{g^2_{\tilde{X}}} + \frac{3}{5} \frac{1}{g_2^2} - \frac{1}{40 \pi^2}.
\end{align}
using $c_X^2 = 2/5$, $c_R^2 = 3/5$, and $C_2({\bf{3}}) = 2$.

Above $\mu = M_{W_R}$, the two loop beta functions for gauge coupling running are 
\begin{align} \label{eq:beta_LR}
    \frac{d g_i^{-2}}{d\log \mu} = \sum_\chi \theta(\mu - m_\chi) \left(\frac{1}{16\pi^2}a^{(\chi)}_i  + \frac{1}{(16\pi^2)^2} b^{(\chi)}_{ij}  g_j^2\right). 
\end{align}
where $\chi$ labels all fields. We separate contributions from light fields with masses below $v_R$ ($\ell,\, \bar{\ell},\, q,\, \bar{q},\, \HL,\, \HR$) and from heavy fields. The light fields contribute
\begin{align}
    a^{(\rm light)} = 
    \left(\begin{array}{c}
    -9 \\
    \frac{19}{3} \\
    14
    \end{array}\right), 
    \qquad 
    b^{(\rm light)} = 
    \left(
\begin{array}{ccc}
 -\frac{23}{2} & -27 & -8 \\
 -\frac{9}{2} & -\frac{35}{3} & -24 \\
 -1 & -18 & 52 \\
\end{array}
\right).
\end{align}
Parity requires $g_L = g_R = g_2$. We assume mass differences between 3221 components of heavy vector-like fermions are negligible such that
\begin{alignat}{2}
    &a^{(\bm{10})} = 
    \frac{-4}{3}\left(\begin{array}{c}
    1 \\
    1 \\
    1
    \end{array}\right), 
    \qquad 
    &&b^{(\bm{10})} = \frac{-2}{3}
    \left(\begin{array}{ccc}
    1 & 0 & 8 \\
    0 & 29 & 0 \\
    1 & 0 & 38
    \end{array}\right)
    \\
    &a^{(\bm{45})} = 
    \frac{-32}{3}\left(
    \begin{array}{c}
    1 \\
    1 \\
    1
    \end{array}
    \right), 
    \qquad 
    &&b^{(\bm{45})} = 
    \frac{-2}{3} \left(
    \begin{array}{ccc}
    20 & 36 & 64 \\
    6 & 238 & 54 \\
    8 & 36 & 334
    \end{array}
    \right)
\end{alignat}
The masses of each generation are unknown. This doesn't significantly modify the success of unification but does modify the gauge coupling strength at unification. The low and high mass scenarios we consider are described in \SecRef{subsec:Gauge Coupling Unification and Proton Decay}.

\subsubsection*{\emph{Threshold Corrections and Gauge Coupling Unification}}

We match 3221 gauge couplings to the unified coupling using 1-loop threshold corrections. For general symmetry breaking $G_{\rm GUT}\rightarrow G_1\times G_2\times\cdots$, these are 
\begin{align}
    &g_{a}^{-2}(\mu) = g_{\rm GUT}^{-2}(\mu) - \Delta_{a,G} - \Delta_{a,S} - \Delta_{a,F} \\
&\Delta_{a,G}(\mu)= \frac{1}{48 \pi^2} \operatorname{Tr}_V\left[t_{a }^2 \left(1-21\ln \frac{M_{\mathrm{V}}}{\mu}\right)\right]\\
&\Delta_{a,S}(\mu)=\frac{1}{48 \pi^2}\operatorname{Tr}_S\left[t_{a }^2  \ln \frac{M_{\mathrm{S}}}{\mu}\right] \\
 &\Delta_{a,F}(\mu)=\frac{8}{48 \pi^2} \operatorname{Tr}_F\left[t_{a }^2 \ln \frac{M_{F}}{\mu}\right]  .
\end{align}
The $t_a$ are generators of $G_a$ and traces are over heavy vectors, scalars, and fermions.

\section{Further Details on Proton Decay in Higgs Parity GUTs} \label{app:Review of Proton Decay in Higgs Parity GUTs}

In this appendix we review proton decay in {\HPU}. Baryon number is an accidental symmetry of the renormalizable 3221 theory but is broken by higher dimensional operators generated from integrating out gauge bosons in $(\bm{3},\bm{2},\bm{2})_{-1/3}$ whose mass we define as $M_{\rm GUT}$. These are
\begin{align}
    \mathscr{L}_{\slashed{B},3221}& =\sum_i^{N_{LR}} C_i^{(\rm{LR})}(\mu) \mathcal{O}_i^{(\rm LR)} .
\end{align}
Once $SU(2)_R$ breaks, baryon violation is described by a new effective theory
\begin{align}
    \mathscr{L}_{\slashed{B},321}& =\sum_i^{N_{SM}} C_i^{(\rm{SM})}(\mu) \mathcal{O}_i^{(\rm SM)}.
\end{align}

\subsubsection*{\emph{Matching to Baryon Violating Operators}}
Denote gauge bosons as $\mathcal{X}^\mu(\bm{3},\bm{2},\bm{2})_{-1/3}$; before $\mathcal{X}$ is integrated out we have
\begin{align}
    \mathscr{L}_{\mathcal{X},3221} = \frac{g_{\rm GUT}}{\sqrt{2}}\left[\left(\bar{q}^\dagger \bar{\sigma}^\mu q +  q^\dagger \bar{\sigma}^\mu \bar{\ell} 
 + \ell^\dagger \bar{\sigma}^\mu \bar{q}\right)\mathcal{X}_\mu + {\rm h.c.} \right] + M_{\rm GUT} \mathcal{X}^\mu\bar{\mathcal{X}}_\mu 
\end{align}
The equation of motion for $\mathcal{X}^\mu$ and $\bar{\mathcal{X}}^\mu$ gives the effective Lagrangian
\begin{align}
    \mathscr{L}_{\slashed{B},3221}
    &= \frac{g_{\rm GUT}^2}{2M_{\rm GUT}^2}\left[\left(\bar{q}^\dagger \bar{\sigma}^\mu q +  q^\dagger \bar{\sigma}^\mu \bar{\ell} +\ell^\dagger \bar{\sigma}^\mu \bar{q}\right)\left(q^\dagger \bar{\sigma}_\mu \bar{q} +  \bar{q}^\dagger \bar{\sigma}_\mu \ell  +\bar{\ell}^\dagger \bar{\sigma}^\mu q\right)\right]
    \\
    &=\frac{g_{\rm GUT}^2}{M_{\rm GUT}^2}\left[ (q^\dagger q^\dagger)(\bar{q}\bar{\ell})+  (\bar{q}^\dagger \bar{q}^\dagger)(q\ell)  + \rm{h.c.}   \right]
\end{align}
using $(a^\dagger \bar{\sigma}^\mu b)(c^\dagger \bar{\sigma}_\mu d) = 2(a^\dagger c^\dagger)(bd)$. Therefore we identify 
\begin{alignat}{2}
    &\mathcal{O}_1^{(\rm LR)} =  (q^\dagger q^\dagger)(\bar{q}\bar{\ell})+ \rm{h.c.}, &&~~~~~ C_1^{(\rm LR)}(M_{\rm GUT}) = \frac{g_{\rm GUT}^2}{M_{\rm GUT}}\\
    &\mathcal{O}_2^{(\rm LR)} = (\bar{q}^\dagger \bar{q}^\dagger)(q\ell) + \rm{h.c.}, &&~~~~~ C_2^{(\rm LR)}(M_{\rm GUT}) = \frac{g_{\rm GUT}^2}{M_{\rm GUT}}
\end{alignat}
These coefficients run from GUT scale to $SU(2)_R$ breaking scale. At this point we break $\bar{q}$ and $\bar{\ell}$ into $321$ components where the operators become
\begin{alignat}{2}
    &\mathcal{O}_1^{({\rm LR})} \longrightarrow \mathcal{O}_1^{({\rm SM})} = (q^\dagger q^\dagger)(\bar{u}\bar{e}), && C_1^{(\rm SM)}(M_{W_R}) = C_1^{(\rm LR)}(M_{W_R}),\\
    &\mathcal{O}_2^{({\rm LR})} \longrightarrow 2\times \mathcal{O}_2^{({\rm SM})} = 2\times(\bar{u}^\dagger \bar{d}^\dagger)(q \ell), ~~~~&&C_2^{(\rm SM)}(M_{W_R}) = 2C_2^{(\rm LR)}(M_{W_R}).
\end{alignat}

\subsubsection*{\emph{Wilson Coefficient Running}}
The Wilson coefficients evolve to $\mu =m_p$ where the proton decay rate is 
\begin{align}
    \Gamma\left(p \rightarrow \pi^0 e^{+}\right) \simeq \frac{m_p}{32 \pi}\left\{1-\left(\frac{m_{\pi^0}}{m_p}\right)^2\right\}^2 \sum_{i=1}^2\left|C_i^{(\rm SM)}\left(m_p\right) W_0^i\right|^2
\end{align}
where $W_0^i$ are the proton form factors. We set $W_0^i = -0.131~\GeV^2$ following \cite{Hamada:2020isl}.

The Wilson coefficient running in the SM and left-right symmetric model are given in \cite{Buras:1977yy,Caswell:1982fx,Hamada:2020isl}. The result is $C^{(\rm SM)}_i(m_p) = A_i C^{(\rm LR)}_i(M_{\rm GUT})$, where $A_i$ is a RGE factor split into contributions from running in the SM and LR EFT $A_i = A_i^{(\rm SM)}\times A_i^{(\rm LR)}$ 
\begin{align}
A^{(\rm{SM })}_{1} & =\left(\frac{\alpha_3(m_p)}{\alpha_3\left(M_{W_R}\right)}\right)^{4 / a_3}\left(\frac{\alpha_2\left(M_Z\right)}{\alpha_2\left(M_{W_R}\right)}\right)^{\frac{9}{2} / a_2}\left(\frac{\alpha_1\left(M_Z\right)}{\alpha_1\left(M_{W_R}\right)}\right)^{\frac{11}{6} / a_1}, \\
A^{(\rm{SM})}_{2} & =\left(\frac{\alpha_3(m_p)}{\alpha_3\left(M_{W_R}\right)}\right)^{4 / a_3}\left(\frac{\alpha_2\left(M_Z\right)}{\alpha_2\left(M_{W_R}\right)}\right)^{\frac{9}{2} / a_2}\left(\frac{\alpha_1\left(M_Z\right)}{\alpha_1\left(M_{W_R}\right)}\right)^{\frac{23}{6} / a_1}, \\
A^{(\rm LR)}_{1}=A^{(\rm LR)}_{2} & =\left(\frac{\alpha_3\left(M_{W_R}\right)}{\alpha_3\left(M_{\mathrm{GUT}}\right)}\right)^{4 / a_3}\left(\frac{\alpha_2\left(M_{W_R}\right)}{\alpha_2\left(M_{\mathrm{GUT}}\right)}\right)^{ 9/ a_2}\left(\frac{\alpha_1\left(M_{W_R}\right)}{\alpha_1\left(M_{\mathrm{GUT}}\right)}\right)^{\frac{1}{2} / a_1} .
\end{align}
The $a_i$ are beta function coefficients defined in \EqRef{eq:beta_SM} and \EqRef{eq:beta_LR}.

\section{General Discussion of Third Generation Mixing} \label{app:General Discussion of Third Generation Mixing}

In this appendix, we analyze general third generation mixing in {\HPU}.

The seesaw origin of flavor results if eigenvalues of $M$ ($\tilde{M}$) are much larger than those of $x v_R$ ($\tilde{x} v_R$). If not, SM quarks and leptons have sizable components of $\X$ and $\Y$. Each 321 representation of $\psi$ shows up once $\X$ or $\Y$ (except SM singlets as $\Y$ contains two). This implies mixing occurs for each 321 representations. The SM quarks and leptons (the massless states in absence of $SU(2)_L$ breaking) are linear combinations of gauge eigenstates of $\psi$ and $\X$ or $\psi$ and $\Y$.

Gauge eigenstates of $\psi$, $\X$, and $\Y$ are related to mass eigenstates (primed) by
\begin{align}
    \begin{pmatrix}
        q_i\\
        Q_i\\
        \bar{u}_i\\
        \bar{U}_i\\
        \bar{d}_i\\
        \bar{D}_i\\
        \ell_i\\
        L_i\\
        \bar{e}_i\\
        \bar{E}_i\\
        \bar{\nu}_i\\
        S_i\\
        N_i
    \end{pmatrix} 
    =
    \left(
    \begin{array}{cccccccccccccc}
    \Omega^q_{ij} & \Xi^q_{ij} & & & & & & & & & & & &\\
    \Xi^Q_{ij} & \Omega^{Q}_{ij} & & & & & & & & & & & & \\
    & & \Omega^{\bar{u}}_{ij} & \Xi^{\bar{u}}_{ij} & & & & & & & & & & \\
    & & \Xi^{\bar{U}}_{ij} & \Omega^{\bar{U}}_{ij} & & & & & & & & & &\\
    & & & & \Omega^{\bar{d}}_{ij} & \Xi^{\bar{d}}_{ij} & & & & & & & &\\
    & & & & \Xi^{\bar{D}}_{ij} & \Omega^{\bar{D}}_{ij} & & & & & & & &\\
    & & & & & & \Omega^{\ell}_{ij} & \Xi^{\ell}_{ij} & & & & & &\\
    & & & & & & \Xi^{L}_{ij} & \Omega^{L}_{ij} & & & & & &\\
    & & & & & & & & \Omega^{\bar{e}}_{ij} & \Xi^{\bar{e}}_{ij} & & & & \\
    & & & & & & & & \Xi^{\bar{E}}_{ij} & \Omega^{\bar{E}}_{ij} & & & &\\
    & & & & & & & & & &\Omega^{\bar{\nu}}_{ij}  & \Theta^{\bar{\nu}}_{ij} & \Upsilon^{\bar{\nu}}_{ij}\\
    & & & & & & & & & & \Theta^{S}_{ij}& \Omega^S_{ij} &\Lambda^S_{ij}\\
    & & & & & & & & & &\Upsilon^N_{ij} & \Lambda^N_{ij}& \Omega^N_{ij}
\end{array}
\right)
    \left(
    \begin{array}{c}
        q'_j\\
        Q'_j\\
        \bar{u}'_j\\
        \bar{U}'_j\\
        \bar{d}'_j\\
        \bar{D}'_j\\
        \ell'_j\\
        L'_j\\
        \bar{e}'_j\\
        \bar{E}'_j\\
        \bar{\nu}'_j\\
        S'_j\\
        N'_j
    \end{array}
    \right).
\end{align}
The $3 \times 3$ matrices $\Omega$, $\Xi$, $\Theta$, and $\Upsilon$ map between gauge and mass eigenstates.  

We go to a basis where $M$ and $\tilde{M}$ are diagonal (before considering mixing with components of $\psi$) and assume only the third generation of $\X$ and $\Y$ have mass eigenvalues near or below $v_R$. We then rescale $\X_{12}$ and $\Y_{12}$ to make the mass matrices equal to the third mass eigenvalue times the identity. The field redefinition modifies $\x$ and $\y$; we absorb this modification into its definition. 

We integrate out the heaviest two mass eigenstates resulting in seesaw contributions to Yukawa couplings and neutrino masses. We treat the third generation differently. After $SU(2)_R$ is broken we have
\begin{alignat}{4}
    \mathscr{L} \;\;\supset \;\; &\big[q_i \y_{Qi3}  v_R &&+  \tilde{M}_{Q3}Q_3 &&\big] \bar{Q}_3 \notag\\
    +&\big[\bar{u}_i \y_{Ui3}^*   v_R  &&+  \tilde{M}_{U3}\bar{U}_3 &&\big] U_3 \notag\\
    +&\big[\bar{d}_i x^*_{Di3}   v_R  &&+  M_{D3}\bar{D}_3 &&\big] D_3\notag\\
    +&\big[\ell_i x_{Li3} v_R   &&+ M_{L3} L_3 &&\big] \bar{L}_3\notag\\
    +&\big[ \bar{e}_i \y_{Ti3}^*  v_R &&+  \tilde{M}_{E3} \bar{E}_{3} &&\big] E_{3}\\
    +&\big[ \bar{\nu}_i \y_{Ti3}^*  v_R &&+  \frac{1}{2}\tilde{M}_{N3} N_{3} &&\big] N_{3} \\
    +& \big[ \bar{\nu}_i \y_{Si3}^*  v_R &&+  \frac{1}{2}\tilde{M}_{S3} S_{3} &&\big] S_{3}.
\end{alignat}
Heavy mass eigenstates with eigenvalues near $v_R$ are proportional to terms in square brackets. The other three orthogonal directions in field space correspond to the SM fermion mass eigenstates $(q'_i,\,\bar{u}'_i,\,\bar{d}'_i,\,\ell'_i,\,\bar{e}'_i)$ and the right handed neutrino mass eigenstates $\bar{\nu}'_i$. Clearly, heavy mass eigenstates can consist mostly of either states in $\X,\,\Y$ or $\psi$ depending on the relative size of $x_{i3}v_R$ and $M_3$.

The SM singlet and non-singlet fields are handled differently since there are three singlets, while non-singlets always come in pairs. Let us first consider non-singlet pairs $(q,Q),\,(\bar{u}, \bar{U}),\,(\bar{d}, \bar{D}),\,(\ell,L),\,(\bar{e}, \bar{E})$. We define 
\begin{align} \label{eq:delta_epsilon}
    \delta_{i} = \frac{x_{i3} v_R}{\sqrt{x_{i3}^* x_{i3} v_R^2 + M_3^2}},\qquad \epsilon = \frac{M_3}{\sqrt{x_{i3}^* x_{i3} v_R^2 + M_3^2}},\qquad \sum_i |\delta_i|^2 +\epsilon^2 = 1,
\end{align}
where values of $\delta_i$ and $\epsilon$ depend on the pair choice. In general $x_{i3}$ are complex and the transformation between gauge and mass eigenstates is complicated. However, if $x_{i3}$ are real (as expected in {\HPU} if dimension-5 operators of \EqRef{eq:Dim-5 Terms} are sufficiently suppressed) the transformation from gauge to mass eigenstates is
\begingroup
\renewcommand{\arraystretch}{1.3}
\begin{align}
\left( 
\begin{array}{c}
    \psi'_1\\
    \psi'_2\\
    \psi'_3\\
    X'_3
\end{array}
\right)
=
\left(
\begin{array}{cccc}
 \frac{\delta _1^2 \epsilon +\delta _2^2+\delta _3^2}{\delta _1^2+\delta _2^2+\delta _3^2} & -\frac{\delta _1 \delta _2 (1-\epsilon)}{\delta _1^2+\delta _2^2+\delta
   _3^2} & -\frac{\delta _1 \delta _3 (1-\epsilon)}{\delta _1^2+\delta _2^2+\delta _3^2} & -\delta _1 \\
 -\frac{\delta _1 \delta _2 (1-\epsilon)}{\delta _1^2+\delta _2^2+\delta _3^2} & \frac{\delta _2^2 \epsilon +\delta _1^2+\delta _3^2}{\delta _1^2+\delta _2^2+\delta
   _3^2} & -\frac{\delta _2 \delta _3 (1-\epsilon)}{\delta _1^2+\delta _2^2+\delta _3^2} & -\delta _2 \\
 -\frac{\delta _1 \delta _3 (1-\epsilon)}{\delta _1^2+\delta _2^2+\delta _3^2} & \frac{\delta _2 \delta _3 (1-\epsilon)}{\delta _1^2+\delta _2^2+\delta _3^2} &
   -\frac{\delta _3^2 \epsilon +\delta _1^2+\delta _2^2}{\delta _1^2+\delta _2^2+\delta _3^2} & -\delta _3 \\
 \delta _1 & \delta _2 & \delta _3 & \epsilon  \\
\end{array}
\right)
\left( 
\begin{array}{c}
    \psi_1\\
    \psi_2\\
    \psi_3\\
    X_3
\end{array}
\right).
\end{align}
\endgroup
Here $\psi$ and $X$ are dummy labels for any pair that mixes (e.g. $\psi \rightarrow q$ and $\X \rightarrow Q$).

\subsection{A Direct Third Generation from $\Y_3$ in $u/\nu$ Sectors.}

Since the top quark Yukawa coupling is near unity, $\tilde{M}_3$ must be close to $v_R$, as shown in \FigRef{fig:M_X}. Therefore it is likely the top quark has sizable components of $\psi_3$ and $\Y_3$. This also implies the third generation right-handed neutrino mass eigenstate, $\bar{\nu}'_3$, will likely have sizable contributions from $\bar{\nu}_3$, $S_3$ and $N_3$.

We assume $\delta_{1,2}$ of \EqRef{eq:delta_epsilon} can be ignored in the $u/\nu$ sector, as occurs if $SO(10)$-breaking terms involving $\Y$ are subdominant. We choose a basis where $\y$ is diagonal. Let us first consider the mixing between $(Q_3,q_3)$, $(\bar{U}_3,\bar{u}_3)$, and $(\bar{E}_3,\bar{e}_3)$. The transformation from gauge to mass (primed) eigenstates is
\begin{align}
&\begin{pmatrix}
        \psi_1'\\
        \psi'_2\\
        \psi'_3\\
        {\Y}'_3
    \end{pmatrix}
    =
    \begin{pmatrix}
        1&0 & 0&0 \\
        0& 1&  0 &0 \\
        0& 0& c_{\Y}& -s_{\Y}\\
        0& 0& s_{\Y}& c_{\Y} 
    \end{pmatrix}
    \begin{pmatrix}
        \psi_1\\
        \psi_2\\
        \psi_3\\
        {\Y}_3
    \end{pmatrix}.
\label{eq:Yrotation}
\end{align}
We use the notation where $(\Y, \psi)$ denotes any pair $(Q,q)$, $(\bar{U},\bar{u})$, and $(\bar{E},\bar{e})$ and we simplify the $\epsilon$ and $\delta_i$ notation to $c_{\Y} = \epsilon_{\Y}$ and $s_{\Y} = \delta_{\Y3}$, the cosine and sine of the single mixing angle in these $2 \times 2$ spaces where
\begin{align}
    t_{\Y} =\frac{s_{\Y}}{c_{\Y}} =   \frac{\y_3 v_R}{\tilde{M}_3}. \label{eq:t}
\end{align}
The transformation from gauge to mass eigenstates in the low energy theory after integrating out heavy mass eigenstates is 
\begin{alignat}{3}
&\psi =  {\cmat}^\psi \psi', \hspace{0.5in} 
    &&\cmat^\psi = 
    \begin{pmatrix}
        1 & 0 & 0  \\
        0 & 1 & 0  \\
        0 & 0 & c_{\Y}  
    \end{pmatrix},\label{eq:c_Y}\\
&\Y =  \Xi^{\Y} \psi' , \hspace{0.5in} 
    &&\Xi^{\Y} = 
     -s_{\Y} P_3 - (v_R/\tilde{M}_{\Y 3})P_{12}\y_{\Y}^{\top (\dagger)} \Omega^\psi ,
\end{alignat}
using $(\Y, \psi)$ to denote any pair $(Q,q)$, $(\bar{U},\bar{u})$, and $(\bar{E},\bar{e})$; $\y_{\Y}^{\top }$ is used for $\Y = Q$ and $\y_{\Y}^{\dagger }$ is used for $\Y = \bar{D},\, \bar{E}$. The matrix form of $\Xi^{\Y}$ is 
\begin{align}
    \Xi^{\Y} = 
    -\frac{v_R}{\tilde{M}_{\Y 3}}
    \begin{pmatrix}
        \x_{\Y11}^{(*)}& \x_{\Y21}^{(*)}& \x_{\Y31}^{(*)}c_{\Y}\\
        \x_{\Y12}^{(*)}& \x_{\Y22}^{(*)}&\x_{\Y32}^{(*)}c_{\Y} c_{\Y}\\
        0&0 & s_{\Y}
    \end{pmatrix},
\end{align}
where $(*)$ is for $\Y = \bar{D},\, \bar{E}$. 

Using the above results, the SM Yukawa couplings are
\begin{align}
    &Y_u =   -\Omega^{q\top} \y_U \Xi^{\bar{U}} - \Xi^{Q\top }\y_Q \Omega^{\bar{u}}\\
    &Y_d = -\Omega^{q\top} \x_D \Xi^{\bar{D}}\\
    &Y_e  = -  \Xi^{L\top} \x_L^\dagger  \Omega^{\bar{e}}.
\end{align}
Inserting expressions for the $\Xi$ transformations gives 
\begin{align}
    Y_u &= \left(\Omega^{q\top} \y_U P_3 s_{\bar{U}} + \frac{v_R}{\tilde{M}_{U3}}\Omega^{q\top} \y_U P_{12} \y_U^\dagger \Omega^{\bar{u}}\right) \notag \\
    &+ \left(s_{Q} P_3 \y_Q \Omega^{\bar{u}} + \frac{v_R}{\tilde{M}_{Q3}}\Omega^{q\top} \y_Q P_{12} \y_Q^\dagger \Omega^{\bar{u}}\right) \label{eq:Y_u_General} \\
    Y_d &=  \Omega^{q\top} \x_D P_3 +   \frac{v_R}{M_{D3}}\Omega^{q\top} \x_D P_{12} x_D^\dagger  \Omega^{\bar{d}} \label{eq:Y_d_General_App}\\
    Y_e &=  P_3 x_L^\dagger \Omega^{\bar{e}} +  \frac{v_R}{M_{L3}}\Omega^{\ell \top } x_L P_{12} x_L^\dagger \Omega^{\bar{e}}. \label{eq:Y_e_General_App}
\end{align}

The SM singlets have the following mass matrix:
\begin{align}
    \mathcal{L} \supset \frac{1}{2}
    \begin{pmatrix}
    \bar{\nu}_3 & S_3 & N_3    
    \end{pmatrix}
    \begin{pmatrix}
    0 & \y_{S33}v_R & \y_{T33}v_R\\
    \y_{S33}v_R & \tilde{M}_{S3} & 0\\
    \y_{T33}v_R & 0 & \tilde{M}_{T3}
    \end{pmatrix}
    \begin{pmatrix}
    \bar{\nu}_3 \\ S_3 \\ N_3    
    \end{pmatrix}.
\end{align}
It is convenient to introduce the notation $\mathcal{N} = (\bar{\nu}_3,S_3,N_3)^\top$ and the relation between gauge and mass eigenstates as $\mathcal{N} = \mathcal{R} \mathcal{N}'$. We calculate $\mathcal{R}$ in \SecRef{sec:The SO(10) Preserving Limit for the (u,nu) Sectors} for the $SO(10)$ preserving limit where it takes a simple form.

The effective theory of singlets is interesting at several energy scales; for example, the EFT for leptogenesis contains $\bar{\nu}'_\alpha, \,\, \alpha = 1,2$ but integrates out $S'_i$, $N'_i$, and $\bar{\nu}'_3$. 
\begin{align}\label{eq:L_Lepto}
    &\mathcal{L}_{\rm Lepto} = \frac{\mathcal{M}_{\nu,{\rm Lepto}}^{ij}}{v_L^2}(\ell'_i \HL)(\ell'_j \HL)  + Y_{\nu,{\rm Lepto}}^{i\alpha }\ell'_i \HL \bar{\nu}'_\alpha + \mathcal{M}_{\bar{\nu},{\rm Lepto}}^{\alpha \beta}\bar{\nu}'_\alpha \bar{\nu}'_\beta.
\end{align}
The EFT describing SM neutrinos after $\bar{\nu}_1'$ and $\bar{\nu}_2'$ are integrated out is
\begin{align} \label{eq:L_nu_SM}
    &\mathcal{L}_{\nu,\, \rm SM} = \frac{\mathcal{M}_{\nu}^{ij}}{v_L^2}(\ell'_i \HL)(\ell'_j \HL),\\
    &\mathcal{M}_{\nu} = \mathcal{M}_{\nu,{\rm Lepto}} - Y_{\nu,{\rm Lepto}} \mathcal{M}_{\bar{\nu},{\rm Lepto}}^{-1} Y_{\nu,{\rm Lepto}}^\dagger v_L^2. \label{eq:M_nu_Formula}
\end{align}
We match from the UV theory to the effective theory to calculate $\mathcal{M}_{\bar{\nu},{\rm Lepto}}^{\alpha \beta}$, $Y_{\nu,{\rm Lepto}}^{i\alpha}$, and $\mathcal{M}_{\nu,{\rm Lepto}}^{ij}$ in the next subsection for the $SO(10)$ preserving limit.

\begin{figure}[t]
  \centering
    \begin{tikzpicture}
    \draw[black, ultra thick, ->] (0,-0.5) -- (0,4.5) ;
    \node[rotate = 90] at (-0.5,4) {Energy};

    \draw[ ultra thick, dashed, MyGreen] (.5,3.8)--(5.95,3.8) node[right,yshift = 6pt,xshift = -90pt]  {};
    \draw[ ultra thick, MyBlue] (.5,3.6)--(5.95,3.6) node[right,yshift = -6pt,xshift = 5pt]  {$M_{\bar{\nu}'3}$};
    \draw[ ultra thick, MyBlue] (.5,2)--(5.95,2) node[right,yshift = 0pt,xshift = 5pt]  {$M_{\bar{\nu}'2}$};
    \draw[ ultra thick, MyBlue] (.5,1)--(5.95,1) node[right,yshift = 0pt, xshift = 5pt]  {$M_{\bar{\nu}'1}$};
    \draw[ ultra thick, MyOrange] (.5,3.9)--(5.95,3.9) node[right,yshift = 0pt,xshift = 5pt]  {$M_{S'3}$};
    \draw[ ultra thick, MyOrange] (.5,4.2)--(5.95,4.2) node[right,yshift = 6pt,xshift = 5pt]  {$M_{N'3}$};

    \draw[black, ultra thick, <->] (1,-.5) -- (1,.9) ;
    \draw[black, ultra thick, <->] (1,2.1) -- (1,3.5) ;
    \draw node at (1.8,0.1) {$\mathcal{L}_{\nu, \rm SM}$};
    \draw node at (1.8,2.75) {$\mathcal{L}_{\rm Lepto}$};
    \end{tikzpicture}
  \caption{Schematic illustration of SM singlet mass spectrum. (For clarity, $ N'_{1,2}$ and $S'_{1,2}$ are not shown.) The dashed line is $\y_3v_R$. Also shown are ranges of validity for EFTs relevant for leptogenesis \EqRef{eq:L_Lepto} and SM neutrino masses \EqRef{eq:L_nu_SM}.}
  \label{fig:M_neutrinos}
\end{figure}
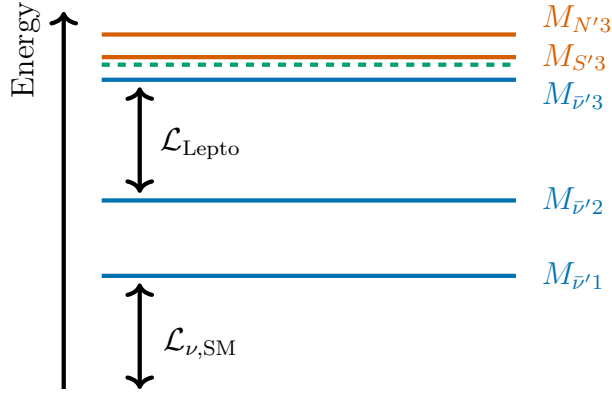

\subsection{$SO(10)$ Preserving $(u,\nu)$ Sectors with Direct Third Generations} \label{sec:The SO(10) Preserving Limit for the (u,nu) Sectors}

We now study the modifications to the $(u,\nu)$ sectors from a direct third generation while ignoring $SO(10)$ breaking effects where the above results for $Y_{u,d,e}$ and $\mathcal{M}_{\nu}$ have simple forms. In this limit $\Xi^{\Y}$ is
\begin{align}
    \Xi^{\Y} = 
    -
    \begin{pmatrix}
        \y_1 & 0 & 0\\
        0 & \y_2 & 0 \\
        0 & 0 & s
    \end{pmatrix},\qquad \Y = Q,\,\bar{U},\,\bar{D}
\end{align}
where $s$ is defined in \EqRef{eq:t} and we have dropped the subscript $\Y$ from $\y$ and $s$, as they are now identical for $Q,\,\bar{U},\,\bar{D}$. The up-type quark Yukawa coupling becomes
\begin{align} \label{eq:Y_u_With_Mixing}
    &Y_u = 
    \frac{v_R}{\tilde{M}_3}\begin{pmatrix}
    2\y_1^2& 0 & 0 \\
    0&2\y_2^2 & 0\\
    0 & 0 &   2 \y_3^2 c^2   
    \end{pmatrix}.
\end{align}

The $SO(10)$ preserving limit has $\y_S = -\frac{\sqrt{3}}{2}\y$, $\y_T = \frac{1}{\sqrt{2}}\y$ and $\tilde{M}_{S3}=\tilde{M}_{T3} = \tilde{M}_3$ such that the rotation from the gauge to mass eigenstate basis for SM singlets is
\begin{align}
&\mathcal{R}= 
\begin{pmatrix}
1 & 0 & 0\\
0 & \cos(\theta_{23}) & \sin (\theta_{23})\\
0 & -\sin (\theta_{23}) & \cos (\theta_{23})
\end{pmatrix}
\begin{pmatrix}
 \cos(\theta_{12}) & \sin (\theta_{12}) & 0\\
 -\sin (\theta_{12}) & \cos (\theta_{12}) & 0\\
 0 & 0 & 1
\end{pmatrix}
\end{align}
where $\tan(\theta_{23}) = \y_{S3}/\y_{T3}$ and $\tan(2\theta_{12}) = 2\sqrt{\y_{S3}^2+\y_{T3}^2}v_R/\tilde{M}_3$.
The mass eigenvalues are 
\begin{align}
    &\tilde{M}_{\bar{\nu}'_3\,(N'_3)} = \tilde{M}_3\left(\frac{1 \mp \sqrt{1+4(\y_{S3}^2+\y_{T3}^2)v_R^2/\tilde{M}_3^2} }{2}\right),\qquad 
    \tilde{M}_{S'_3} = \tilde{M}_3.
\end{align}
We can match the UV theory with the effective theory described by $\mathcal{L}_{\rm Lepto}$ from \EqRef{eq:L_Lepto}. The mass matrix for $(\bar{\nu}_1, \bar{\nu}_2)$ from integrating out $S'_\alpha$ and $N'_\alpha$ is
\begin{align}\label{eq:M_nu_bar_Lepto}
    \mathcal{M}_{\bar{\nu},{\rm Lepto}} &= 
    \frac{5}{4}\frac{v_R^2}{\tilde{M}_3}
    \begin{pmatrix}
    \y_1^2 &                  \\
                      &\y_2^2 
    \end{pmatrix}.
\end{align}
The effective neutrino Yukawa coupling is obtained in the same way and is 
\begin{align}\label{eq:Y_nu_Lepto}
    Y_{\nu,{\rm Lepto}} = 
    \frac{3v_R}{4\tilde{M}_3}\Omega_\ell^\top\begin{pmatrix}
    \y_1^2 & 0 \\
    0 & \y_2^2\\
    0 & 0
    \end{pmatrix} = \frac{3v_R}{4\tilde{M}_3} \begin{pmatrix}
    \y_1^2 & 0 \\
    0 & \y_2^2\\
    0 & 0
    \end{pmatrix}.
\end{align}
Finally, the neutrino mass matrix reduces to
\begin{align} \label{eq:M_nu_Lepto}
    \mathcal{M}_{\nu,{\rm Lepto}} &= \frac{5}{4}\frac{v_L^2}{\tilde{M}_3}\begin{pmatrix}
    1& & \delta_1\\
    &1 & \delta_2 \\
    & &\epsilon 
    \end{pmatrix}
    \begin{pmatrix}
    \y_1^2& & \\
    &\y_2^2 & \\
    & & \y_3^2   f_{\mathcal{N}}
    \end{pmatrix}
    \begin{pmatrix}
    1& &  \\
    &1 &  \\
    \delta_1&\delta_2 &\epsilon 
    \end{pmatrix}
\end{align}
where the factor $f_\mathcal{N}$ originates from mixing of the third generation
\begin{align} \label{eq:f_N}
    f_{\mathcal{N}}&=\frac{4}{5}\left(\y_{T3}^2+ \tilde{M}_3\y_{S3}^2  (\tilde{M}_{\mathcal{N}}^{-1})_{22} \right)
    = \frac{4}{5}\left(\y_{T3}^2 + \frac{\y_{T3}^2 \y_{S3}^2}{\y_{T3}^2 + \y_{S3}^2}\right) =\frac{16}{25} .
\end{align}

Finally, using \EqRef{eq:M_nu_bar_Lepto}, \EqRef{eq:Y_nu_Lepto}, and \EqRef{eq:M_nu_Lepto}, we can calculate the neutrino masses at low energy using \EqRef{eq:M_nu_Formula}. The result simplifies to
\begin{align}\label{eq:M_nu_Full}
    \mathcal{M}_\nu &= \frac{4}{5}\frac{v_L^2}{\tilde{M}_3}\begin{pmatrix}
    1& & \delta_1\\
    &1 & \delta_2 \\
    & &\epsilon 
    \end{pmatrix}
    \begin{pmatrix}
    \y_1^2  & & \\
    &\y_2^2    & \\
    & & \y_3^2  
    \end{pmatrix}
    \begin{pmatrix}
    1& &  \\
    &1 &  \\
    \delta_1&\delta_2 &\epsilon 
    \end{pmatrix}.
\end{align}
Therefore, the low energy neutrino mass is exactly the same as in the case where the third generation of $(u,\nu)$ sector comes from a pure seesaw. Since the top Yukawa is generated directly, relations between the $u$ and $\nu$ sectors are broken for the third generation. In the see-saw limit where $\tilde{M}_3\gg \y_3 v_R$ we return to the relation 
\begin{align}
    \mathcal{M}_\nu \; \xrightarrow[\tilde{M}\gg \y_3 v_R]{} \; \frac{2}{5} \frac{v_L}{v_R} \; \Omega_\ell^\top \mathcal{M}_u \,\Omega_\ell.
\end{align}

\section{Threshold Corrections to $\lambda(v_R)$ } \label{app:Threshold Corrections}

In this appendix, we describe 1-loop threshold corrections to the Higgs quartic coupling which are essential to accurately prediction the scale of $SU(2)_R$ breaking, $v_R$, in the Higgs Parity mechanism. Additionally, we discuss the model dependence of these threshold corrections. We argue the flavor structure in the up-quark and neutrino sectors greatly reduces the model dependence. 

%\subsection{Matching the Coleman-Weinberg Effective Potential}

In \SecRef{subsec:Higgs Parity Mechanism}, we described how the Higgs Parity mechanism requires the SM Higgs quartic coupling to vanish at the scale $v_R$ by matching the tree-level scalar potentials of the UV and IR theories at the scale $v_R$. More precisely, we must match the potential including 1-loop corrections. The Coleman-Weinberg 1-loop effective potentials in $\overline{\rm MS}$ for the UV and IR theories are 
\begin{align}
    &\Delta V =\frac{1}{64 \pi^2} \sum_i(-1)^{F_i} \, n_i \; \mathcal{M}_i^4 \; \left[\log \left(\frac{\mathcal{M}_i^2}{\mu^2}\right) - c_i\right].
\end{align}
Here $i$ enumerates fields that get masses from $H_{L,R}$, and $F_i=0$ $(1)$ for scalars (fermions), $\mathcal{M}_i$ are the field-dependent effective masses, and $n_i$ are the number of degrees of freedom. The coefficients $c_i $ are renormalization scheme dependent constants which in $\overline{\rm MS}$ are $ 3/2$ for both scalars and fermions and $ 5/6$ for vectors. 

We now proceed to match parameters of the UV and IR effective actions.

\subsubsection*{\emph{Correction from Gauge Bosons}}
We first consider 1-loop corrections from gauge bosons. In the UV theory the vacuum values of $H_L$ and $H_R$ will generate effective masses for left and right handed charged gauge bosons and lead to a correction

\begin{align}
    \Delta V_{{\rm UV},\, W} = a |H_L|^4 \left[\log\left(\frac{b|H_L|^2}{\mu^2}\right) - \frac{5}{6}\right]+ (L\leftrightarrow R).
\end{align}
where $a = \frac{6}{64\pi^2} \left(\frac{g^2}{2}\right)^2$ and $b = g^2/2$. %\KH{$a=\frac{1}{64\pi^2}*6*(g^2/2)^2=3g^4/(128\pi^2)$. $b=g^2/2$.} 
We replace $H_R \rightarrow v_R + h_R/\sqrt{2}$ and integrate out $h_R$ using the equations of motion. We can solve for $\lambda'$ by setting the mass term for $H_L$ in the effective Lagrangian to zero. To leading order in $a$ we get
\begin{align}
     V_{{\rm eff},\,W}(H_L)/|H_L|^4  \approx \frac{a}{2}\left[1+2 
 \log\left(\frac{|H_L|^2}{v_R^2}\right) \right]. 
\end{align}
We match at scale $\mu = v_R$ to the IR theory with potential
\begin{align}
    V_{{\rm IR},\, W} (H_L)/|H_L|^4 &\approx  \lambda_{\rm SM}(v_R) + a\left(\ln \frac{b|H_L|^2}{v_R^2}-\frac{5}{6}\right),
\end{align}
which results in the SM quartic
\begin{align}
    \lambda_{{\rm SM},\, W}(v_R) = a \log\left(\frac{e^{4/3}}{b}\right) = \frac{3 g^4}{64 \pi^2} \log\left(\frac{\sqrt{2}\, e^{2/3}}{g}\right) 
\end{align}
Heavy neutral gauge bosons contribute to threshold corrections with squared masses  
\begin{align}
   \mathcal{M}^2_{Z,\pm} = \frac{(g^2+g_{B-L}^2) \left(|\HL|^2+|\HR|^2\right)}{4} \left(1 \pm \sqrt{1-\frac{4 \left(g^4+2g^2
   g_{B-L}^2\right)  }{\left(g^2+g_{B-L}^2\right)^2
   } \frac{|\HL|^2 |\HR|^2}{\left(|\HL|^2+|\HR|^2\right)^2}}\right)
\end{align}
Performing the same procedure leads to 
\begin{align}
     V_{{\rm eff},\,Z}(H_L)/|H_L|^4  \approx \frac{3\left(g^2+g'^2\right)^2}{256 \pi^2}    \log \left(\frac{e^{1/2}|\HL|^2 \left(g^4-g'^4\right)}{g^4 v_R^2}\right). 
\end{align}
where we used $g' = g g_{B-L}/\sqrt{g^2 + g_{B-L}^2}$. We match with
\begin{align}
    V_{{\rm IR},\, Z} (H_L)/|H_L|^4 &\approx  \lambda_{\rm SM}(v_R) + \frac{3(g^2+{g'}^2)^2}{256\pi^2}\left[\ln \left(\frac{(g^2+{g'}^2)|H_L|^2}{2v_R^2}\right)-\frac{5}{6}\right].
\end{align}
which leads to the threshold correction 
\begin{align}
    \lambda_{{\rm SM},\, Z}(v_R) = \frac{3 \left(g^2+{g'}^2\right)^2 }{256 \pi^2} \log \left(\frac{2 e^{4/3} \left(g^4-{g'}^4\right)}{g^4 \left(g^2+{g'}^2\right)}\right)
\end{align}

\subsubsection*{\emph{Corrections from Top quarks}}

\begin{figure}
    \centering
    \includegraphics[width=1\linewidth]{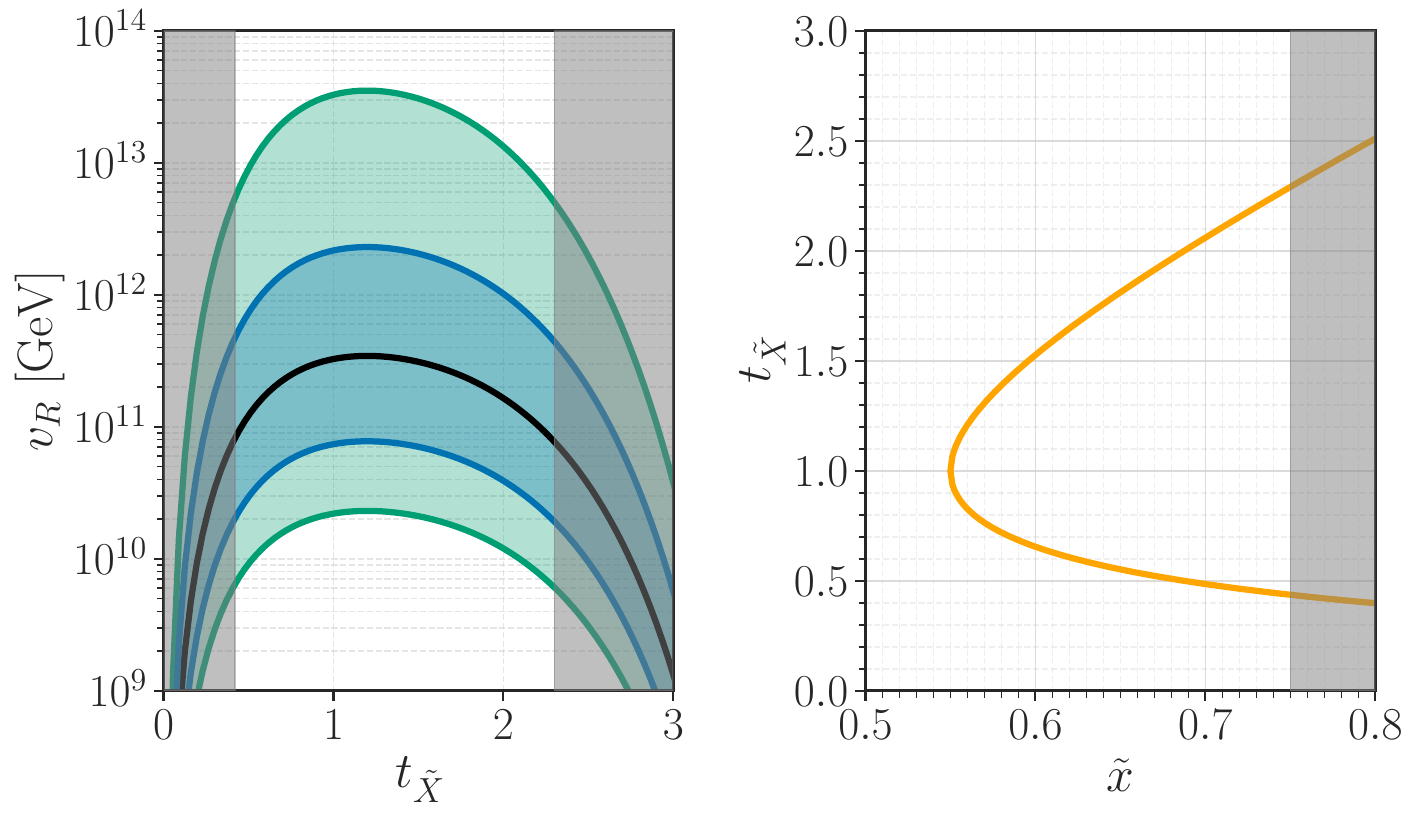}
    \caption{On the right we show the two possible values of $t_{\tilde{X}}$ as a function of $\tilde{x}$. If $\tilde{x}_{Q,U} \gtrsim 0.75$ there is a Landau pole before the Planck scale. }
    \label{fig:vR_vs_t}
\end{figure}

The large value of top quark Yukawa coupling requires $\tilde{x}_{33}$ to be large; this leads to additional threshold corrections to the Higgs quartic which we now discuss. We assume unified relations $\y_U= \y_Q\equiv \y$ and $\tilde{M}_U=\tilde{M}_Q$ and drop the $33$ subscript from now on as this is the only entry which is relevant. The threshold correction coming from the top Yukawa coupling is
\begin{align}
    \lambda_{{\rm SM},\,t}(v_R) = & - \frac{3 y_t^4}{8\pi^2} \left( {\rm ln} \frac{e}{y_t} + f(t_{\tilde{X}}) \right), \\
    f(t) =& -\frac{ \left(t^2+2\right) \left(t^2+1\right)^4 \log \left(t^2+1\right)}{32t^6}+\frac{
   \left(t^2+1\right)^4}{16t^4}+ \frac{1}{2} \log \left(\frac{4 t^4}{\left(t^2+1\right)^3}\right)+\frac{5}{12}, \nonumber \\
   t_{\tilde{X}} \, \equiv & \, \frac{\y_3 v_R}{\tilde{M}_3} =  \; \frac{y_t}{\y} \; \frac{1}{1 \pm \sqrt{1 - y_t^2 /\y^2}}. 
\end{align}
Here we adapted results in~\cite{Hall:2019qwx}. Note $f(1)=0$, so results in the main text corresponds to $t_{\tilde{X}}=1$.
In \FigRef{fig:vR_vs_t}, we show the prediction
for $v_R$ as a function of $t_{\tilde{X}}$, fixing $\alpha_s(m_Z) =0.1179$ and $m_t=172.57$ GeV. 

\subsubsection*{\emph{Uncertainty from the top sector}} \label{app:Uncertainty from the top sector}

\begin{figure}[t]
    \centering
    \includegraphics[width=1\linewidth]{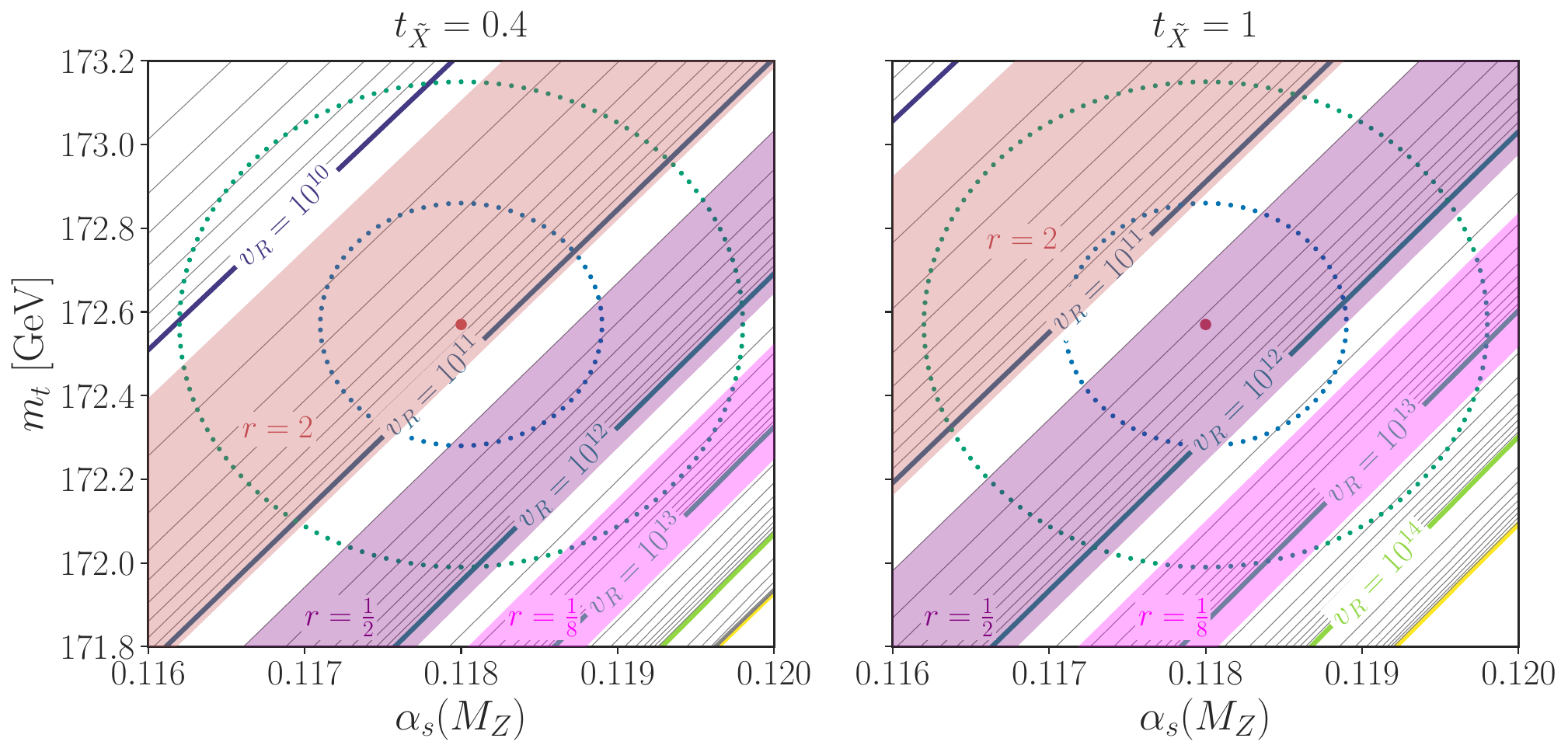}
    \caption{Modification to the prediction for $v_R$ in {\HPU} from $t_{\tilde{X}}$. The right side for $t_{\tilde{X}} = 1$ predicts essentially the largest values of $v_R$ varying over just $t_{\tilde{X}}$. The left side for $t_{\tilde{X}} = 0.4$ predicts essentially the smallest values of $v_R$ varying over just $t_{\tilde{X}}$ while still resulting in perturbative $\tilde{x}$ up to the Planck scale.  } 
    \label{fig:t_dependence}
\end{figure}

Values of $t_{\tilde{X}}$ far from $1$ require large $\tilde{x}$ and may lead to a Landau pole below the Planck scale, or even below the unification scale. 
Ignoring terms involving the $x$ couplings, the RGE of the Yukawa couplings $\tilde{x}_Q$, $\tilde{x}_U$, $\tilde{x}_T$, and $\tilde{x}_S$ are 
\begin{align}
    16\pi^2 \frac{{\rm d} \log \tilde{x}_Q}{{\rm d ln} \mu} &= \frac{15}{2} \tilde{x}_Q^2 + \frac{7}{2} \tilde{x}_U^2+ 3 \tilde{x}_T^2 +  \tilde{x}_S^2 - 8 g_3^2 - \frac{27}{4}g_2^2 - \frac{5}{8} g_{B-L}^2, \\
    16\pi^2 \frac{{\rm d} \log \tilde{x}_U}{{\rm d ln} \mu} &=  7 \tilde{x}_Q^2 + \frac{9}{2} \tilde{x}_U^2+ 3 \tilde{x}_T^2 +  \tilde{x}_S^2 - 8 g_3^2 - \frac{9}{4}g_2^2 - \frac{17}{8} g_{B-L}^2,\\
   16\pi^2 \frac{{\rm d} \log \tilde{x}_T}{{\rm d ln} \mu} &= 6 \tilde{x}_Q^2 + 3 \tilde{x}_U^2+ \frac{11}{2} \tilde{x}_T^2 +  \frac{11}{6}\tilde{x}_S^2  - \frac{33}{4}g_2^2 - \frac{9}{8} g_{B-L}^2, \\
    16\pi^2\frac{{\rm d} \log \tilde{x}_S}{{\rm d ln} \mu} &= 6 \tilde{x}_Q^2 + 3 \tilde{x}_U^2+ \frac{11}{2} \tilde{x}_T^2 +  \frac{7}{2}\tilde{x}_S^2  - \frac{9}{4}g_2^2 - \frac{9}{8} g_{B-L}^2,
\end{align}
where $g_{B-L}$ is the coupling associated with the properly normalized $U(1)_{B-L}$ gauge symmetry. The RGE above the unification scale depends on some assumptions. We assume couplings of the form $\X \Sigma \X$, $\X \mathcal{S} \X$ and $\X \rightarrow \Y$ are negligible. This is justified as they are assumed to generated masses well below the GUT scale. 
Therefore, we assume the only large Yukawa coupling is $\tilde{x}$. Running in the $SO(10)$ theory can be obtained by evaluating running of a single term in the $SO(10)$ limit and include contributions from the colored Higgs fields. For example, the running of the Yukawa and gauge couplings are of the form
\begin{align}
     \frac{16\pi^2}{\tilde{x}} \frac{{\rm d} \tilde{x}}{{\rm d ln} \mu} &=  \frac{151}{8} \tilde{x}^2   - \frac{327}{8}g_{10}^2.\\
     16\pi^2 \frac{{\rm d} g_{10}}{{\rm d ln} \mu} &=  18\, g_{10}^3,
\end{align}
where for the running of $g_{10}$ we include contributions from all scalars above the GUT scale; the final result is not very sensitive to exactly what scale these are included as the fermions dominate the running. Requiring $\tilde{x}< \pi$ at the Planck scale, we find $\tilde{x}_{Q,U}\lesssim 0.75$ at $v_R = 10^{12-13}$ GeV. This bound corresponds to $t_{\tilde{X}} \gtrsim 0.4$.
In \FigRef{fig:t_dependence}, we show the prediction for $v_R$ as a function of $m_t$ and $\alpha_s(m_Z)$ for $t_{\tilde{X}} = 0.4$ and $t_{\tilde{X}} = 1$. 
Reducing $t_{\tilde{X}}$ from 1 to 0.4 reduces the Higgs Parity prediction for $v_R$ by about a factor of 4. 
The prediction for $\alpha_s(M_Z)$ from gauge coupling unification is increased by about 0.001.

\bibliographystyle{JHEP}
\bibliography{biblio.bib}

\end{document}